\documentclass[prb,aps,twocolumn,floatfix,showpacs]{revtex4}
\usepackage{graphics}
\begin{document}
\title{Superfluidity of $p$-wave and $s$-wave atomic Fermi gases in optical lattices}
\author{M. Iskin and C. A. R. S{\'a} de Melo}
\affiliation{School of Physics, Georgia Institute of Technology, Atlanta, Georgia 30332, USA}
\date{\today}

\begin{abstract}
We consider $p$-wave pairing of single hyperfine state and
$s$-wave pairing of two hyperfine states ultracold atomic gases trapped 
in quasi-two-dimensional optical lattices. 
First, we analyse superfluid properties of $p$-wave and $s$-wave 
symmetries in the strictly weak coupling BCS regime where we discuss the order parameter, 
chemical potential, critical temperature, atomic compressibility and superfluid density 
as a function of filling factor for tetragonal and orthorhombic optical lattices.
Second, we analyse superfluid properties of $p$-wave and $s$-wave superfluids 
in the evolution from BCS to BEC regime at low temperatures ($T \approx 0$), where
we discuss the changes in the quasi-particle excitation spectrum, chemical potential, 
atomic compressibility, Cooper pair size and momentum distribution
as a function of filling factor and interaction strength for tetragonal and orthorhombic optical lattices.

\end{abstract}
\maketitle

\section{Introduction}
\label{sec:introduction}

Tunable optical lattices have been extensively used to study phase transitions 
in bosonic atomic gases~\cite{opt4,greiner2}, 
since they allow the controlled manipulation of the particle density $n$, 
and of the ratio between the particle transfer energy $t$, and the interparticle interaction 
strength $V$~\cite{regal1,ohara}. 
This kind of control is not fully present in standard fermionic condensed 
matter systems, and has hindered the development of experiments that could probe systematically 
the effects of strong correlations as a function of $n$ and $t/V$. However, fermionic atomic gases like
$^6{\rm Li}$ and $^{40}{\rm K}$ have been succesfully trapped,
and their normal state and superfluid properties are beginning 
to be studied~\cite{modugno,kohl,jaksch,hofsfetter,torma1}. 
Because of the greater tunability of experimental parameters, 
novel superfluid phases may be more easily
accessible in the experiments involving ultracold atomic gases. 
For instance, single hyperfine state (SHS) ultracold atomic systems 
are ideal candidates for the observation of novel triplet superfluid phases
and for testing theoretical models that were proposed earlier.
Thus, it is only natural to propose that optical lattices could be used to study the normal state
and superfluid properties of ultracold fermionic systems as a function of $n$, $t/V$ and
lattice symmetry. These systems are of broad interest not only for the
atomic physics community but also for the nuclear, condensed matter 
and more generally for the many-body physics communities,
where models for superfluidity have been investigated in various contexts.

The interaction between induced dipole moments of atoms and the electric field 
of laser beams is used to trap atoms, particularly alkali atoms, in optical lattices.
Alkali atoms have only one electron ($S = 1/2$) out of closed shells. This electron is in a zero orbital 
angular momentum $(L = 0)$ channel, and its total angular momentum
$\mathbf{J}=\mathbf{L}+\mathbf{S}$ gives $J = 1/2$.
The nuclear angular momentum $I$ and electron angular momentum $J$ are combined in a hyperfine state with total 
angular momentum $\mathbf{F} = \mathbf{I} + \mathbf{J}$ which gives $F=I\pm 1/2$ for alkalis.
Furthermore, the electron and nuclear spins are coupled by the hyperfine interaction that splits 
the atomic levels in the absense of magnetic field $H_{\rm{hf}} \propto \mathbf{I}\cdot\mathbf{J}$.
A weak magnetic field causes Zeeman splitting of the hyperfine levels $|F,m_F>$ with different $m_F$, 
which can be made to correspond to pseudo-spin labels.
Trapping one~\cite{botelho1,iskin,botelho2,ohashi,skyip,iskin3D}, two~\cite{one1,one3,gurarie,tlho}, three~\cite{modawi} 
and four~\cite{torma} hyperfine states were considered by several authors.
However, it is also known that trapping more hyperfine states increases the 
number of channels through which the gas can decay. 
Therefore, trapping one or two hyperfine states of ultracold fermionic gases
is experimentally more plausible.

For paired identical fermions, the Pauli exclusion principle requires the total pair
wave function to be anti-symmetric. The total orbital angular momentum 
should be odd for pseudo-spin symmetric pairs and even for pseudo-spin anti-symmetric ones.
Therefore in the case of trapping two hyperfine states (THS), 
$s$-wave scattering of atoms between fermions from different hyperfine states is dominant. 
Thus, one expects that the superfluid ground state of such two-component Fermi gases 
to be $s$-wave and pseudo-spin singlet. 
Presently there is experimental evidence that 
$^{40} {\rm K}$ (Ref.~\cite{regal2,greiner}) and $^6{\rm Li}$ (Ref.~\cite{hulet,litium1,litium2,litium3,kinast}) 
can form weakly and tightly bound atom pairs, when the magnetic field is swept through an $s$-wave Feshbach resonance.

However, the properties of SHS ultracold fermions 
and their possible superfluid behaviour are beginning 
to be investigated~\cite{regal3,ticknor,zhang,schunck,gunter}. These systems
are probably the next frontier for experiments with ultracold atoms.
When identical fermionic atoms are trapped in a single hyperfine state, 
the interaction between them is strongly influenced by the Pauli exclusion principle, 
which prohibits $s$-wave scattering of atoms in identical pseudo-spin states.
As a result, in SHS degenerate Fermi gases, two fermions 
can interact with each other at best via $p$-wave scattering. 
Thus, one expects that the superfluid ground state of such SHS Fermi 
gases to be $p$-wave and pseudo-spin triplet. 

In the $p$-wave channel, if the atom-atom interactions are effectively attractive then 
the onset for the formation of Cooper-pairs in three dimensions occurs at a temperature~\cite{john,iskin3D}  
$T_c\approx \epsilon_F\exp[-\pi/2(k_F a_{\rm{sc}})^3]$ in the BCS regime,
where $\epsilon_F$ is the Fermi energy and $a_{\rm{sc}}$ is the $p$-wave scattering length. 
Unfortunately, this temperature is too low to be observed experimentally. However, 
in the presence of Feshbach resonances~\cite{john,regal3}, $p$-wave interactions can be enhanced, and
the critical temperature for superfluidity is expected to increase to experimentally accessible values. 
On the other hand, we show that the pseudo-spin triplet ($p$-wave) weak coupling limit in optical lattices 
(like in the singlet case~\cite{hofsfetter}) may be sufficient to produce a superfluid critical temperature 
that is accessible experimentally. Furthermore, several interesting superfluid properties can be investigated
in $p$-wave systems as the system is tuned from the BCS to the 
BEC regimes~\cite{john,botelho1,iskin,botelho2,ohashi,skyip,iskin3D,gurarie,tlho}. 
In this paper, we address superfluid properties of ultracold 
fermionic atoms in optical lattices for both SHS $p$-wave and THS $s$-wave states 
as a function of atom filling factor, interaction strength, temperature and lattice symmetry.

The rest of the paper is organized as follows. In Sec.~\ref{sec:hamiltonian}, 
we discuss the effective action method and the saddle point approximation for 
a lattice Hamiltonian with attractive interactions in the $p$-wave and $s$-wave  
channels. We also derive the order parameter, critical temperature and number equations
in the saddle point approximation and discuss pseudo-spin triplet states in the
$d$-vector formalism in this section.
In Sec.~\ref{sec:bcs}, we analyse superfluid properties of SHS $p$-wave and 
THS $s$-wave symmetries 
in the strictly weak coupling BCS regime. There, we discuss the order parameter, chemical potential, 
critical temperature, atomic compressibility, and superfluid density as a function of filling factor 
for tetragonal and orthorhombic optical lattices.
In Sec.~\ref{sec:BCS-BEC}, we analyse superfluid properties of SHS $p$-wave and $s$-wave superfluids 
in the evolution from BCS to BEC regime at low temperatures ($T \approx 0$). There,
we discuss  changes in quasi-particle excitation spectrum, chemical potential, 
atomic compressibility, Cooper pair size and momentum distribution as a 
function of filling factor and interaction strength 
for tetragonal and orthorhombic optical lattices. 
Finally, a summary of our conclusions is given in Sec.~\ref{sec:conclusions}.

\section{Lattice Hamiltonian}
\label{sec:hamiltonian}

In this manuscript, we consider quasi-two-dimensional optical lattices
with a periodic trapping potential of the form $U(r)=\sum_{i} U_{0,i}\cos^2(k_ix_i)$,
with $U_{0,z}\gg \min \{U_{0,x},U_{0,y} \}$, which strongly suppresses tunneling along the $\hat{\mathbf{z}}$ direction. This is 
a non-essential assumption, which just simplifies the calculations, but still describes an experimentally relevant
situation. Here  $x_i = x, y,~{\rm or}~ z$ labels the spatial coordinates, $k_i=2\pi/\lambda_i$ is the wavelength, 
and $U_{0,i}$  is the potential well depth along direction $\hat{\mathbf{x}}_i$, respectively.
The parameters $U_{0,i}$ are proportional to the laser intensity along each direction, and it is typically several 
times the one photon recoil energy $E_R$ such that tunneling is small and the tight-binding approximation can be used.

Thus, in the presence of magnetic field $\mathbf{h}$, we consider the following quasi-two-dimensional lattice 
Hamiltonian (already in momentum space) for an SHS $p$-wave Fermi gas
\begin{eqnarray}
\label{eqn:hamiltonian}
H=\sum_{\mathbf{k}}\xi(\mathbf{k})a_{\mathbf{k}\uparrow}^\dagger a_{\mathbf{k}\uparrow} + 
\frac{1}{2}\sum_{\mathbf{k},\mathbf{k'},\mathbf{q}}V_p(\mathbf{k},\mathbf{k'}) b_{\mathbf{k},\mathbf{q}}^\dagger b_{\mathbf{k'},\mathbf{q}}, 
\end{eqnarray}
where the pseudo-spin $\uparrow$ labels the trapped hyperfine state represented by 
the creation operator $ a_{\mathbf{k}\uparrow}^\dagger$, and
$b_{\mathbf{k},\mathbf{q}}^\dagger=a_{\mathbf{k}+\mathbf{q}/2,\uparrow}^\dagger a_{-\mathbf{k}+\mathbf{q}/2,\uparrow}^\dagger$.
Furthermore, 
$\xi(\mathbf{k})=\varepsilon(\mathbf{k})-\tilde{\mu}$ describes the tight-binding dispersion 
$\varepsilon(\mathbf{k}) = - t_x \cos (k_xa_x) - t_y \cos (k_ya_y) - t_z \cos (k_za_z) $, 
with $\tilde{\mu}=\mu+g\mu_Bh - V_H$.  Here, $t_i$ is a tranfer energy along 
the $i$-th direction, $\mu$ is the chemical potential, $g\mu_B h$ is
a Zeeman energy, $V_H$ is a possible Hartree energy shift. 
We also assume that $\min \{t_x, t_y\} \gg t_z$, and that $V(\mathbf{k},\mathbf{k'})$
is the pseudo-spin triplet pairing interaction between fermions.

In the nearest neighbour approximation, the lattice interaction in the single hyperfine state case has
only a $p$-wave (triplet) component, which is given by
\begin{equation}
V_p(\mathbf{k},\mathbf{k'})= - 2\sum_{i=x,y}V_{0,i}\sin (k_ia_i) \sin (k_i'a_i),
\label{eqn:pwave-interaction}
\end{equation}
where $V_{0,i}> 0$ is the effective interaction strength and 
$a_i$ is the corresponding lattice length along the $i^{\rm{th}}$ direction. 
Notice that Eq.~(\ref{eqn:pwave-interaction}) has the necessary symmetry under the Parity operation, 
where either $\mathbf{k} \to -\mathbf{k}$ or similarly $\mathbf{k'} \to -\mathbf{k'}$
leads to $-V_p(\mathbf{k}, \mathbf{k'})$. 
Furthermore, the interaction $V_p(\mathbf{k}, \mathbf{k'})$ is invariant under the 
transformation $(\mathbf{k},\mathbf{k'}) \to (-\mathbf{k},-\mathbf{k'}) $.
This is a necessary condition for an SHS $p$-wave interaction
since the interaction characterizes a triplet channel, and it has to reflect the Pauli principle. 
This interaction can be written in a separable form as
\begin{equation}
V_p(\mathbf{k},\mathbf{k'})=\Gamma^\dagger(\mathbf{k})\mathbf{V}\Gamma(\mathbf{k'})
\end{equation}
where the interaction strength matrix has elements $V_{ij} = -2V_{0,i}\delta_{ij}$
and the symmetry vector is $\Gamma^\dagger(\mathbf{k})=[\sin (k_xa_x), \sin (k_ya_y)]$.
Notice that, the interaction $V_p(\mathbf{k},\mathbf{k'})$ is not separable as a product of scalar functions of 
$\mathbf{k}$ and $\mathbf{k'}$ in the usual sense, but it is separable in terms of vector functions.

In addition, we compare SHS $p$-wave superfluids with THS $s$-wave superfluids.
In the nearest neighbour approximation, the lattice interaction term in the THS case leads
to singlet pairing channels $s$, extended-$s$ and $d$-wave terms, and a separate triplet pairing 
THS $p$-wave channel. 
However, the $s$-wave channel is expected to dominate in the absence of any exotic mechanism 
as it corresponds to the case of minimal free energy.
The Hamiltonian for a THS $s$-wave Fermi gas is given by
\begin{eqnarray}
\label{eqn:hamiltonian.swave}
H=\sum_{\mathbf{k},\sigma}\xi(\mathbf{k})a_{\mathbf{k}\sigma}^\dagger a_{\mathbf{k}\sigma} + 
\frac{1}{2}\sum_{\mathbf{k},\mathbf{k',\sigma},\mathbf{q}}V_{s}b_{\mathbf{k},\mathbf{q}}^\dagger b_{\mathbf{k'},\mathbf{q}}, 
\end{eqnarray}
where the pseudo-spins $\sigma = (\uparrow, \downarrow)$ label the trapped hyperfine states represented by 
the creation operator $ a_{\mathbf{k}\sigma}^\dagger$, and
$b_{\mathbf{k},\mathbf{q}}^\dagger=a_{\mathbf{k}+\mathbf{q}/2,\sigma}^\dagger a_{-\mathbf{k}+\mathbf{q}/2,-\sigma}^\dagger$.
Here, $V_{s} = -|V_{0,s}|$ is the attractive $s$-wave interaction.

In the following section, we describe the effective action method for the SHS $p$-wave Hamiltonian in Eq.~(\ref{eqn:hamiltonian}) and 
we discuss its saddle point order parameter and number equations, as well as the critical temperature.
This method can be applied to the $s$-wave Hamiltonian in Eq.~(\ref{eqn:hamiltonian.swave}), 
and similar equations can be derived for $s$-wave case, but we do not repeat this derivation here.

\subsection{Effective Action Formalism}
\label{sec:effective-action-formalism}

In the imaginary-time functional integration formalism~\cite{popov} ($\beta=1/T$ and units $\hbar=k_B=1$), 
the partition function can be written as 
\begin{equation}
Z=\int D[a^\dagger,a]e^{-S}
\end{equation}
with an action given by
\begin{eqnarray}
S=\int_0^\beta d\tau\left[ \sum_{\mathbf{k}}a_{\mathbf{k}\uparrow}^\dagger(\tau)(\partial_\tau) a_{\mathbf{k}\uparrow}(\tau) 
+ H(\tau) \right].
\end{eqnarray}
The Hamiltonian given in Eq.~(\ref{eqn:hamiltonian}) can be rewritten in the form
\begin{eqnarray}
H(\tau)=\sum_{\mathbf{k}}\xi({\mathbf{k}}) a_{\mathbf{k}\uparrow}^\dagger(\tau) a_{\mathbf{k}\uparrow}(\tau)+
\sum_{\mathbf{q}}\mathbf{B}_{\mathbf{q}}^\dagger(\tau)\frac{\mathbf{V}}{2}\mathbf{B}_{\mathbf{q}}(\tau)
\end{eqnarray}
with $\mathbf{B}_{\mathbf{q}}(\tau)=\sum_{\mathbf{k}}\Gamma(\mathbf{k}) b_{\mathbf{k},\mathbf{q}}(\tau)$.
We then introduce Nambu spinor 
$\psi^\dagger(p)=( a_{p\uparrow}^\dagger , a_{-p\uparrow} )$, 
where we use $p=(\mathbf{k},w_\ell)$ 
to denote both momentum and fermionic Matsubara frequency $w_\ell=(2\ell+1)\pi/\beta$ 
and use a Hubbard-Stratonovich transformation to decouple fermionic and bosonic degrees of freedoms.
Performing the integration over the fermionic part ($D[\psi^\dagger,\psi]$) leads to 
\begin{eqnarray}
S_{\rm eff}& = &2\beta\sum_{q} \Phi^\dagger(q)\mathbf{V}^{-1} \Phi(q) \nonumber \\
&+& \sum_{p,p'}\left[\frac{\beta}{2}\xi(\mathbf{k})\delta_{p,p'}
- \rm{Tr}\ln\frac{\beta}{2}\mathbf{G}^{-1} \right], 
\end{eqnarray}
where we use $q=(\mathbf{q},v_\ell)$ with bosonic Matsubara frequency $v_\ell=2\ell\pi/\beta$ and define the bosonic vector field $\Phi(p-p')$.
Here,
\begin{eqnarray}
\mathbf{G}^{-1} &=& \Phi^\dagger(q)\Gamma(\frac{p+p'}{2})\sigma_- 
+ \Gamma^\dagger(\frac{p+p'}{2})\Phi(-q)\sigma_+ \nonumber \\
&+& \left[iw_\ell \sigma_0-\xi(\mathbf{k})\sigma_3\right]\delta_{p,p'}
\end{eqnarray}
is the inverse Nambu propagator and $\sigma_{\pm}=(\sigma_1\pm\sigma_2)/2$ and $\sigma_{i}$ is the Pauli spin matrix.
The bosonic vector field 
\begin{equation}
\Phi(q) = \Delta_0\delta_{q,0}+ \Lambda(q)
\end{equation}
has $\tau$-independent and $\tau$-dependent parts where 
$ \Delta_0^\dagger=(\Delta_{0,x}^*, \Delta_{0,y}^*)$ and 
$\Lambda^\dagger(q) = [\Lambda_x(q), \Lambda_y(q)]$, respectively.

Performing an expansion in $S_{\rm eff}$ to quadratic order in the vector $\Lambda$, we obtain
\begin{equation}
S_{\rm{gauss}} = S_0 + \frac{1}{2} \sum_{q} \bar{\Lambda}^\dagger(q) \mathbf{F}^{-1}(q) \bar{\Lambda}(q),
\end{equation}
where $S_0$ is the saddle point action, the 4-component vector $\bar{\Lambda}^\dagger(q)$ is such that
$\bar{\Lambda}^\dagger(q) = [\Lambda^\dagger(q), \Lambda(-q)]$ and $\mathbf{F}^{-1}(q)$ is the
inverse fluctuation propagator. The fluctuation term in the action leads to a correction
to the thermodynamic potential, which can be written as 
$\Omega_{\rm{gauss}} = \Omega_0 + \Omega_{\rm{fluct}}$ with 
$\Omega_{\rm{fluct}} = \beta^{-1}\sum_{q}\ln\det[\mathbf{F}^{-1}(q)/2]$.
In weak coupling for all temperatures and for all couplings at low temperatures, 
it can be shown that~\cite{jan} Gaussian corrections $S_{\rm{fluct}}$ 
to the gaussian action $S_{\rm gauss}$ are small in the quasi-two-dimensional case, and therefore,
we consider only the saddle point action
\begin{eqnarray}
S_0=2\beta\Delta_0^\dagger\mathbf{V}^{-1} \Delta_0
+ \sum_{p}\left[\frac{\beta}{2}\xi(\mathbf{k}) - \rm{Tr}\ln\frac{\beta}{2}\mathbf{G_0}^{-1}\right], 
\end{eqnarray}
where the inverse Nambu propagator becomes
\begin{equation}
\mathbf{G}_0^{-1}=iw_\ell \sigma_0-\xi(\mathbf{k})\sigma_3 + 
\Delta_0^\dagger \Gamma(\mathbf{k})\sigma_- + \Gamma^\dagger(\mathbf{k}) \Delta_0 \sigma_+.
\end{equation}
Notice that, this will not be the case close to the critical temperatures $T_c$ for 
intermediate and strong interactions, since the 
inclusion of fluctuation corrections change the number equation considerably 
and are necessary to produce correct behaviour~\cite{carlos}.

\subsection{Saddle-Point Equations}
\label{sec:saddle-point-equations}

The saddle point condition $\delta S_0 /\delta \Delta_0^* = 0$ leads to a matrix equation for the order parameter
\begin{equation}
\Delta_0 = \mathbf{M}\Delta_0,
\label{eqn:gap}
\end{equation}
where matrix $\mathbf{M}$ has the following matrix elements 
\begin{equation}
M_{ij}=\sum_{\mathbf{k}}\frac{V_{0,i}\sin (k_ia_i) \sin (k_ja_j)}{E(\mathbf{k})} \tanh\frac{\beta E(\mathbf{k})}{2}.
\end{equation}
Here, we introduce the quasi-particle energy
\begin{equation}
E(\mathbf{k})=(\xi^2(\mathbf{k})+|\Delta(\mathbf{k})|^2)^{\frac{1}{2}}
\end{equation}
and the scalar order parameter 
\begin{equation}
\Delta(\mathbf{k})= \Gamma^\dagger(\mathbf{k})\Delta_0,
\end{equation}
which is naturally separable in temperature $T$ and momentum $\mathbf{k}$. 
The critical temperature, $T_c = \max\{ T_{c,x}, T_{c,y} \}$, 
is determined from the condition $\det\,\mathbf{M}=1$ in Eq.~(\ref{eqn:gap}), and can be written as
\begin{eqnarray}
0=\prod_{i=x,y}\left( 1-V_{0,i}\sum_{\mathbf{k}}\frac{\sin ^2 (k_ia_i)}{\xi(\mathbf{k})}
\tanh \frac{\xi(\mathbf{k})}{2T_{c,i}} \right) \label{tc}.
\label{eqn:tc}
\end{eqnarray}

Both the order parameter and critical temperature equations have to be solved simultaneously 
with the number equation 
$N=-\partial \Omega/\partial {\mu}$ where $\Omega$ is the full thermodynamic potential. 
This leads to two contributions to the number equation $N \approx N_{\rm gauss} = N_0 + N_{\rm{fluct}}$. 
Here, $N_0=-\partial \Omega_0/\partial {\mu}$ is the saddle point number equation, where
$\Omega_0=S_0/\beta$ is the saddle point thermodynamic potential, 
and $N_{\rm{fluct}} = -\partial \Omega_{\rm{fluct}}/\partial \mu$ is the 
fluctuation contribution.
In the low temperature limit ($T \approx 0$) for any coupling, 
or in the weak coupling limit for any temperature $(T \le T_c)$ 
the fluctuation contribution 
\begin{equation}
N_{\rm{fluct}} = -T \sum_{q}\frac{\partial [\det \mathbf{F}^{-1}(q)]} {\partial {\mu}}
\frac{1}{\det \mathbf{F}^{-1}(q)}
\end{equation}
is small and negligible 
compared to the saddle point value $N_0$~\cite{carlos}. Thus, the number equation becomes 
$N \approx N_0$, given by
\begin{equation}
N = \sum_{\mathbf{k}}n(\mathbf{k}),
\label{eqn:numbereqn}
\end{equation}
where $n(\mathbf{k})$ is the momentum distribution defined by
\begin{equation}
\label{eqn:md}
n(\mathbf{k})=\frac{1}{2}\left[ 1 - \frac{\xi(\mathbf{k})}{E(\mathbf{k})}\tanh\frac{\beta E(\mathbf{k})}{2} \right].
\end{equation}

For comparison, we discuss next the $s$-wave case.
Using the $s$-wave Hamiltonian in Eq.~(\ref{eqn:hamiltonian.swave}), we follow a similar procedure and derive
the order parameter, critical temperature and number equations in the saddle point approximation. 
The order parameter equation is given by
\begin{equation}
\Delta_{0,s} = V_{0,s} \sum_{\mathbf{k}}\frac{\Delta_{0,s}}{2E(\mathbf{k})} \tanh\frac{\beta E(\mathbf{k})}{2}
\label{eqn:gap.swave}
\end{equation}
where $E(\mathbf{k}) = \sqrt{\xi^2(\mathbf{k}) + |\Delta_{0,s}|^2}$ is the quasi-particle energy spectrum.
The critical temperature is determined when $\Delta_{0,s} = 0$, which leads to
\begin{equation}
1 = \sum_{\mathbf{k}}\frac{V_{0,s} } {2\xi(\mathbf{k})} \tanh\frac{\xi(\mathbf{k})}{2T_c}.
\label{eqn:tc.swave}
\end{equation}
Both order parameter and critical temperature equations have to be solved simultaneously 
with the number equation 
\begin{equation}
N = \sum_{\mathbf{k}\sigma}n(\mathbf{k}),
\label{eqn:numbereqn.swave}
\end{equation}
where the momentum distribution $n(\mathbf{k})$ is defined in Eq.~(\ref{eqn:md}).

\subsection{Spin Triplet (SHS $p$-wave) States}
\label{sec:critical-temperatures}

In general, the pseudo-spin triplet order parameter can be written 
in the standard form~\cite{rice}
\begin{equation}
\mathbf{O}(\mathbf{k})
= \left( \begin{array}{cc} -d_1(\mathbf{k})+id_2(\mathbf{k}) & d_3(\mathbf{k}) 
\\ d_3(\mathbf{k}) & d_1(\mathbf{k})+id_2(\mathbf{k}) \end{array}\right),
\end{equation}
where $d(\mathbf{k})$ is an odd vector function of $\mathbf{k}$.
In our SHS $p$-wave interaction ${\mathbf{O}}_{\uparrow\uparrow}(\mathbf{k}) = \Delta (\mathbf{k})$, 
therefore, $d_3(\mathbf{k})=0$ and $d_1(\mathbf{k})=-id_2(\mathbf{k})$ which leads to 
$d(\mathbf{k})=d_1(\mathbf{k})(1,i,0)$.
Within the irreducible representations of the D$_{4h}$ (D$_{2h}$) group in the tetragonal (orthorhombic) lattices,
~\cite{annett} our exotic SHS $p$-wave state corresponds to the $^3E_u(n)$ representation with a $d$-vector given by 
\begin{equation}
d(\mathbf{k})=f(\mathbf{k})(1,i,0) \label{dvector}, 
\end{equation}
where $f(\mathbf{k})=AX+BY$, and $X$ and $Y$ 
are $\sin (k_xa_x)$ and $\sin (k_ya_y)$, respectively. Notice that, 
this state also breaks time-reversal symmetry, as expected from a fully spin-polarized state.

In the tetragonal lattice, the stable solution for our model corresponds to the case $A=B\ne0$, and thus
to the $^3E_u(d)$ representation,  
where spin-orbit symmetry is preserved, but both spin and orbit symmetries are independently 
broken. In the orthorhombic lattice, the stable solutions correspond to either $A\ne0,B=0$
or $A=0, B\ne0$, thus leading to the $^3E_u(b)$ representation. 
Notice that, depending on the lattice anisotropy, there are three 
distinct solutions to the order parameter equation Eq.~(\ref{eqn:gap}) in relation to Eq.~(\ref{dvector}).
The first solution (case 1) occurs when $\Delta_{0,x}\ne0$ and $\Delta_{0,y}=0$, 
which corresponds to a $d$-vector with $A \ne 0$ and $B = 0$.
Similarly, a second solution (case 2) occurs when $\Delta_{0,x}=0$ and $\Delta_{0,y}\ne0$, and
it corresponds to a $d$-vector with $A = 0$ and $B \ne 0$). 
Third and final solution (case 3) occurs when $\Delta_{0,x}= \Delta_{0,y}\ne 0$, 
which corresponds to a $d$-vector with $A = B \ne 0$.
In a tetragonal lattice both directions are degenerate (case 3), 
but even small anisotropies in the optical lattice spacings, transfer energies
$(t_x, t_y)$, or optical lattice potentials $(U_{0,x}, U_{0,y})$ 
lifts the degeneracy and throws the system into either case 1 or 2.

\section{Superfluid Properties in the weak coupling BCS regime}
\label{sec:bcs}

Our main interest is in tetragonal (square) lattices, however, we also want to investigate the effects of 
small anisotropies in optical lattice lengths. For this purpose 
we investigate five different cases that may be encountered experimentally. 

Case (I) corresponds to lattice spacings $a_x = a_y$, to
transfer integrals $t_x =  t_y$, and to interaction strengths $V_{0,x} = V_{0,y}$. 
Case (II) corresponds to $a_x = a_y$, $t_x \ne t_y$ and $V_{0,x} = V_{0,y}$. 
Case (III) corresponds to $a_x = a_y$, $t_x \ne t_y$ and $V_{0,x} \ne V_{0,y}$. 
Case (IV) corresponds to $a_x \ne  a_y$, $t_x \ne t_y$ and $V_{0,x} = V_{0,y}$. 
Case (V) corresponds to $a_x \ne a_y$, $t_x \ne t_y$ and $V_{0,x} \ne V_{0,y}$. 
Notice that cases (II) and (IV) are equivalent except for a unit cell volume normalization.
The same is true for cases (III) and (V).
Case (I) is effectively a tetragonal lattice, while cases (II) through (V) are effectively 
orthorhombic lattices. Notice that even though the optical lattice spacings $a_x = a_y$ in cases (II) and (III), 
these cases correspond effectively to "orthorhombic" lattices because the transfer energies
$t_x$ and $t_y$ are different. 

Experimentally it may be easier to use only one type of laser with specified wavelength, and thus
generate a lattice with $a_x = a_y$, and change the focus width in one direction (say along $\hat{\mathbf{x}}$) while
keeping the width in the other direction fixed (say along $\hat{\mathbf{y}}$). 
The change in focus width along the $\hat{\mathbf{x}}$ direction
will modify the transfer integral $t_x$ by either reducing it or enlarging it with respect to $t_y$.
This type of experiment can cover cases (I), (II) and (III).
A second type of experiment (a bit harder) could use two types of lasers with different wavelengths
and produce a lattice with $a_x \ne a_y$. Further control of the focal widths can produce the situations 
encountered in cases (IV) and (V). 

We concentrate here on cases (I), (II) and (III), where we set $a_x = a_y = a$.
For case (I) we choose $t_x = t_y = t$, and $V_{0,x} = V_{0,y} = V_0$.
For case (II) we choose $t_x = (1 + \alpha)t$, $t_y = t$, and $V_{0,x} = V_{0,y} = V_0$.
For case (III) we choose $t_x = (1 + \alpha)t$, $t_y = t$, and $V_{0,x} = (1 + \alpha) V_0$, $V_{0,y} = V_0$.
Here we take $V_0 = 0.6E_0$ with $E_0 = 2t$.
In these cases, the transfer energies and the parameters $\alpha$ can be determined by 
using exponentially decaying on-site Wannier functions and the WKB approximation.
When the lattice spacings $a_x = a_y = a$ are the same but 
the optical lattice potentials  $U_{0,x} > U_{0,y} = U_0$ are slightly different, 
the transfer integral $t_x/t$ is proportional to $\exp{[-\pi(U_{0,x} - U_0)/\sqrt{U_{0,x} E_R}]}$,
in the limit of $U_{0,x} - U_0 \ll U_{0,x}$.
Here, $E_R$ is the one photon recoil energy, while
$U_0$ and $U_{0,x}$ are defined with respect to 
$E_g$, which is the energy of the lowest reference state of the trapping potential.
In the following sections, we discuss two possibilities where 
$U_{0,x} - U_0 = 0.05E_R$ corresponding to $\alpha \approx 0.1$ 
and $U_{0,x} - U_0 = 0.2E_R$ corresponding to $\alpha \approx 0.6$
for $U_{0} \approx 2.5E_R$.

For cases (IV) and (V) we set $a_x = a (1 - \delta)$ and $a_y = a$, with $0 < \delta \ll 1$, respectively.
For case (IV) we choose $t_x = (1 + \alpha)t$, $t_y = t$, and $V_{0,x} = V_{0,y} = V_0$.
For case (V) we choose $t_x = (1 + \alpha)t$, $t_y = t$, and $V_{0,x} = (1 + \alpha) V_0$, $V_{0,y} = V_0$.
In these cases, the parameter $\alpha$ can also be determined by 
using exponentially decaying on-site Wannier functions and the WKB approximation.
When the optical lattice potentials $U_{0,x} = U_{0,y} = U_0$ are the same but
the lattice spacings $a_x < a_y = a$  are different, 
the transfer integral $t_x/t$ is proportional to $\exp{(-2\pi\delta\sqrt{U_{0}/E_R})}$,
for small $\delta$.
In the following sections, we discuss two possibilities where 
$\delta = 0.01$ corresponding to $\alpha \approx 0.1$ and
$\delta = 0.05$ corresponding to $\alpha \approx 0.6$
for $U_{0} \approx 2.5E_R$.
As we increase (decrease) the ratio $\delta$, both interactions 
and tunneling rate increase (decrease) in $\hat{x}$
direction. 

Notice that a different choice of on-site Wannier functions 
does not change our general conclusions, the only qualitative difference 
is that $t_x$ and $V_{0,x}$ are different functions of $\alpha$.

\subsection{Density of Fermionic States}
\label{sec:density-of-fermionic-states}

\begin{figure} [ht]
\centerline{\scalebox{0.36}{\includegraphics{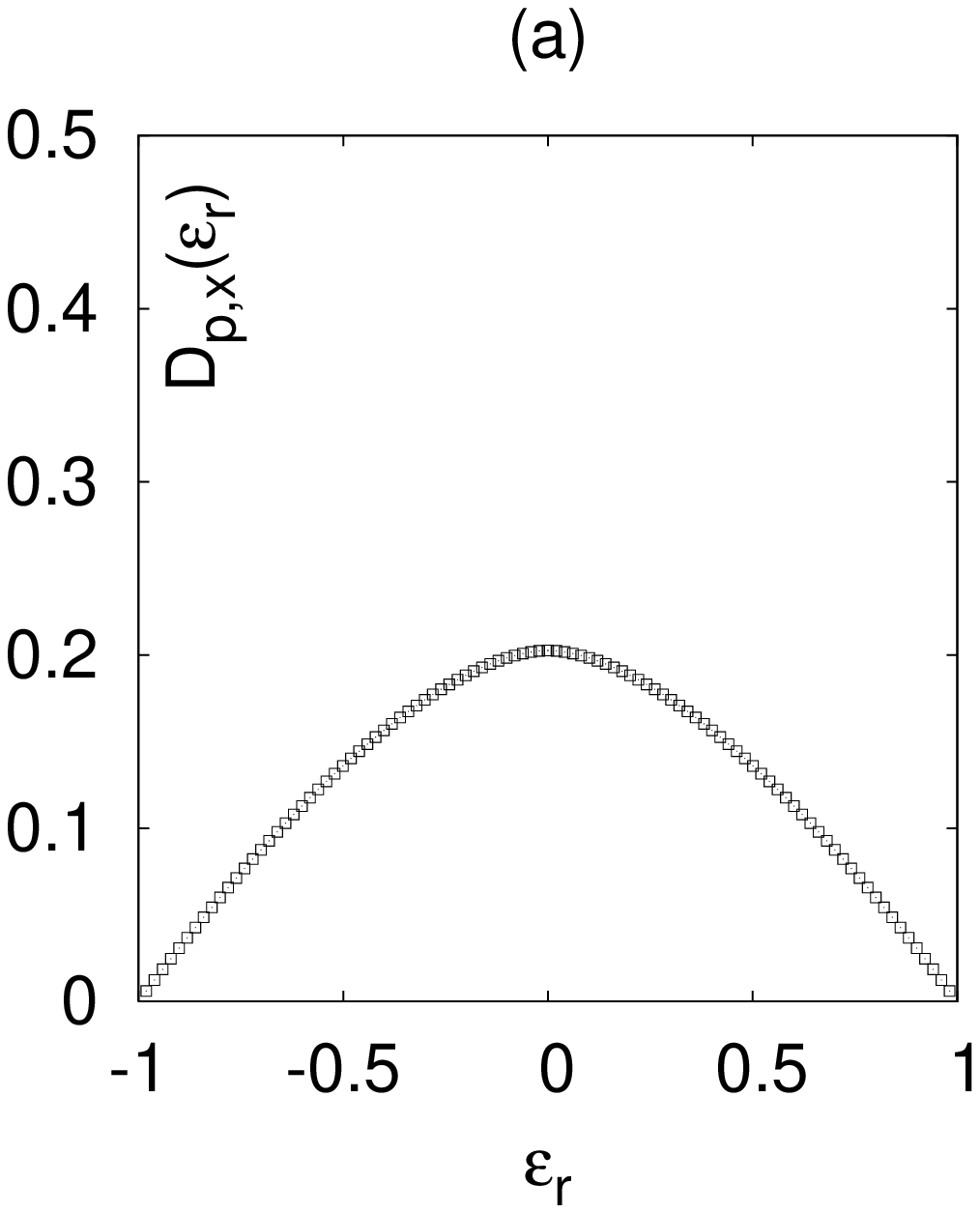} \includegraphics{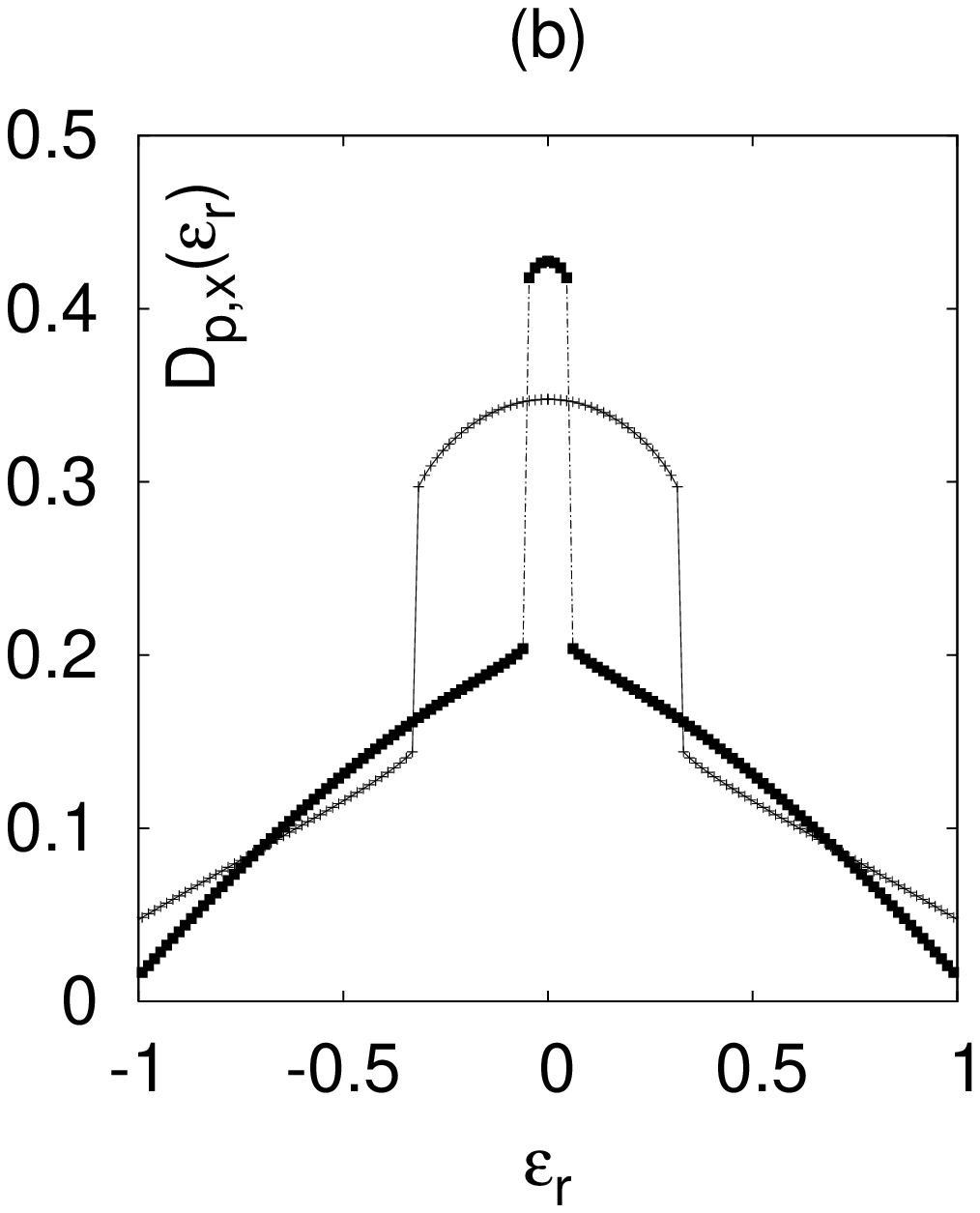}}}
\centerline{\scalebox{0.36}{\includegraphics{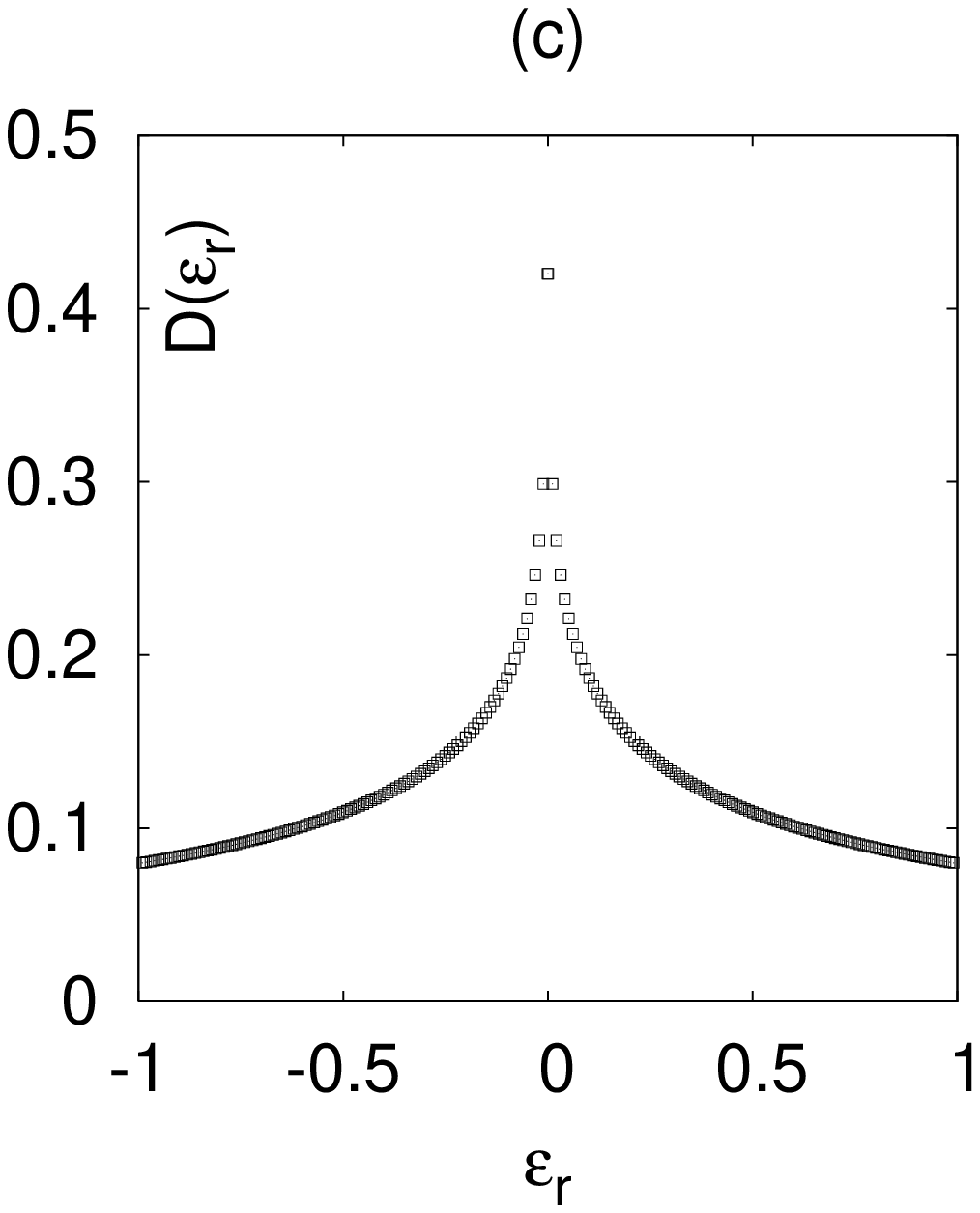} \includegraphics{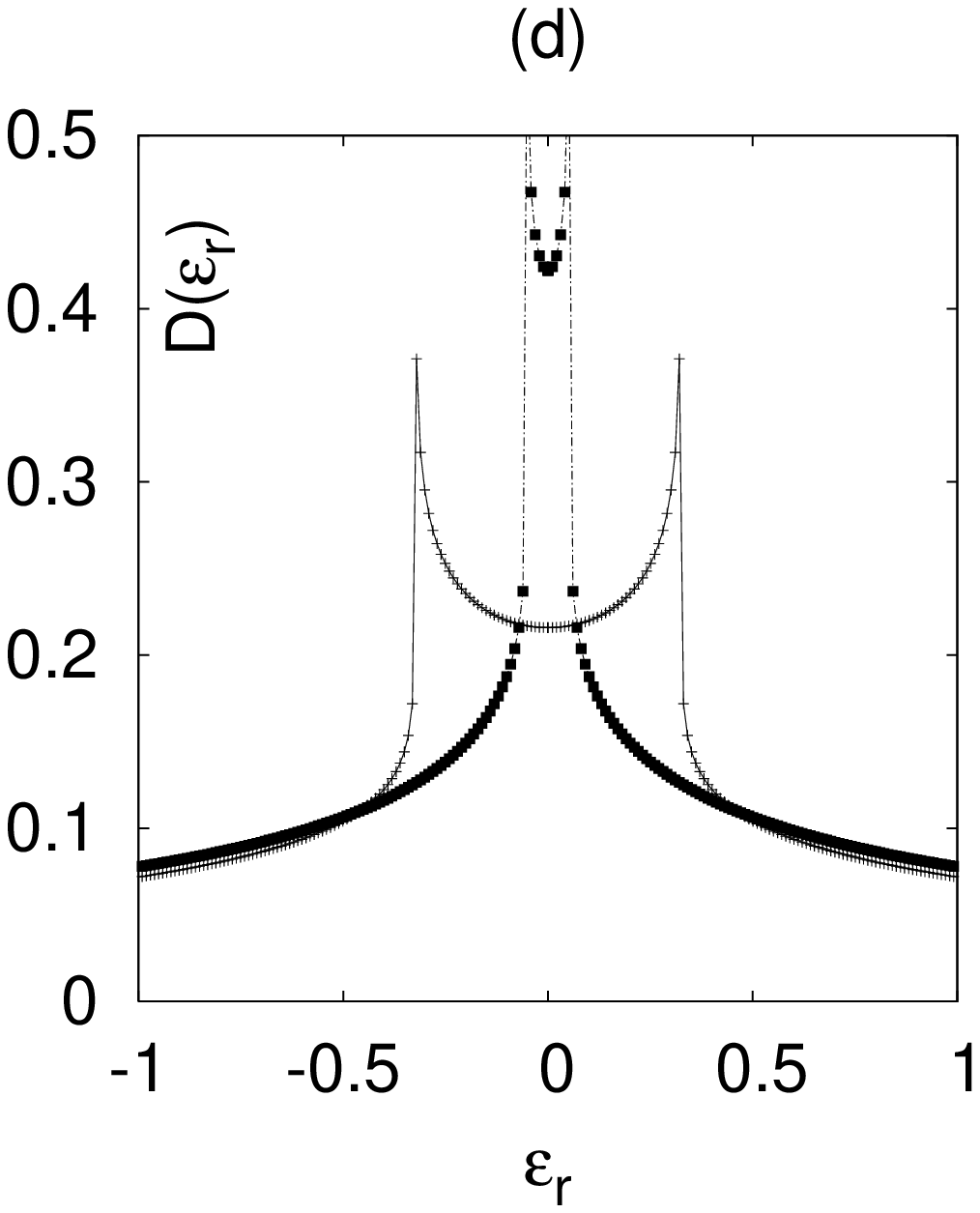}}}
\caption{\label{fig:dos} Plots of EDOS versus reduced energy $\varepsilon_r = \varepsilon / E_0$ 
for SHS $p$-wave symmetry when
a) $\alpha = 0$ (hollow square), and
b) $\alpha = 0.1$ (solid square) and $\alpha = 0.6$ (line). 
In addition plots of DOS for $s$-wave symmetry when
c) $\alpha = 0$ (hollow square), and
d) $\alpha = 0.1$ (solid square) and $\alpha = 0.6$ (line) are shown.
}
\end{figure}

The critical temperature and superfluid properties of 
Fermi systems depend highly on the order parameter symmetry,
as can be seen by rewriting Eq.~(\ref{eqn:tc}) as
\begin{equation}
\label{eqn:tc-dos}
1 = \prod_{i=x,y}V_{0,i} \int_{-t_x-t}^{t_x+t} d\varepsilon 
\frac{\tanh\frac{\varepsilon - \tilde{\mu}}{2T_{c,i}}}{2(\varepsilon - \tilde{\mu})}
D_{p,i}(\varepsilon),
\end{equation}
where we define an effective density of states (EDOS)
\begin{equation}
D_{p,i}(\varepsilon) = \sum_{\mathbf{k}} \delta[\varepsilon - \varepsilon(\mathbf{k})][\sqrt{2}\sin (k_ia_i)]^2.
\label{eqn:edos}
\end{equation}
Here $\sqrt{2}\sin (k_ia_i)$ is the symmetry factor related to the SHS $p$-wave order parameter.
We transformed discrete summations over $\mathbf{k}$-space to continuous integrations to obtain 
\begin{equation}
D_{p,x}(\varepsilon) = \int_{i}^{f} \frac{dk_ya}{\pi^2t_x^2} \left[ t_x^2 -(\varepsilon -t\cos (k_ya))^2 \right]^{1/2},
\end{equation}
where the integration region is restricted by $\vert \frac{\varepsilon - t\cos (k_ya)} {t_x} \vert \le 1$
in the first Brillouin zone (1BZ). 
Plots of EDOS are shown in Figs.~\ref{fig:dos}a and~\ref{fig:dos}b for three cases: 
$\alpha = 0$ (hollow squares), $\alpha = 0.1$ (solid squares), and $\alpha = 0.6$ (line).

In Fig.~\ref{fig:dos}a, we plot EDOS which is smooth and maximal 
at half-filling in square lattices ($\alpha = 0$). 
For the SHS $p$-wave symmetry discussed, $V_{0,i} D_{p,i}(\varepsilon)$
plays the role of a dimensionless coupling parameter which controls the critical temperature.
This is simply because it is much easier to form Cooper pairs with a small
attractive interaction (and lower the free energy of the Fermi system) when the
density of single fermion states is high.
Additionally, EDOS decreases and finally vanishes at the band edges
where a small ratio of $T_c/T_F$ was predicted from continuum models.
However, this is not the case around half-filling, and we expect weak interactions
to be sufficient for the observation of superfluidity.

In contrast, the symmetry factor is $1$ in the $s$-wave case and EDOS becomes
\begin{equation}
D(\varepsilon) = \sum_{\mathbf{k}} \delta[\varepsilon - \varepsilon(\mathbf{k})],
\label{eqn:dos}
\end{equation}
which is identical to the density of single fermion states (DOS) of normal fermions. 
A similar calculation yields
\begin{equation}
D(\varepsilon) = \int_{i}^{f} \frac{dk_ya}{2\pi^2}\left[ t_x^2 -(\varepsilon -t\cos (k_ya))^2 \right]^{-1/2},
\end{equation}
where the integration region is again restricted by 
$\vert \frac{\varepsilon - t\cos (k_ya)} {t_x} \vert \le 1$
 in the 1BZ.
Plots of DOS are shown in Fig.~\ref{fig:dos}c and Fig.~\ref{fig:dos}d 
for three cases: $\alpha = 0$ (hollow square), $\alpha = 0.1$ (solid square), and $\alpha = 0.6$ (line).

For instance, notice that in the tetragonal case of a two dimensional lattice, 
the DOS at half-filling is very large and tends to infinity logarithmically (Fig.~\ref{fig:dos}c). However, away from
half-filling the DOS decreases rapidly to a constant at the
minimum and maximum band edges. This constant DOS for very low or very high
filling factors is equivalent to the DOS of the continuous (homogeneous) systems since at the
bottom or at the top of the band, the single fermion dispersion energy can be approximated by
a parabola in momentum space. Furthermore, for $s$-wave superfluids,
$V_0 D(\varepsilon)$ is the dimensionless coupling parameter which controls the value of critical
temperature $T_c$. This important quantity ($T_c$) is discussed next.

\subsection{Critical Temperature}
\label{sec:critical-temperature}

\begin{figure} [htb]
\centerline{\scalebox{0.37}{\includegraphics{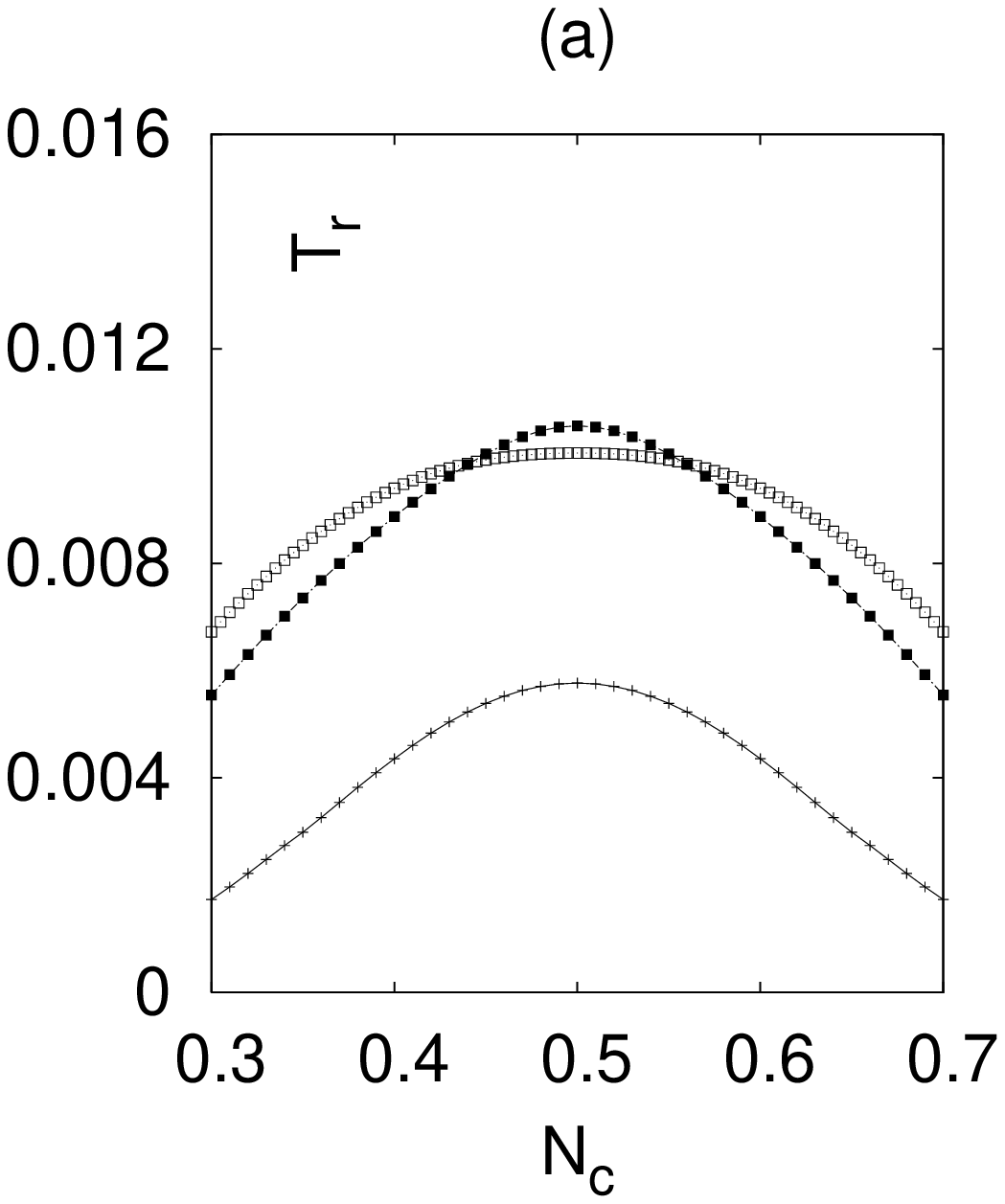} \includegraphics{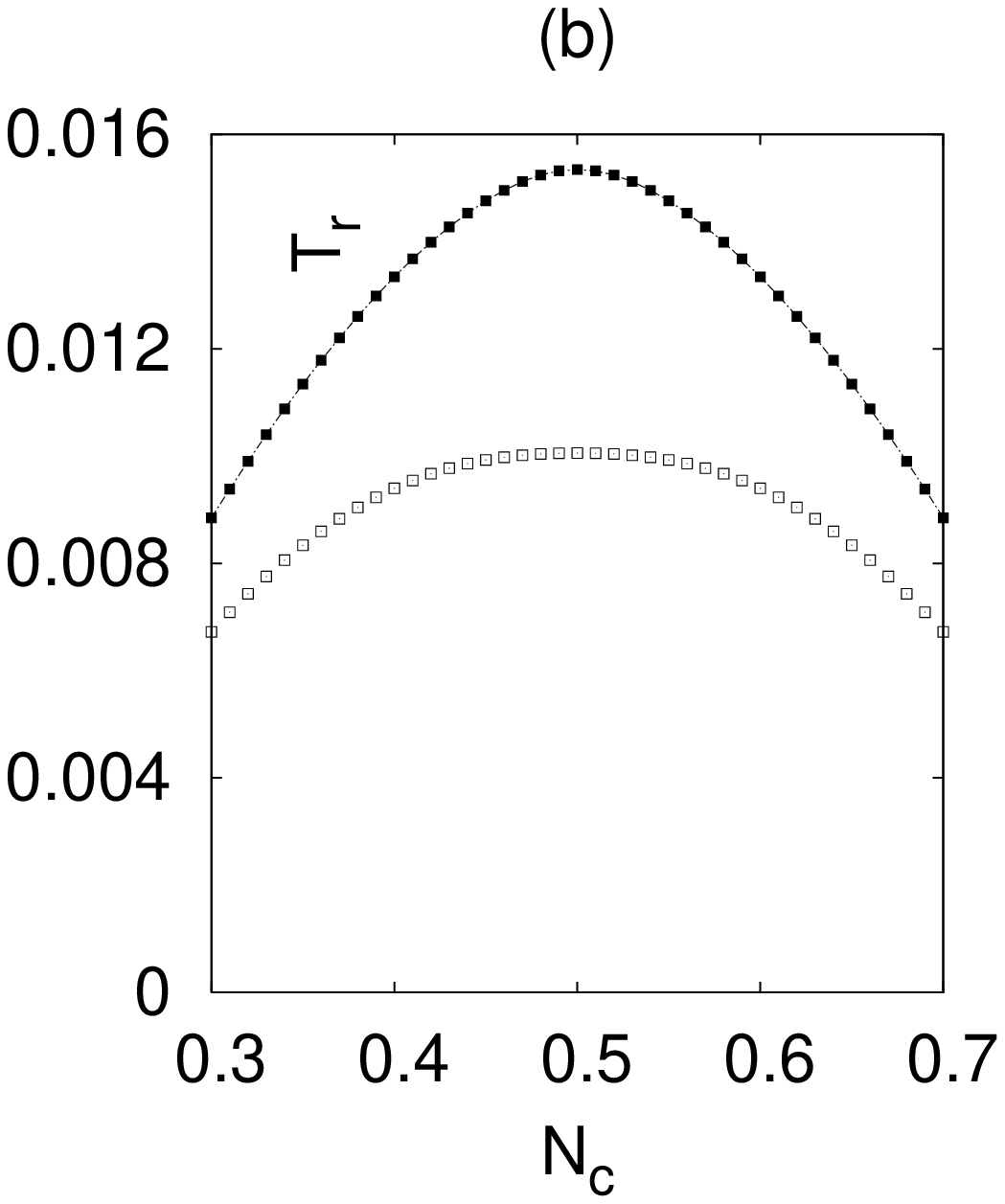}}}
\centerline{\scalebox{0.37}{\includegraphics{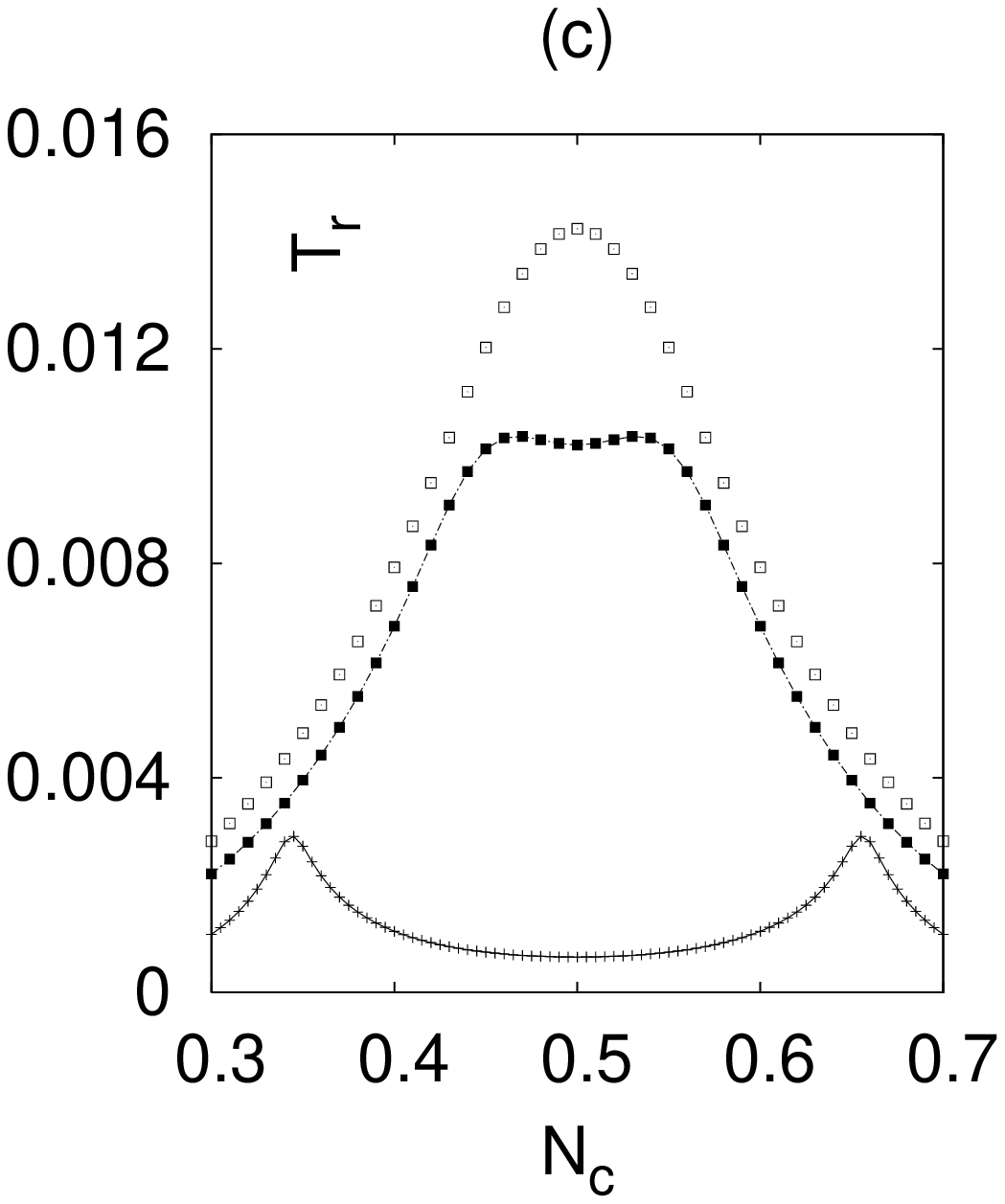} \includegraphics{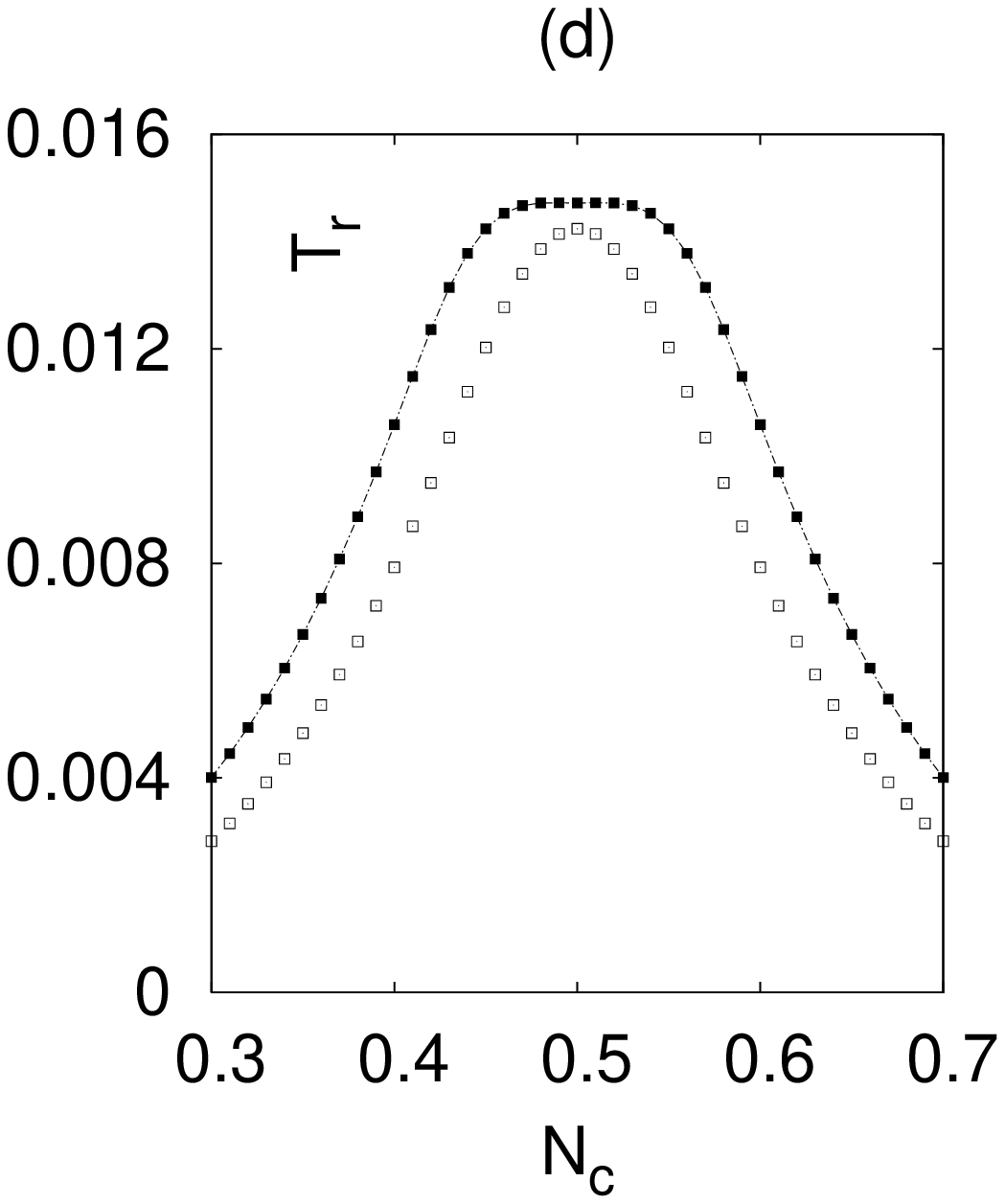}}}
\caption{\label{fig:tc} Plots of reduced critical temperature $T_r = T_c/E_0$ 
versus filling factor $N_c$ for SHS $p$-wave symmetry are shown for cases
a) (I) and (II) and
b) (I) and (III). 
In addition, plots for $s$-wave symmetry is shown for cases
c) (I) and (II), and
d) (I) and (III).
Plots for $\alpha = 0, 0.1$ and $0.6$ are shown with hollow squares, solid squares
and line, respectively.
}
\end{figure}

Plots of reduced critical temperature $T_r = T_c/E_0$ as a function of the number of
atoms per unit cell ($N_c$) are shown in Fig.~\ref{fig:tc} 
for SHS $p$-wave and $s$-wave superfluids in cases (I), (II), and (III).
Here, $T_c$ is the critical temperature calculated from Eqs.~(\ref{eqn:tc}) or (\ref{eqn:tc-dos}) and
$E_0 = 2t$ is the half-filling Fermi energy for the square lattice with respect
to the bottom of the band.

The square lattice case (I) is shown in Figs.~\ref{fig:tc}a and~\ref{fig:tc}c (hollow squares)
for SHS $p$-wave and $s$-wave superfluids, respectively.
Notice that $T_r$ is maximal at half-filling, and that 
it has a value $0.01$, which is much higher than the theoretically predicted $T_c$ from a continuum model,
and comparable to experimentally attainable temperatures $T/T_F \approx 0.04$~\cite{jthomas}. 
This implies that the superfluid regime of
SHS fermion gases may be observed experimentally in a lattice,
even in the limit of weak interactions (BCS regime). The observability of a superfluid transition in
SHS $p$-wave Fermi systems is clearly enhanced when the system is driven through 
a Feshbach resonance (in a lattice or in the
continuum), as $T_c$ is expected to increase further in this case, however our calculations indicate 
that the weak interaction (BCS) limit may be sufficient in the lattice case.
This possibility for Fermi gases in optical lattices has a parallel in the experimental observation
of superfluid-insulator transitions in weakly interacting Bose systems in optical lattices~\cite{greiner2}.

In case (II), while we keep interaction strengths $V_{0,x} = V_{0,y} = V_0$ the same, we vary only
the transfer energy along $\hat x$ direction $t_x = (1 + \alpha)t$.
This case is shown in Fig.~\ref{fig:tc}a and Fig.~\ref{fig:tc}c for SHS $p$-wave and $s$-wave 
superfluids, respectively. 
Here, solid squares (line) correspond to $\alpha = 0.1$ ($\alpha = 0.6$).

For SHS $p$-wave superfluids, an infinitesimally small anisotropy in the transfer energy ratio $t_x/t_y$ 
causes a discontinous jump in EDOS at half-filling because of the symmetry factor enhancement,
however $T_c$ at half-filling is the same as in case (I), because the discontinuity at
EDOS is limited to a single point.  
As the anisotropy ratio increases ($\alpha = 0.1$), 
the discontinuous point in EDOS at half-filling expands into a region where EDOS is higher than in case (I).
Thus over a narrow range around half-filling, $T_c$ is larger than in case (I), 
but varies smoothly as a function of $\alpha$ or $N_c$. Further increase of $\alpha$ 
reduces EDOS near half-filling pushing $T_c$ down. 
This behavior is characteristic of the SHS $p$-wave state through its symmetry factors. 
In contrast, for $s$-wave superfluids shown in Fig.~\ref{fig:tc}c, $T_c$ decreases 
for fixed $N_c$ around half-filling with increasing anisotropy (at least for $ 0 < \alpha < 0.6)$. Furthermore, $T_c$ develops
two maxima which are symmetric around half-filling. This behavior in $T_c$ is also
related to the DOS of the normal fermions which is shown in Fig.~\ref{fig:dos}.
Notice here that, when there is an anisotropy in the transfer energy,
$T_c$ near half-filling corresponding to an SHS $p$-wave pairing is considerably higher than
the $T_c$ of an $s$-wave pairing for the same parameters of interest (assuming that 
$V_{0,s} = V_{0,p} = V_{0}$).

In case (III), we vary both the interaction strength $V_{0,x} = (1 + \alpha)V_0$ and
the transfer energy $t_x = (1 + \alpha)t$ along $\hat{x}$ direction.
This case is shown in Figs.~\ref{fig:tc}b and~\ref{fig:tc}d
for SHS $p$-wave and $s$-wave superfluids, respectively. In these figures it is assumed
that $V_{0,s} = V_{0,p} = V_{0,x}$. 
In case (III), both SHS $p$-wave and $s$-wave superfluids have critical 
temperatures that are much larger than in case (II), because of the increase in
$V_{0,x}$, in addition to the density of states effect, 
discussed above.
Notice again that  $T_c$ near half-filling corresponding to an 
SHS $p$-wave pairing is also considerably higher than
$T_c$ for an $s$-wave pairing for the same interaction parameters.

In this section, we established that the critical temperatures for SHS $p$-wave superfluids in
optical lattices can be comparable or larger to the critical temperatures of $s$-wave superfluids, 
depending on the lattice anisotropy and interaction strength. Thus, we expect that
SHS $p$-wave superfluids may be experimentally attainable. 
In the next sections, we discuss some experimentally relevant
observables which could be used to identify the superfluid phase of our
exotic SHS $p$-wave state.

\subsection{Order Parameter and Chemical Potential}
\label{sec:order-parameter-and-chemical-potential}

\begin{figure} [htb]
\centerline{\scalebox{0.37}{\includegraphics{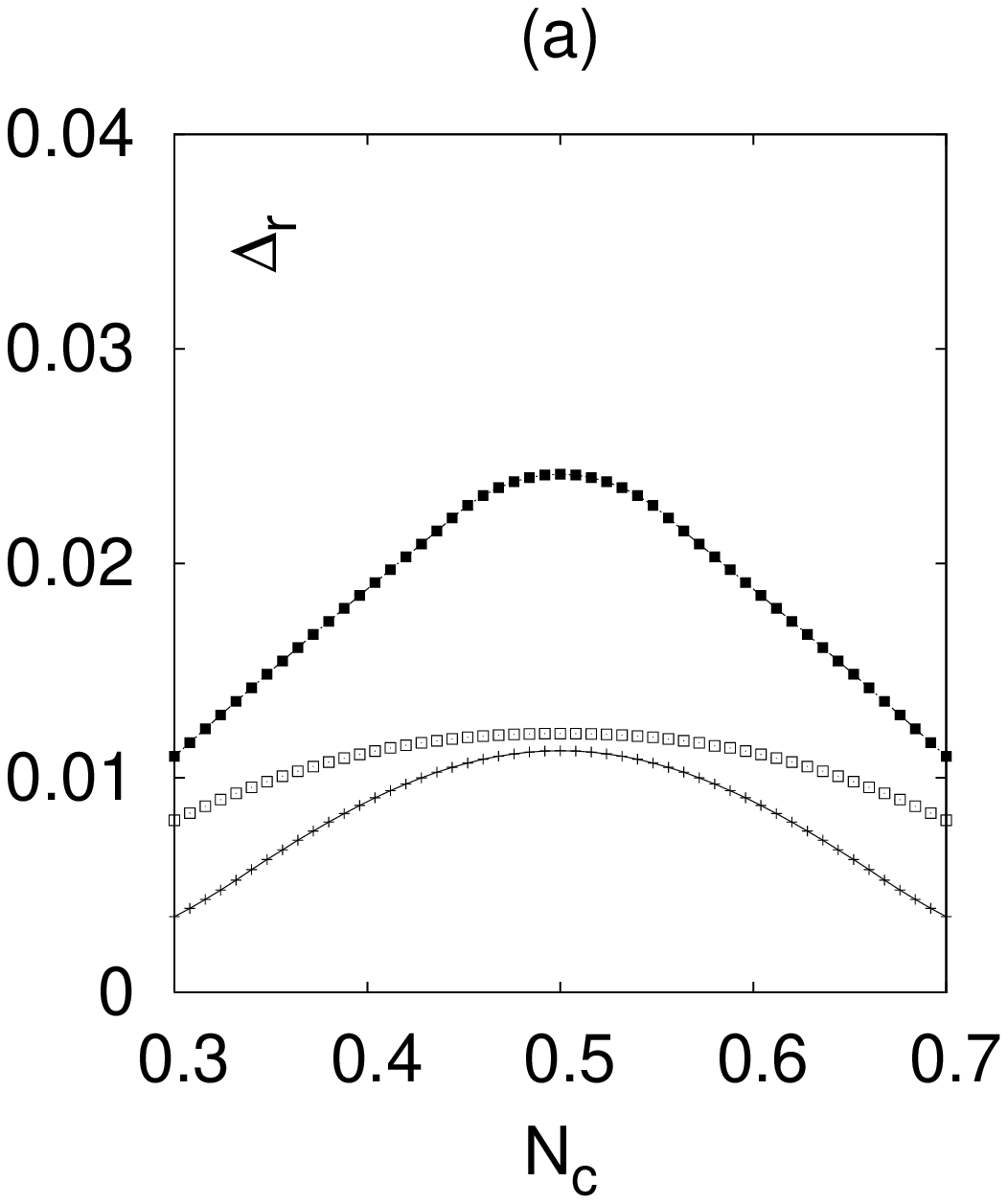} \includegraphics{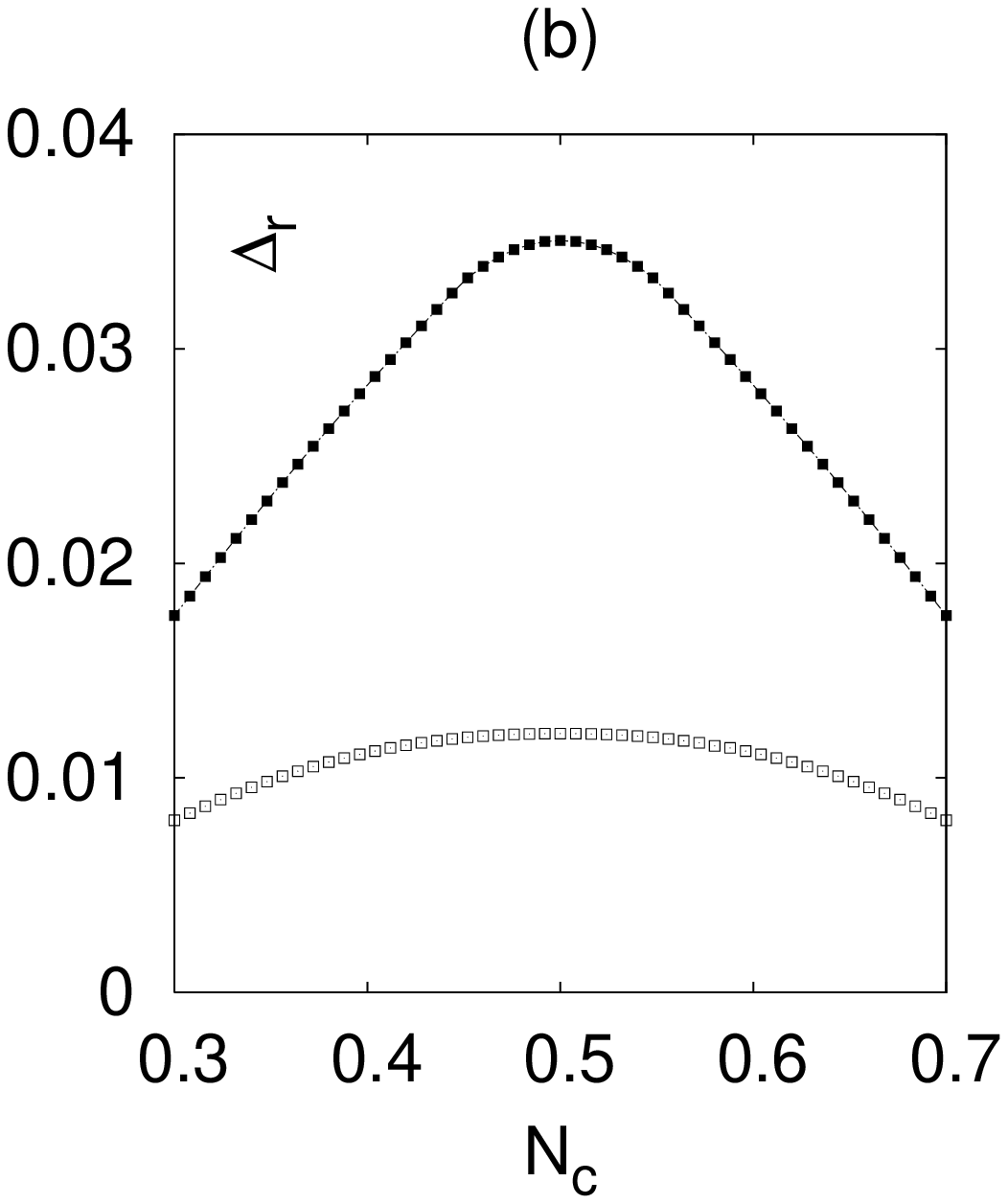}}}
\centerline{\scalebox{0.37}{\includegraphics{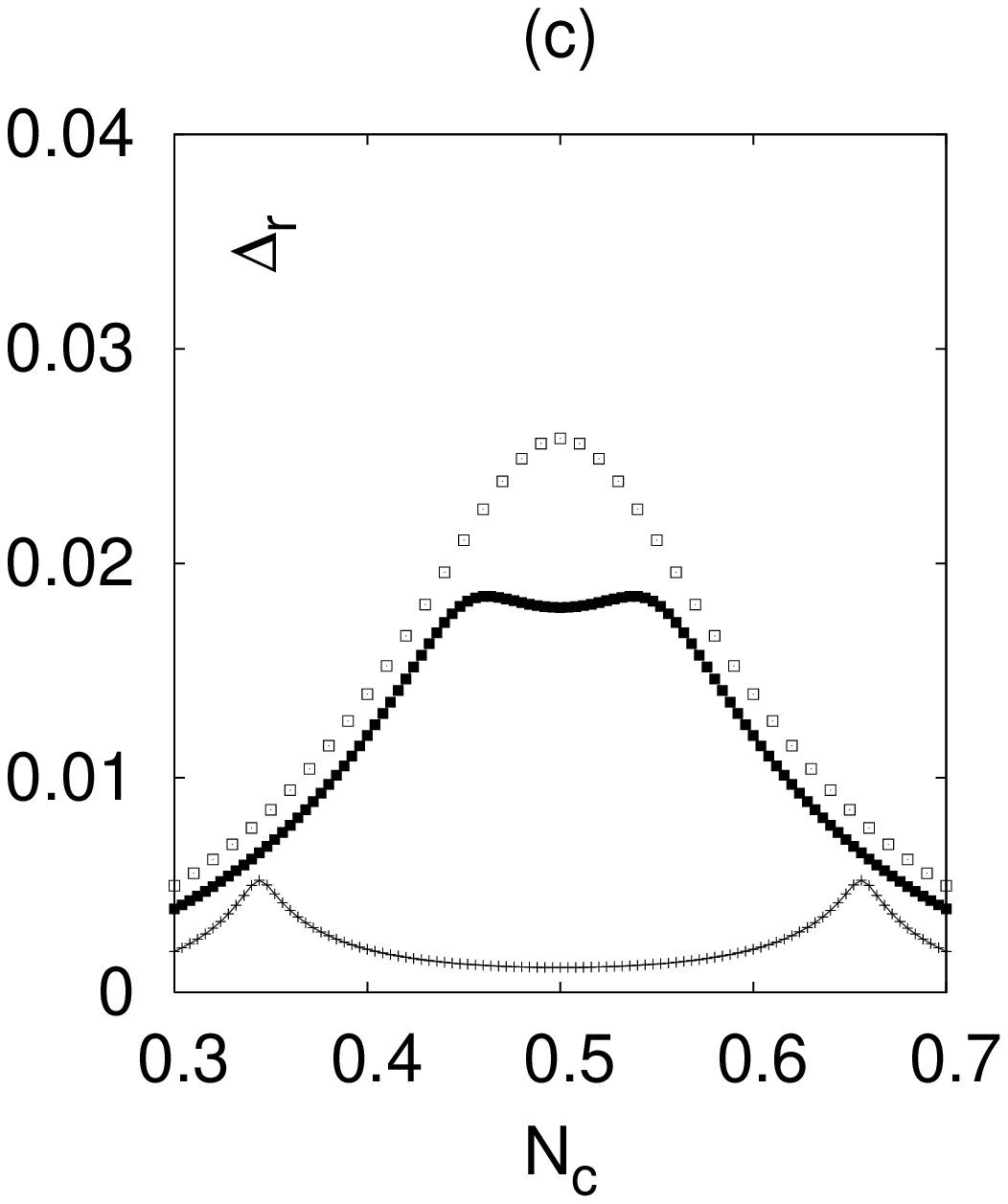} \includegraphics{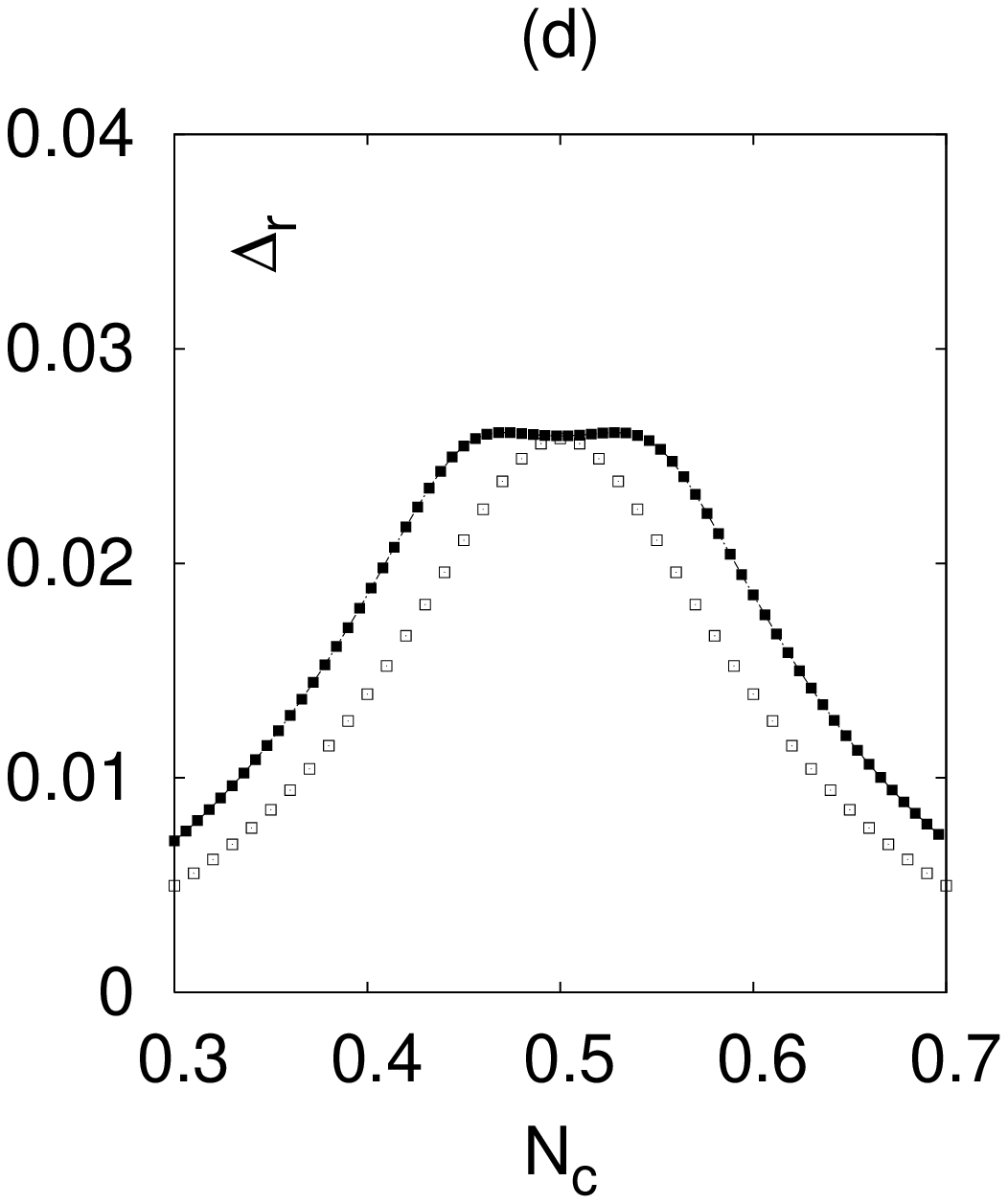}}}
\caption{\label{fig:gap} Plots of zero temperature reduced order parameter
amplitude $\Delta_r = \Delta_{0,x}/E_0$ versus filling factor $N_c$ for SHS $p$-wave symmetry for cases
a) (I) and (II) and
b) (I) and (III). 
In addition, plots for $s$-wave symmetry $\Delta_r = \Delta_{0,s}/E_0$ is shown for cases
c) (I) and (II), and
d) (I) and (III).
Plots for $\alpha = 0, 0.1$ and $0.6$ are shown with hollow squares, solid squares
and line, respectively.
}
\end{figure}

In this section, we discuss the low-temperature behavior of the order
parameter $\Delta ({\bf k}) = \Delta_{0,x} \sin (k_x a_x) + \Delta_{0,y} \sin (k_y a_y)$
defined in Eq.~(\ref{eqn:gap}) for the SHS $p$-wave state.
In case (I) $\Delta_{0,x} = \Delta_{0,y} \ne 0$, while in cases (II) and (III)
$\Delta_{0,x} \ne 0$ and $\Delta_{0,y} = 0$.
Here, we also discuss for comparison the singlet $s$-wave case where
$\Delta ({\bf k}) = \Delta_{0,s}$, with $\Delta_{0,s} \ne 0$ for cases (I), (II) and (III).
The reduced amplitudes of the SHS $p$-wave order parameter
$\Delta_{r} = \Delta_{0,x}/E_0$ versus $N_c$ are shown in Figs.~\ref{fig:gap}a and~\ref{fig:gap}b. 
In addition, we plot $s$-wave $\Delta_{r} = \Delta_{0,s}/E_0$ in 
Figs.~\ref{fig:gap}c and~\ref{fig:gap}d for cases (I), (II), and (III).
Notice that, the qualitative behaviour of $\Delta_r$ as a function of $N_c$ is very different 
from that of the chemical potential $\mu_r = \tilde{\mu}/E_0$, which is discussed next.

\begin{figure} [htb]
\centerline{\scalebox{0.37}{\includegraphics{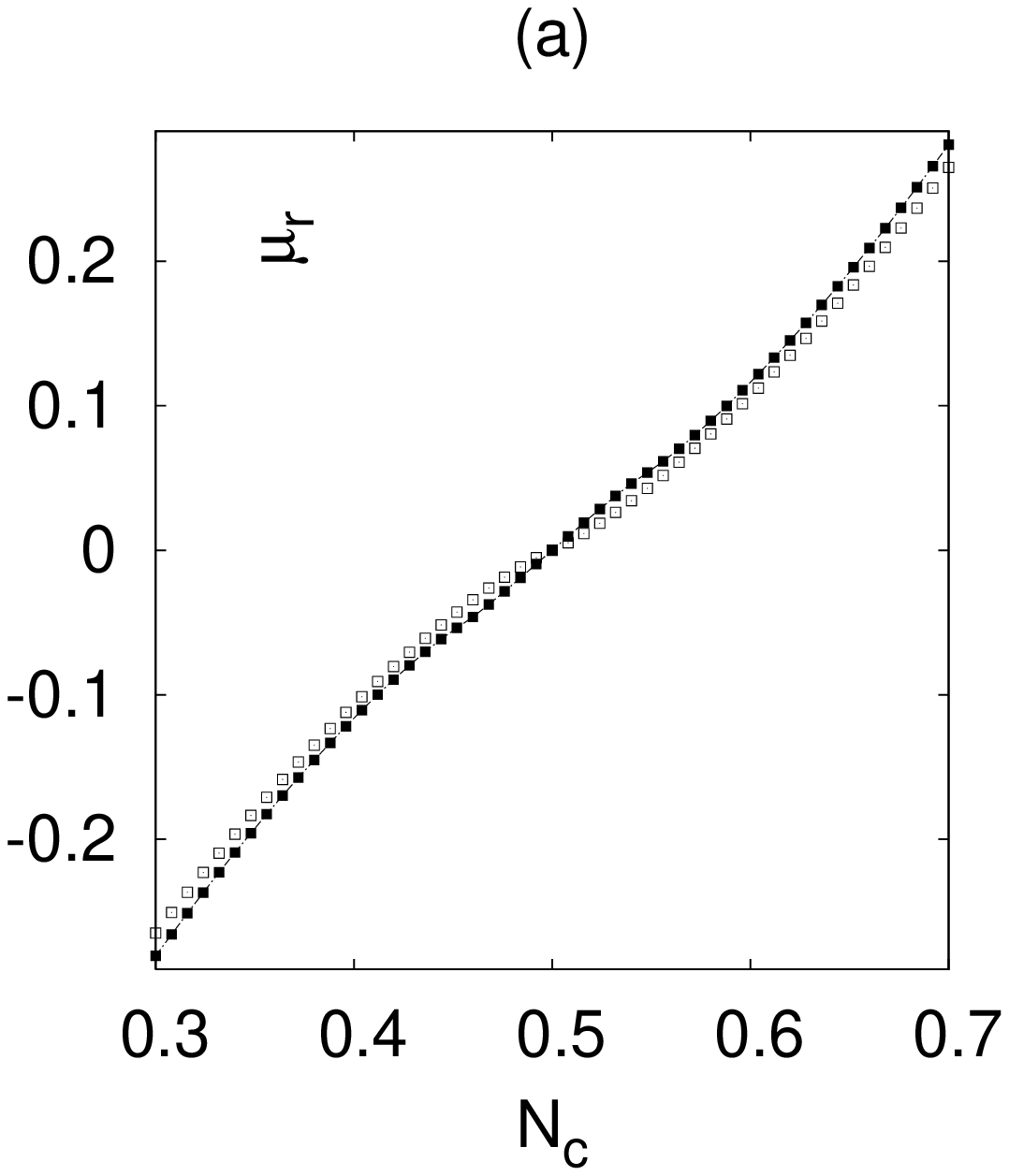} \includegraphics{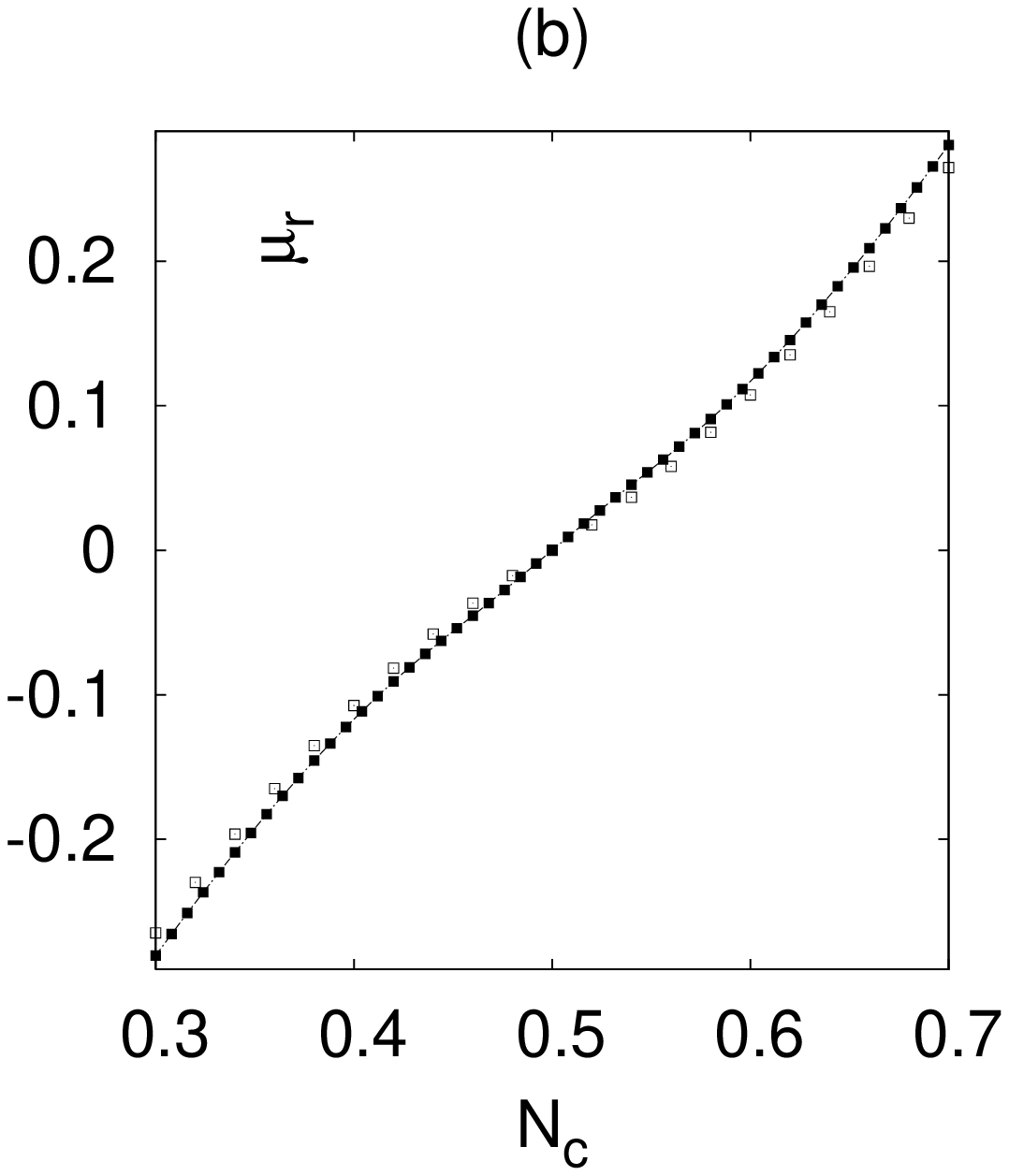}}}
\caption{\label{fig:mu} Plots of zero temperature reduced chemical potential $\mu_r = \tilde{\mu}/E_0$
versus filling factor $N_c$ for 
a) SHS $p$-wave, and 
b) $s$-wave superfluids.
Plots for $\alpha = 0$ and $0.1$ are shown with hollow and solid squares, respectively.
}
\end{figure}

Plots of the low-temperature $\mu_r$ versus $N_c$ are shown in Fig.~\ref{fig:mu} 
for SHS $p$-wave and $s$-wave superfluids, respectively.
Here, we present only the corresponding chemical potentials for cases (I) (hollow squares) and (II) (solid squares), 
where $\alpha = 0$ and $\alpha = 0.1$, respectively.
Chemical potentials for case (III) are very similar to those of case (II).
Notice that $\tilde{\mu}$ is always within the limits of the 
energy dispersion of the optical lattice,
$- t_x - t < \tilde{\mu} <  t_x + t$, which characterizes the weak coupling (BCS) limit.

Although the chemical potentials of SHS $p$-wave and the singlet $s$-wave cases 
look similar at first glance, they are not.
The major qualitative differences between the chemical potentials of the 
SHS $p$-wave and the singlet $s$-wave cases are more clearly seen in the derivative
$\partial \tilde \mu / \partial N_c$. The appropriate thermodynamic quantity that is
related to this derivative is the atomic compressibility to be discussed next.

\subsection{Atomic Compressibility}
\label{sec:atomic-compressibility}

The isothermal atomic compressibility at finite temperatures is defined by 
$\kappa_T(T) = -(\partial V / \partial P)_{T, N}/V$ where $V$ is the volume and $P$ is the pressure 
of the gas. This can be rewritten as
\begin{equation}
\kappa_T(T)=-\frac{1}{N_c^2} \left( \frac{\partial^2\Omega} {\partial \tilde{\mu}^2} \right)_{T,V}
=\frac{1}{N_c^2} \left( \frac{\partial N_c}{\partial \tilde{\mu}} \right)_{T,V}.
\end{equation}
The partial derivative $\partial N_c / \partial \tilde \mu$ is 
\begin{equation}
\label{eqn:dndmu}
\frac{\partial N_c}{\partial \tilde{\mu}} = 
\sum_{\mathbf{k}} \frac{\Delta^2(\mathbf{k})}{2E^3(\mathbf{k})} \tanh \frac{\beta E(\mathbf{k})}{2} 
+ \sum_{\mathbf{k}} Y(\mathbf{k})\frac{\xi^2(\mathbf{k})}{E^2(\mathbf{k})},
\end{equation}
where 
$Y(\mathbf{k})=(\beta/4)\rm{sech}^2[\beta E(\mathbf{k})/2]$
is the Yoshida function.

\begin{figure} [ht]
\centerline{\scalebox{0.36}{\includegraphics{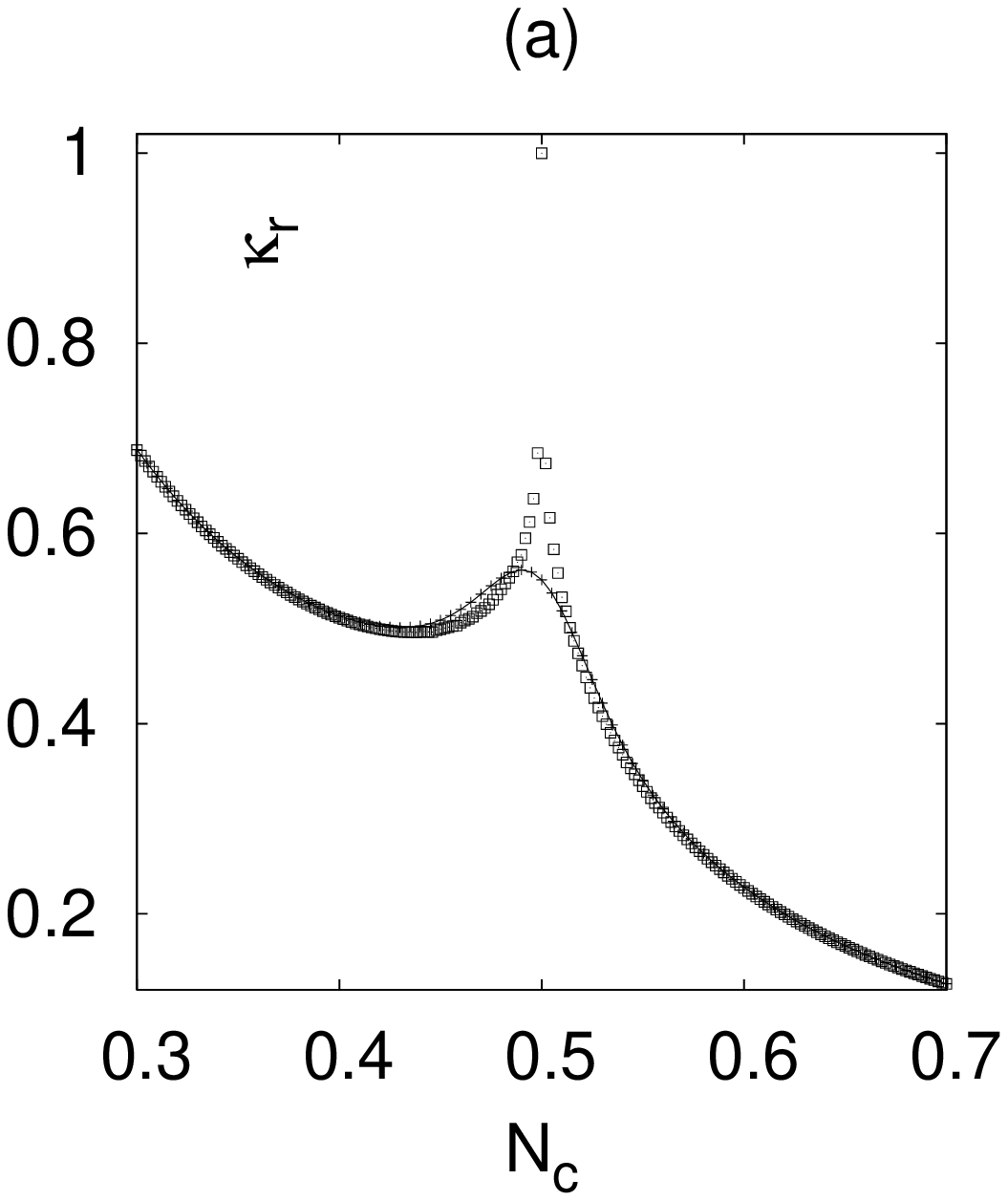} \includegraphics{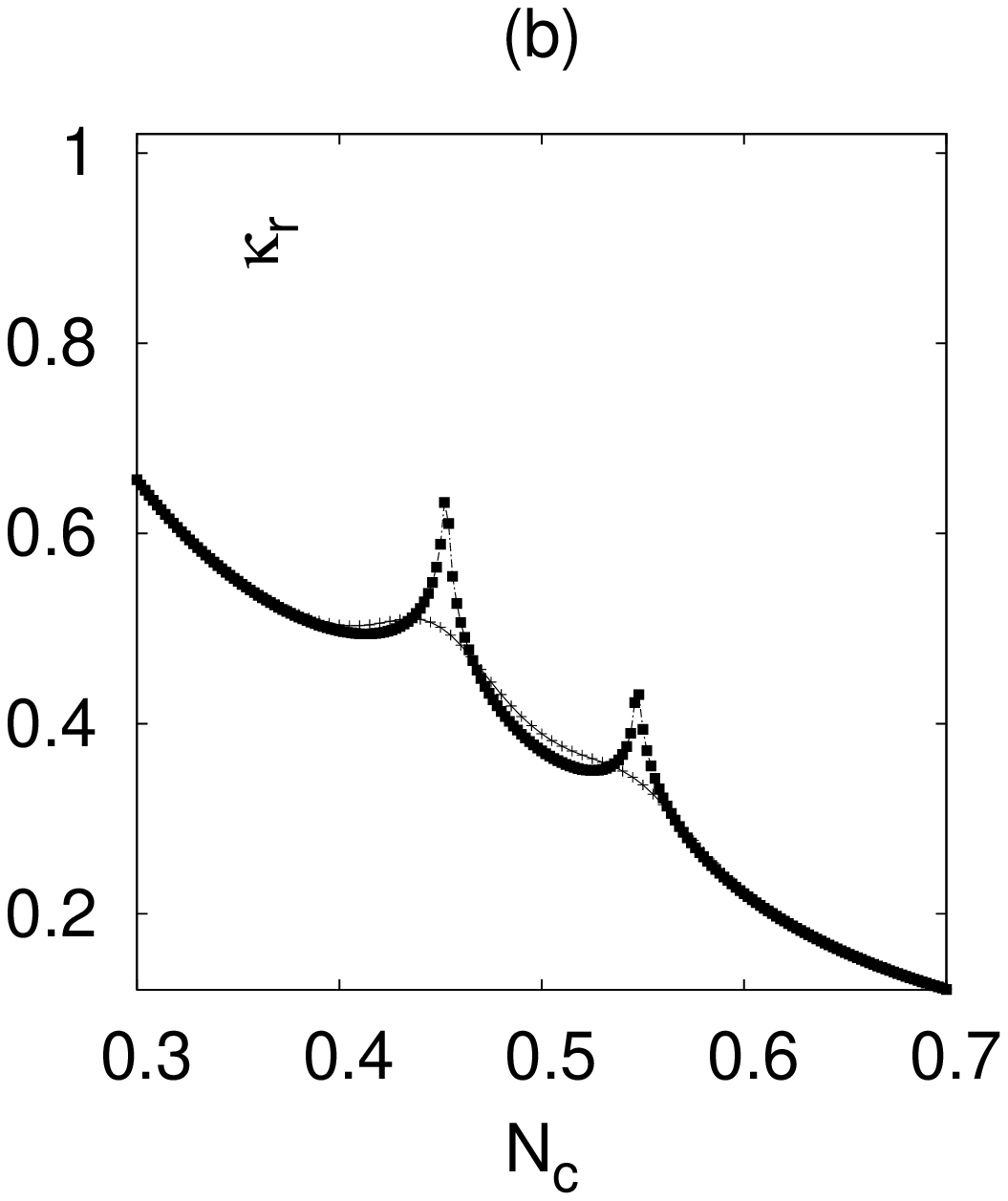}}}
\centerline{\scalebox{0.36}{\includegraphics{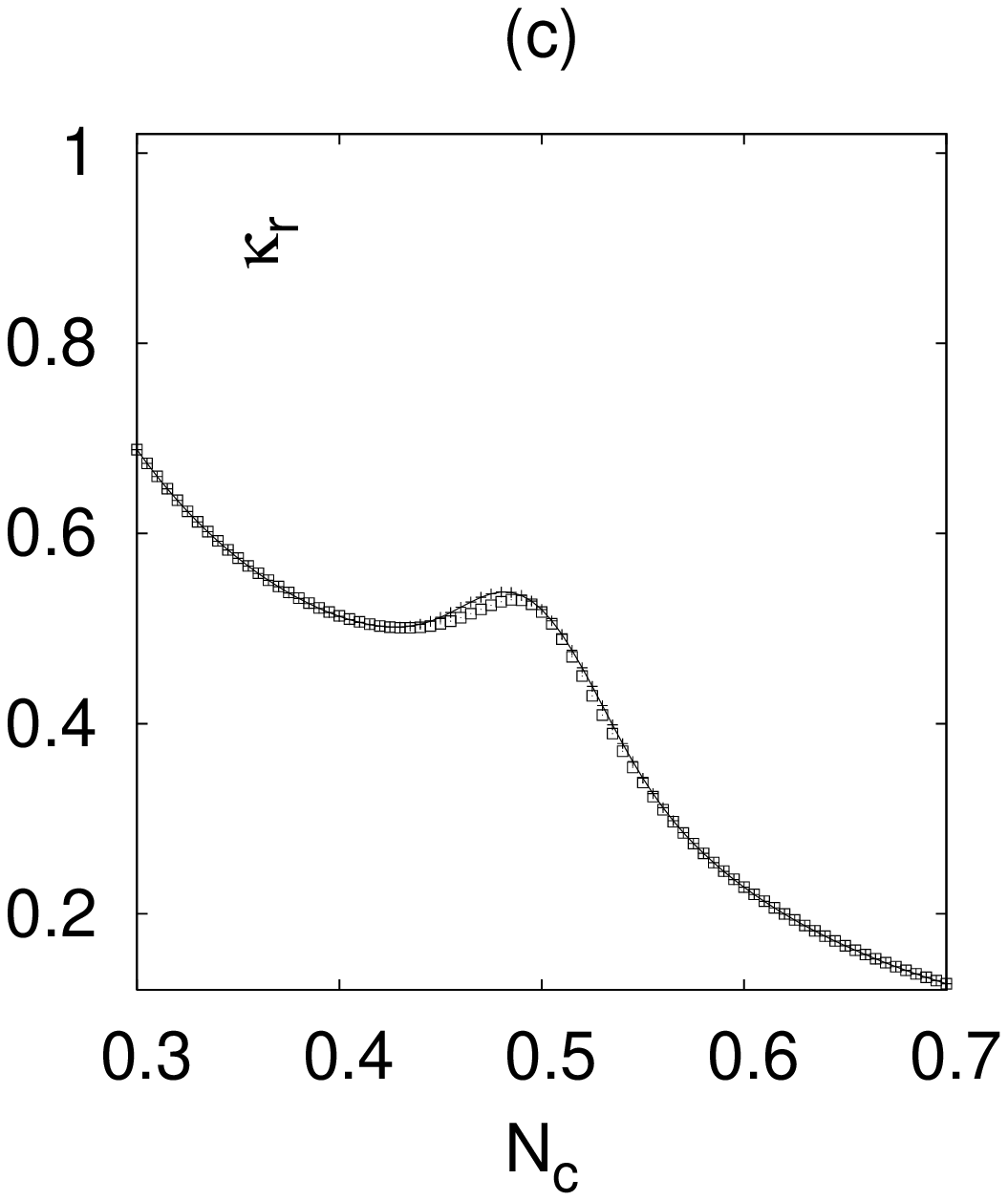} \includegraphics{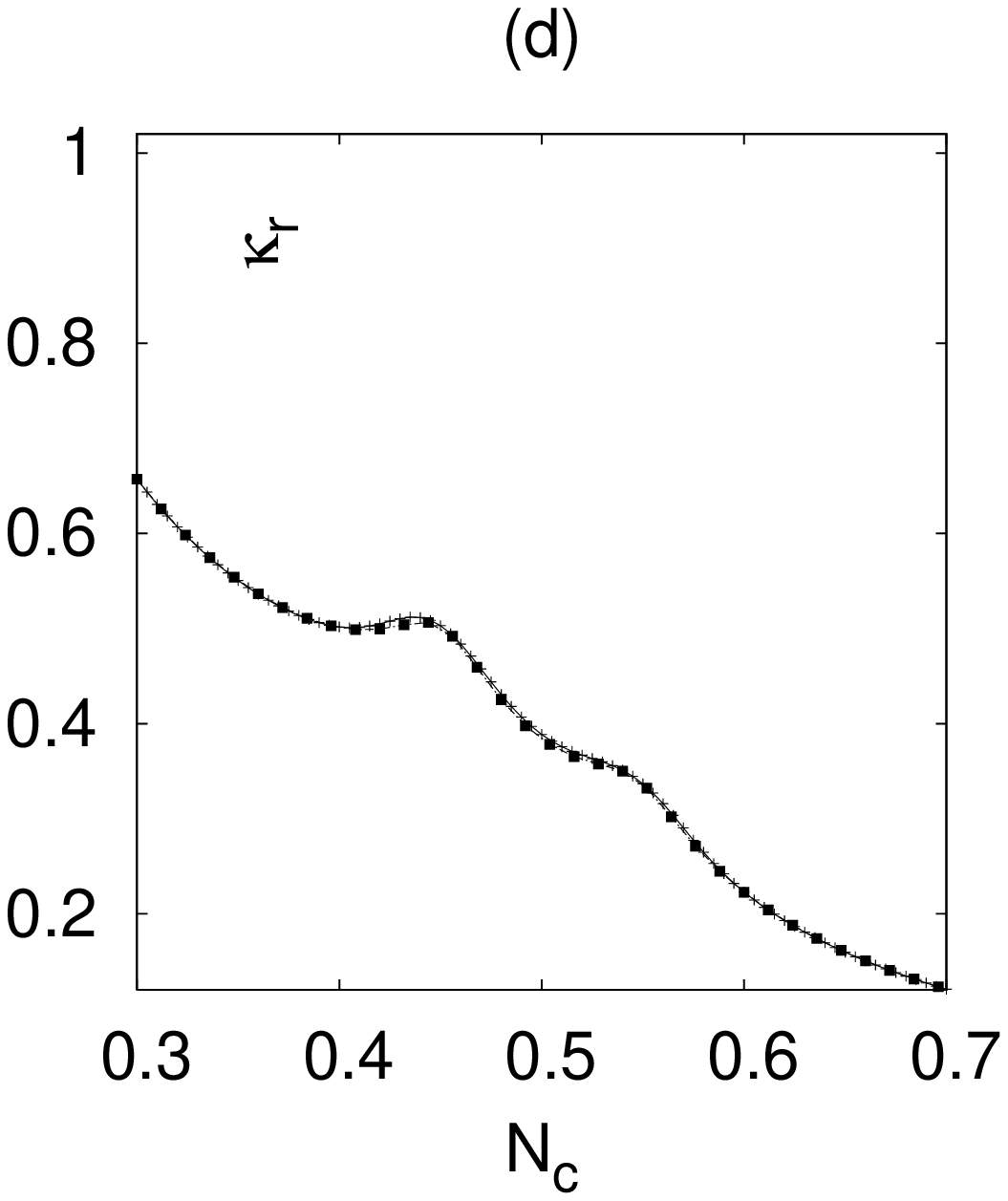}}}
\caption{\label{fig:ac} Plots of reduced compressibility $\kappa_r = \kappa_T(T)/\kappa_0$
versus filling factor $N_c$ for SHS $p$-wave symmetry in cases
a) (I) and
b) (II).
In addition, plots for $s$-wave symmetry are shown in cases 
c) (I) and
d) (II)
at temperatures $T = 0$ (squares), and $T = T_c$ (line).
Plots for $\alpha = 0$ and $0.1$ are shown with hollow and solid squares, respectively.
}
\end{figure}

Plots of the reduced isothermal atomic compressibility 
$\kappa_r = \kappa_T(T)/\kappa_0$ are shown in Fig.~\ref{fig:ac} for both SHS $p$-wave and $s$-wave superfluids 
for cases (I) and (II). For case (III), $\kappa_r$ is both qualitatively and quantitatively 
similar to case (II), and it is not shown here. The normalization constant
$\kappa_0$ is the SHS $p$-wave isothermal compressibility evaluated at $T=0$ and half-filling.

In case (I), $\kappa_r$ has a peak at half-filling
and low temperatures, and a hump at $T_c$ for SHS $p$-wave superfluids. Notice the strong temperature
dependence of $\kappa_r$ near half-filling. In the $s$-wave case there is only a hump both at 
$T_c$ and at low temperatures, and a very weak temperature dependence for all filling factors shown.
In case (II), $\kappa_r$ has the additional feature that the central peak (hump) splits into two 
due to degeneracy lifting, for the SHS $p$-wave case. Notice again the drastic difference between 
the values of $\kappa_r$ for $N_c \approx 0.45$ and $N_c \approx 0.55$ for 
$T \approx 0$ and $ T \approx T_c$. 
However, in the $s$-wave case the original hump also splits into two, but there is very weak
temperature dependence of $\kappa_r$.

To understand the peaks of $\kappa_r$ at $T \approx 0$, and its humps at 
$T = T_c$ limits for the SHS $p$-wave case, 
we indicate that $\partial N_c / \partial \tilde{\mu}$ 
[see Eq.~(\ref{eqn:dndmu})] has two contributions,
where the first term dominates at $T = 0$, and only the second term survives at $T = T_c$.
Therefore, the first term is responsible for the peaks, and the second term is
responsible for the humps.

The peaks at $T \approx 0$, can be understood by noting that
the first contribution to $\kappa_T(T)$ in Eq.~(\ref{eqn:dndmu}) can be written as
\begin{equation}
\kappa_T(T = 0 ) =  \frac{2}{N_c^2}\sum_{\mathbf{k}}
\frac{n(\mathbf{k}) [1 - n(\mathbf{k})]}{E(\mathbf{k})},
\label{eqn:zerokappa}
\end{equation}
where $n(\mathbf{k})$ is the momentum distribution and defined in Eq.~(\ref{eqn:md}).
Therefore, the peaks are due to non-vanishing $n(\mathbf{k}) [1 - n(\mathbf{k})]$
in regions of $\mathbf{k}$-space where $E(\mathbf{k})$ vanishes and
$n(\mathbf{k})$ is rapidly changing.

In case (I), the integrand $n(\mathbf{k}) [1 - n(\mathbf{k})]/ E(\mathbf{k})$ 
has 4 $\mathbf{k}$-space points ($0, \pm \pi$), and ($\pm \pi, 0$)
in the first Brillouin zone (1BZ) where it diverges only 
when the chemical potential $\tilde{\mu} = 0$ ($N_c = 0.5$). 
Similarly in case (II), the integrand diverges  
at 2 $\mathbf{k}$-space points ($\pm \pi, 0$) in the 1BZ 
when $\tilde{\mu} = t_x - t \approx 0.021$ ($N_c \approx 0.55$).
Furthermore,  the integrand diverges at 2 $\mathbf{k}$-space points ($0, \pm \pi$) when
$\tilde{\mu} = -t_x + t \approx -0.021$ ($N_c \approx 0.45$). 
For other values of $\tilde{\mu}$ (other filling factors), the integrand is well-behaved and does not produce 
additional peaks in $\kappa_r$.

In contrast, the compressibility peaks at $T = 0$ do not exist for $s$-wave superfluids
(see Figs.~\ref{fig:ac}c and~\ref{fig:ac}d),
where the quasi-particle excitation spectrum $E(\mathbf{k})$ is always gapped.
Generally speaking, we expect compressibility peaks for nodal superfluids or 
superconductors,
where quasi-particle energy spectrum vanishes in regions of $\mathbf{k}$-space.

At $T = T_c$, $\kappa_T(T)$ is dominated by 
the second term of the integrand (see Eq.~(\ref{eqn:dndmu}))
\begin{equation}
\kappa_T(T = T_c) =  \frac{1}{2T_cN_c^2}
\sum_{\mathbf{k}} \rm{sech}^2\frac{\beta \xi(\mathbf{k})}{2}.
\end{equation}
Therefore, the humps are not related to the order parameter (since at $T = T_c$, 
$\Delta(\mathbf{k}) = 0$),
but are due to the peaks appearing in the single fermion (normal) DOS
(see Fig.~\ref{fig:dos}c and Fig.~\ref{fig:dos}d).
This is simply because $\kappa_T(T = T_c)$ can be written in terms of
DOS (see Eq.~\ref{eqn:dos}) for both SHS $p$-wave and $s$-wave symmetries.
Notice that, while DOS have only one peak at half-filling in case (I) (leading to 
one hump in $\kappa_r$), DOS has two peaks in cases (II) and (III) (leading to two humps
in $\kappa_r$). Thus, at $T = T_c$ the humps correspond to a Van Hove phenomenon~\cite{economou}, 
but the atomic compressibility is smooth.
Furthermore, we expect humps at $T = T_c$ for all pairing symmetries, 
since these humps are related only to normal state properties 
but not to the symmetry of the pairing interaction.

Theoretically, the calculation of isothermal atomic compressibility 
$\kappa_T(T)$ is easier than 
the isentropic atomic compressibility $\kappa_S(T)$.
However, performing measurements of $\kappa_S(T)$ may be simpler in cold
Fermi gases, since the gas expansion upon release from the trap is expected to be 
nearly isentropic. Fortunately, $\kappa_S(T)$ is related to $\kappa_T(T)$ via the 
thermodynamic relation
\begin{equation}
\kappa_S(T) = \frac{C_V(T)}{C_P(T)}\kappa_T(T),
\end{equation}
where $\kappa_T(T) > \kappa_S(T)$ since specific heat capacitites $C_P(T) > C_V(T)$.
Furthermore, at low temperatures ($T \approx 0$) 
the ratio $C_P(T)/C_V(T) \approx \gamma$ is 
a constant, and therefore, $\kappa_S(T \approx 0) \propto \kappa_T(T \approx 0)$. 
Thus, we expect qualitatively similar behaviour
in both the isentropic and isothermal compressibilities at low temperatures $(T \approx 0)$, where 
appearence of such peaks could be used as a signature of SHS $p$-wave superfluidity.

The measurement of the atomic compressibility could also be performed via
an analysis of particle density fluctuations. As it is well know from thermodynamics~\cite{reif}
the isothermal atomic compressibility is connected to density fluctuations via
the relation
\begin{eqnarray}
\langle n^2 \rangle -\langle n \rangle^2 = \frac{\langle n \rangle^2 T}{\langle V \rangle}\kappa_T(T),
\end{eqnarray}
where $\langle n \rangle$ and $\langle V \rangle$ are average density of atoms and
volume of the condensate, respectively. From the measurement of density fluctuations the 
isothermal compressibility can be extracted at any temperature $T$.

Furthermore, the spin susceptibility tensor component $\chi_{\eta\eta}(T)$
can be written as a spin-spin response function
to a magnetic field ${\bf h} = h {\bf \hat \eta}$ applied along
an arbitrary $\eta$-direction~\cite{leggettHe}. 
In the case of SHS $p$-wave superfluids, 
\begin{equation}
\chi_{\eta\eta}(T)=-\frac{\partial^2 \Omega}{\partial h^2}
= g^2\mu_B^2 \frac{\partial N_c}{\partial \tilde{\mu}}.
\end{equation}
Therefore, for the SHS $p$-wave case, 
$\chi_{\eta \eta}(T)$ is directly related to the isothermal 
atomic compressibility and is given by $[N_c^2/(g\mu_B)^2]\kappa_T(T)$.
Thus, we expect similar effects in the magnetic spin susceptibility
as in the case of isothermal atomic compressibility discussed above.
In cold atoms, the measurement of the spin susceptibility will be most likely achieved
by using techniques similar to nuclear magnetic resonance (NMR), where
$\chi_{\eta\eta}(T)$ can be measured via the Knight shift~\cite{leggettHe}. 

Specifically, while we expect only one peak at exactly half-filling in the tetragonal lattices at $T = 0$,
two peaks will appear in the orthorhombic lattices for a SHS $p$-wave superfluid.
However at $T = T_c$, these peaks disappear and turn into humps.
Notice that the filling factor dependence of $\chi_{\eta\eta}(T)$ will be qualitatively different from 
$\kappa_T(T)$. This is because while $\chi_{\eta \eta}(T)$ is symmetric around half-filling
($N_c = 0.5$), $\kappa_T(T)$ is not.
The relation between the particle compressibility and the spin susceptibility given above 
is not valid for the $s$-wave case. Their relationship is more complicated in this case, and
we do not discuss it here. Another important property is the superfluid
density to be discussed next.

\subsection{Superfluid Density Tensor}
\label{sec:superfluid-density-tensor}

The superfluid density tensor is defined as a response function 
to phase twists of the order parameter~\cite{feynman}. 
In the approximation used in this paper, 
the temperature dependence of its components is given by
\begin{equation}
\rho_{ij}=\frac{1}{2V}\sum_{\mathbf{k}}\left[ n(\mathbf{k})\partial_{i}\partial_{j}\xi(\mathbf{k}) 
-Y(\mathbf{k})\partial_{i}\xi(\mathbf{k})\partial_{j}\xi(\mathbf{k})\right],
\end{equation}
where $n(\mathbf{k})$ is the momentum distribution, 
$Y(\mathbf{k})$ is the Yoshida function defined in the previous section,
and $\partial_i$ denotes the partial derivative with respect to $k_i$.
Generally speaking, there are two components to the reduction of the superfluid density at non-zero 
temperatures, one coming from fermionic (quasi-particle excitations) and
the other bosonic (collective modes) degrees of freedom. 
Here, we do not discuss the bosonic contribution, except to say that 
at low temperatures the dominant terms come from Goldstone modes 
associated with the phase of the order parameter~\cite{anderson}, which 
are underdamped in our case due to sub-critical Landau damping. 
Furthermore, in the case of tetragonal symmetry,
Goldstone modes do not contribute to the off-diagonal component of the superfluid density, 
which is the main focus of the analysis that follows.
\begin{figure} [ht]
\centerline{\scalebox{0.55}{\includegraphics{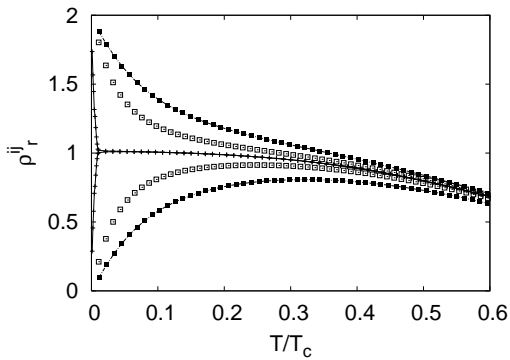}}}
\caption{\label{fig:sd1} Plots of reduced diagonal ($\rho^{xx}_r = \rho^{yy}_r$; upper curves) 
and off-diagonal ($\rho^{xy}_r = \rho^{yx}_r$; lower curves) 
superfluid density tensor components $\rho^{ij}_r = \rho_{ij} / \rho_{xy}^{\rm{max}}$
versus temperature $T/T_c$ for SHS $p$-wave superfluids in tetragonal lattices (case (I)).
$N_c=0.5, 0.45$ and $0.4$  are shown with line, hollow and solid squares.
}
\end{figure}
\begin{figure} [ht]
\centerline{\scalebox{0.55}{\includegraphics{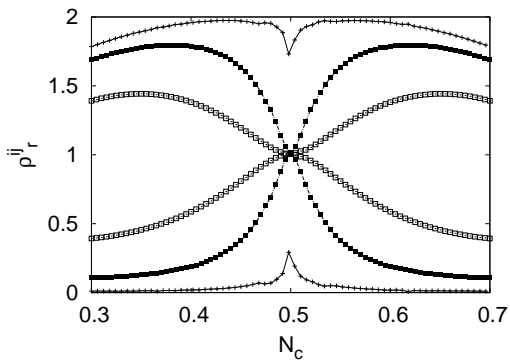}}}
\caption{\label{fig:sd2} Plots of reduced diagonal ($\rho^{xx}_r = \rho^{yy}_r$; upper curves) 
and off-diagonal ($\rho^{xy}_r = \rho^{yx}_r$; lower curves) 
superfluid density tensor components $\rho^{ij}_r = \rho_{ij} / \rho_{xy}^{\rm{max}}$
versus number of atoms per unit cell $N_c$ for SHS $p$-wave superfluids in tetragonal lattices (case (I)).
$T_1=0.002T_0$, $T_2=0.02T_0$ and $T_3=0.1T_0$ are shown with line, solid and hollow squares.
Here $T_0$ is the critical temperature at half-filling.
}
\end{figure}

We discuss first $\rho_{ij}$ for SHS $p$-wave superfluids. 
Plots of $\rho^{ij}_r = \rho_{ij} / \rho_{xy}^{\rm{max}}$
are shown in Fig.~\ref{fig:sd1} as a function of temperature for case (I), 
and three filling factors:
$N_c = 0.5$ (line), $N_c = 0.45$ (hollow squares), and $N_c = 0.4$ (solid squares).
The normalization constant 
$\rho_{xy}^{\rm{max}}$ is the maximum value of the off-diagonal component in a
square lattice. This maximum occurs at half-filling and 
$T \approx 0.02T_0$ in a square lattice, where $T_0$ is the critical temperature at
half-filling.
It is important to emphasize that square lattices 
[case (I)] have identical diagonal elements $\rho_{xx}=\rho_{yy}$ 
due to tetragonal symmetry, but have nonzero $\rho_{xy}$ component 
as a result of the absence of reflection 
symmetry in the $yz$-plane ($x \to -x$) or the $xz$-plane ($y \to -y$) 
for the $d$-vector defined in Eq.~(\ref{dvector}). 
In cases (II) and (III), reflection symmetry is restored and
$\rho_{xy}$ vanishes identically in the orthorhombic case for any temperature $T$.

In the case of tetragonal symmetry [case (I)], $\rho_{xy}$ has a peak as a function of $N_c$,
but the diagonal components have a dip of the same size at exactly half-filling.
In Fig.~\ref{fig:sd2}, we plot $\rho_{ij}/\rho_{xy}^{\rm{max}}$ as a function of
$N_c$ for three temperatures.

Furthermore, notice that at $T = 0$ the superfluid density tensor is reduced to 
\begin{equation}
\rho_{ij} = 
\frac{1}{2V}\sum_{\mathbf{k}} n(\mathbf{k})\partial_{i}\partial_{j}\xi(\mathbf{k}),
\end{equation}
which is just the integral of the momentum distribution weighted by the 
the curvature tensor of the dispersion $\xi(\mathbf{k})$. Thus, 
$\rho_{xy}$ is zero at $T = 0$ for any filling factor. However, as $T$ increases $\rho_{xy}$
increases and reaches approximately the same values as $\rho_{xx}$ and $\rho_{yy}$. Further 
increase of $T$ leads all tensor components of $\rho_{ij}$ to vanish at $T = T_c$, as expected.

In contrast, for $s$-wave superfluids, the off-diagonal tensor elements ($\rho_{xy}, \rho_{yx}$) 
are zero both in square (case (I)) and orthorhombic lattices (cases (II) and (III)). 
However, similar to SHS $p$-wave superfluids, 
$\rho_{xx} = \rho_{yy}$ in the square and $\rho_{xx} \ne \rho_{yy}$
in the orthorhombic lattices. We do not plot $\rho_{xx}$ and $\rho_{yy}$ for the $s$-wave case,
because our main interest here is the analysis of $\rho_{xy}$ which is strictly zero in
this case. We note in passing that the main difference between the 
diagonal components of $s$-wave 
and SHS $p$-wave superfluid densities is their temperature dependence. In the $s$-wave case at low
temperatures, $\rho_{xx}$ and $\rho_{yy}$ deviate from their zero temperature value only by a
small exponential correction, while in the SHS $p$-wave case the deviation is a power law.

Measurements of the superfluid density in cold atoms might be performed 
through the rotation of the condensate in combination with experimental
techniques that are sensitive to rotations. Through this rotational sensitivity
it may be possible to extract the velocity fields and their correlations,
as it is currently possible with positional (density) sensitive 
techniques~\cite{jin-correlations}. 
This experiment in cold atoms would be the analogue of the rotating bucket (or cylinder) 
experiments in liquid Helium~\cite{leggettHe}.

In this section, we have discussed some of the experimentally relevant quantities
of SHS $p$-wave superfluids and compared them with the $s$-wave superfluids in the
weak coupling BCS regime. Furthermore, we presented signatures of the SHS $p$-wave superfluid 
state, which can be used to identify superfluidity in this limit.
In the next section, we study some of these thermodynamic quantitites
as a function of interaction strength from weak (BCS) to strong (BEC) coupling regimes at low filling factors, 
and complete our analysis with a discussion of signatures of a possible 
BCS-to-BEC quantum phase transition for SHS $p$-wave superfluids.

\section{Superfluid properties in the BCS-to-BEC Evolution}
\label{sec:BCS-BEC}

At very low temperatures $(T \approx 0)$, the saddle point order parameter
and the number equations are approximatelly valid for all couplings from weak
to strong coupling regimes, where a small correction to the number
equation is negligible~\cite{carlos}. 

For low filling factors $0 < N_c < 0.5$, the chemical potential 
$\tilde{\mu} = -|\tilde{\mu}|$ decreases as a function of interaction 
strength $V_{0,x}$ and crosses the bottom of
the energy band ($\epsilon_{\rm{min}} = - t_x - t_y$) at some critical value of $V_{0,x}$. The decrease 
of $\tilde{\mu}$ is associated with the formation of bound particle pairs which are pulled out of the
two-particle continuum.
Similarly, for high filling factors $0.5 < N_c < 1$, $\tilde{\mu} = |\tilde{\mu}|$ 
increases with increasing $V_{0,x}$ and crosses the top of the energy band $\epsilon_{\rm{max}} = t_x + t_y$ at some critical value of $V_{0,x}$. 
The increase of $\tilde{\mu}$ is associated with the formation of bound hole pairs which are pulled out
of the two-hole continuum.  
However, exactly at half-filling $N_c = 0.5$, $\tilde{\mu}$ is pinned at zero for any interaction strength
due to perfect particle-hole symmetry.

Notice that the situation in lattices is very different from the continuum, because of particle-hole
symmetry. In the continuum the chemical potential is only pulled down below the bottom of the band since
there is no energy upper bound for a parabolic dispersion.
In the lattice case, as discussed above, the chemical potential can move below (above) the bottom (top) 
of the band for filling factors below (above) half-filling.
Notice in addition, that for  optical lattices of cold fermion
atoms at high filling factors, superfluidity is associated with correlated motion of holes (atom voids),
rather than particles (atoms).

\subsection{Phase Diagram}
\label{sec2:phase-diagram}

\begin{figure} [htb]
\centerline{\scalebox{0.37}{\includegraphics{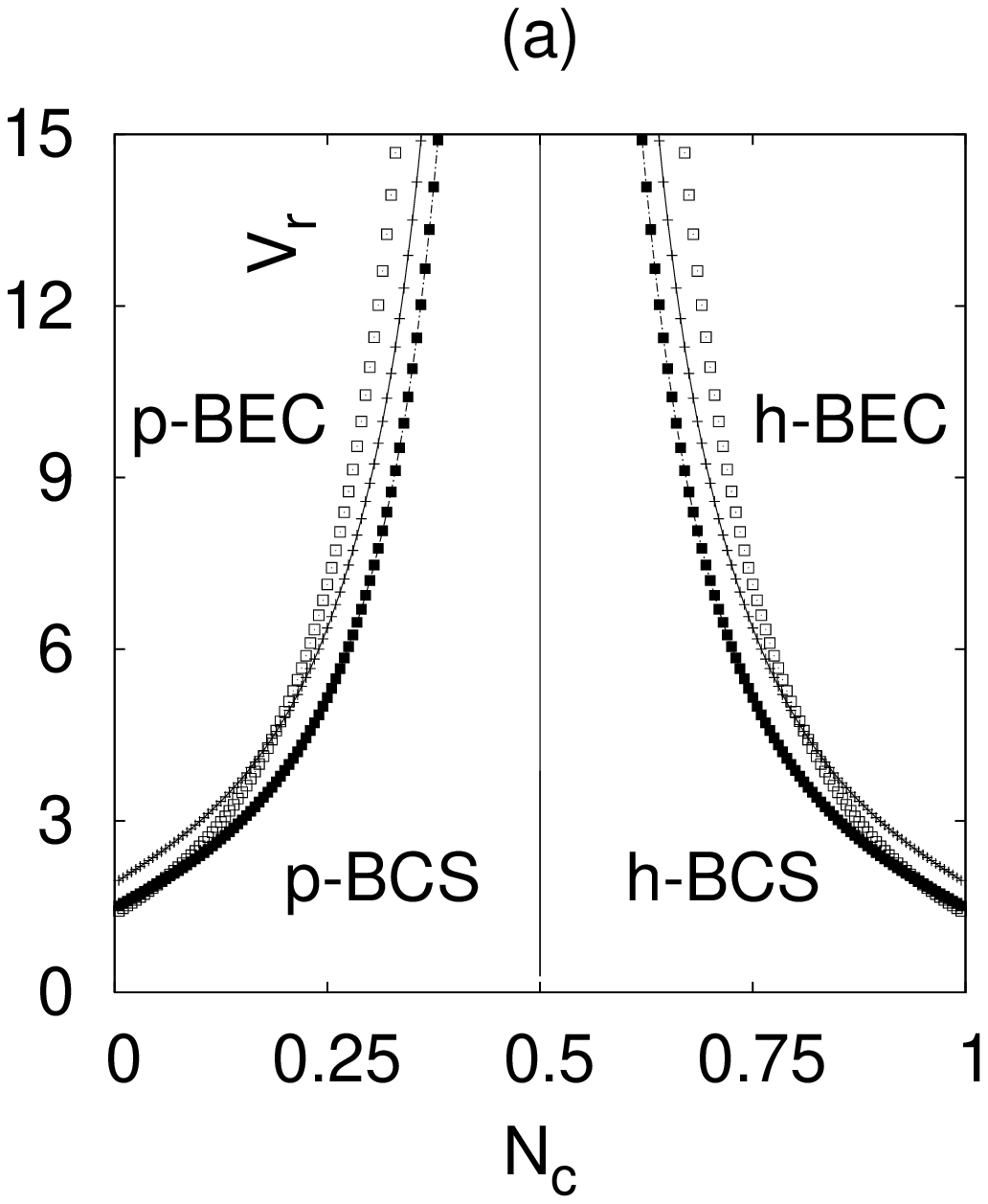} \includegraphics{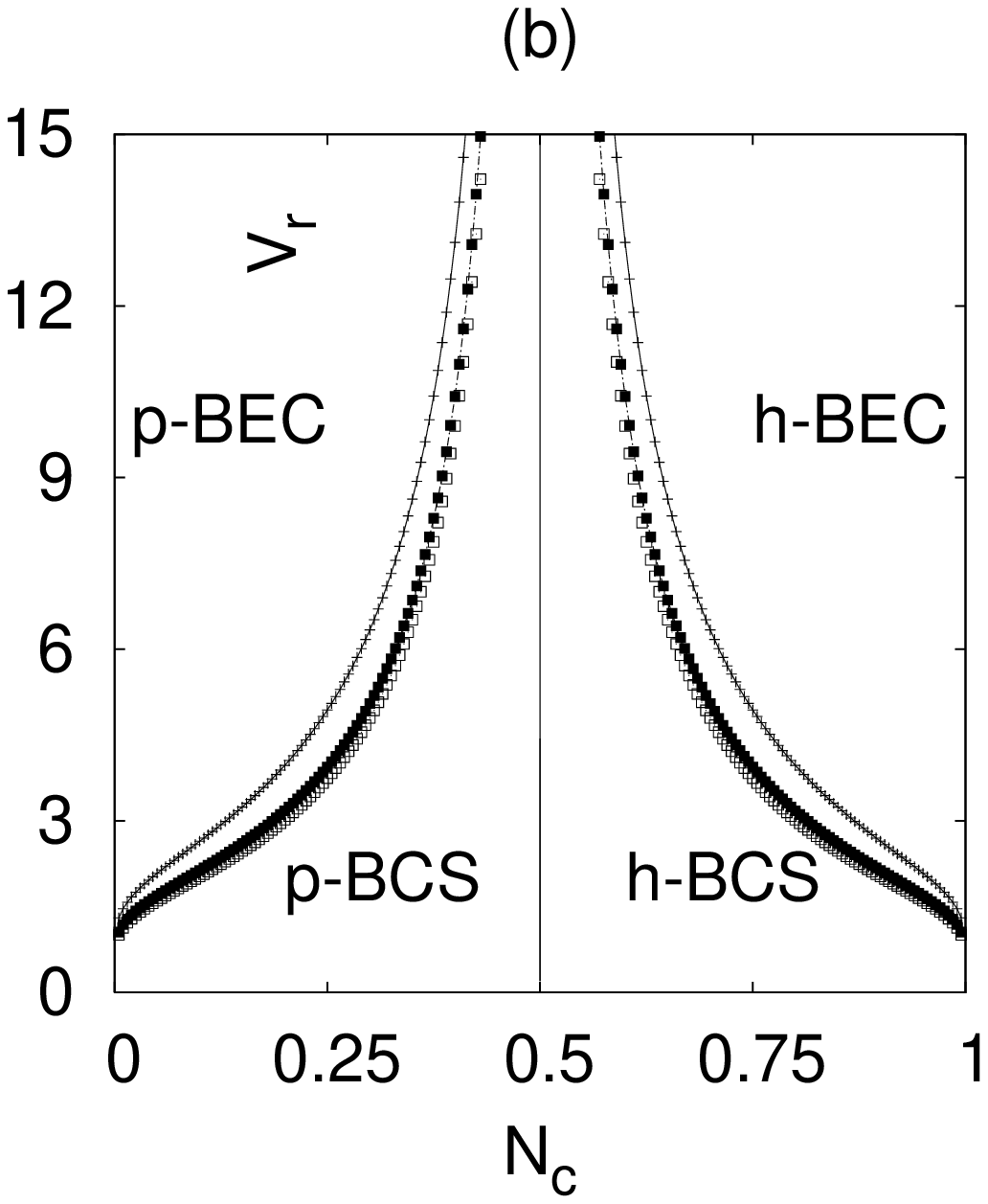}}}
\caption{\label{fig:phase} Plots of critical interaction strength $V_r = V_{0,x}/E_0$
and $V_r = V_{0,s}/E_0$ versus number of atoms per unit cell $N_c$ for
a) SHS $p$-wave, and 
b) $s$-wave superfluids,
respectively at $T = 0$.
Plots for $\alpha = 0, 0.1$ and $0.6$ are shown with hollow squares, solid squares and line, respectively.
}
\end{figure}

Depending on the behavior of $\tilde \mu$, 
the $T = 0$ phase diagram can be plotted as a function of filling factor $N_c$ 
and interaction strength $V_{0,x}$.
The BCS region, where $|\tilde \mu| < t_x + t_y $, corresponds 
to the weaker interaction region for fixed density
as shown in Fig.~\ref{fig:phase}. In the BCS region there is 
a well defined Fermi surface locus where $\xi ({\mathbf k}) = 0$.
The BEC region, where $|\tilde \mu| >  t_x + t_y$, correponds 
to the stronger interaction region for fixed density
as shown in Fig.~\ref{fig:phase}. In the BEC region there is 
no Fermi surface locus, since $\xi ({\mathbf k})$ can not be zero.

In Fig.~\ref{fig:phase} we also compare the phase diagrams of SHS $p$-wave and $s$-wave superfluids. 
In case (I) (hollow squares) of SHS $p$-wave superfluids, 
a wide region of filling factors around half-filling requires very strong critical $V_{0,x}$ to 
reach the BEC regime. For a small anisotropy $\alpha = 0.1$ (solid squares) corresponding
to case (II), this region narrows. However, further anisotropy $\alpha = 0.6$ (line),
this region widens again. For $s$-wave superfluids,
in case (I) (hollow squares), the region around half-filling expands very little for small anisotropy 
$\alpha = 0.1$ (solid squares) corresponding
to case (II). Further anisotropy $\alpha = 0.6$ (line) expands the BCS region
around half-filling, thus making it more difficult to reach the BEC regime at
fixed filling factor.

Notice in addition, that when the filling factor $N_c \to 0$ or $N_c \to 1$
a finite interaction strength is necessary to evolve from the BCS to the BEC regime in both
SHS $p$-wave and $s$-wave superfluids. This indicates that two-particle (or two-hole) bound
states in a two-dimensional lattice require finite energy for the symmetries discussed.
This is in contrast with the situation found in two dimensional continuum models, where
a two-body bound state is found at arbitrarily small attractive interaction for $s$-wave.
However, in two-dimensional continuum models for the SHS $p$-wave case, 
the creation of a two-body bound state requires a finite interaction strength. 

Furthermore, there are two possible ways of investigating the 
evolution from BCS to BEC regime in cold atom experiments. 
The first way is by changing the interaction strength for a fixed filling factor,
while the second is by changing filling factor for a fixed interaction strength. 
Probably both ways are possible in cold atom experiments, where
interaction strength and atom filling factor could be tuned independently. 

In all physical properties to be discussed in the next sections, first
we fix the filling factor to $N_c = 0.25$ (quarter-filling) 
and vary $V_r$ to cross the BCS-BEC boundary. But, in addition,  
we fix the interaction strength to $V_r = 6$, and vary $N_c$ to cross the BCS-BEC boundary.
We present results for tetragonal (case (I) with $\alpha = 0$) 
and orthorhombic (case (II) with $\alpha = 0.1$) lattices.
Notice that cases (II) and (III) have similar behavior, and differ 
only by a scale factor. Thus
we do not present case (III) here.

\subsection{Quasi-particle Excitation Spectrum}
\label{sec2:excitation-spectrum}

The quasi-particle excitation spectrum of SHS $p$-wave and $s$-wave superfluids during the
BCS to BEC evolution are very different because of the symmetry of the order parameter.
As discussed in Sec.~\ref{sec:hamiltonian}, the quasi-particle spectrum of
SHS $p$-wave superfluids is given by 
$E(\mathbf{k}) = \sqrt{\xi^2(\mathbf{k}) + |\Delta(\mathbf{k})|^2}$ 
with $\xi(\mathbf{k}) = \epsilon(\mathbf{k}) - \tilde{\mu}$. 
Since the cross-product 
$d^* ({\mathbf k}) \times d ({\mathbf k}) \ne 0$ is non-zero,
it is expected that triplet superfluids have additional quasi-particle excitation 
branches. However, for the SHS $p$-wave state these branches do not enter
the problem as they are pushed to extremely high energies and are not accessible.
For instance, when a THS $p$-wave superfluid 
is formed from a pseudo-spin 1/2 system, there are 
two accessible quasi-particle branches when time reversal symmetry is broken. 
Thus, in the SHS $p$-wave there is only one quasi-particle energy branch.

For the SHS $p$-wave case, notice that $E(\mathbf{k}) = 0$
when both $\xi ({\mathbf k}) = 0$ and $\Delta ({\mathbf k}) = 0$. This
implies that the momentum space region of zero $E(\mathbf{k})$ 
occurs when $-t_x\cos (k_x a_x) - t_y\cos (k_y a_y) = \tilde \mu$ 
and $\Delta_{0,x}\sin (k_x a_x) + \Delta_{0,y}\sin (k_y a_y) = 0$
for chemical potentials inside the BCS region ($|\tilde{\mu}| < t_x + t_y $).
This means that in the BCS region the quasi-particle excitation spectrum is 
gapless.
However, $E(\mathbf{k})$ never vanishes 
for chemical potentials inside the BEC region ($|\tilde{\mu}| > t_x + t_y $),
since $\xi ({\mathbf k})$ can never be zero. However, the order parameter
$\Delta ({\mathbf k})$ can still be zero. This implies
that in the BEC region the quasi-particle excitation spectrum is 
fully gapped.
Notice that this change in minimum excitation energy accompanies the existence
or non-existence of the Fermi surface locus $\xi ({\mathbf k}) = 0$.
Since quasi-particle excitations are evolving from gapless BCS to gapped BEC regime 
as a function of interaction strength for fixed filling factor, or 
as a function of filling factor for fixed interaction strength, 
the evolution between the BCS and BEC regimes is not smooth and a
quantum phase transition occurs.
As we shall see in Sec.~\ref{sec2:isothermal-atomic-compressibility}, 
where ground state properties like the isothermal compressibility 
are calculated, this quantum phase transition is characterized by the chemical 
potential crossing either the bottom  $(\tilde{\mu}_c = -t_x -t_y)$ or the 
top $(\tilde{\mu}_c = t_x + t_y)$ of the energy band.

In contrast, for the $s$-wave case, $E(\mathbf{k})$ never vanishes and is always gapped
in both the BCS and BEC regimes. This is a major difference between SHS $p$-wave 
(or more generally any nodal superfluids) and $s$-wave superfluids.
In $s$-wave superfluids quasi-particle excitations are always gapped
and the evolution from the BCS to the BEC regime is smooth, 
so that no quantum phase transition occurs.
This gapless to gapped quantum phase transitions were first considered in the
context of $\rm{He^3}$ by Volovik~\cite{volovik}, and later in the context of
high-$T_c$ superconductivity~\cite{borkowski,duncan} and atomic Fermi gases~\cite{botelho1,botelho2}.

\subsection{Chemical Potential}
\label{sec2:chemical-potential}

\begin{figure} [htb]
\centerline{\scalebox{0.37}{\includegraphics{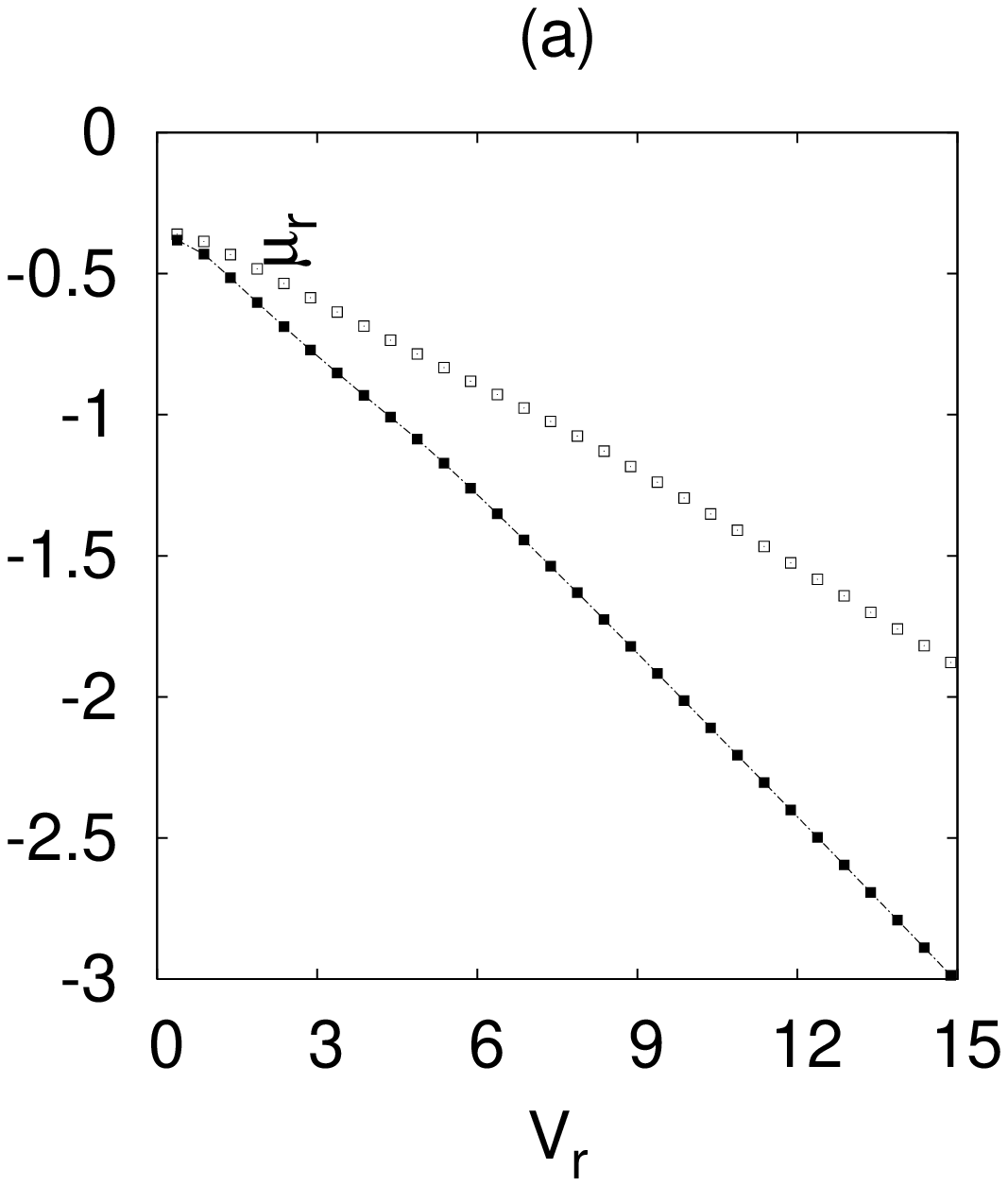} \includegraphics{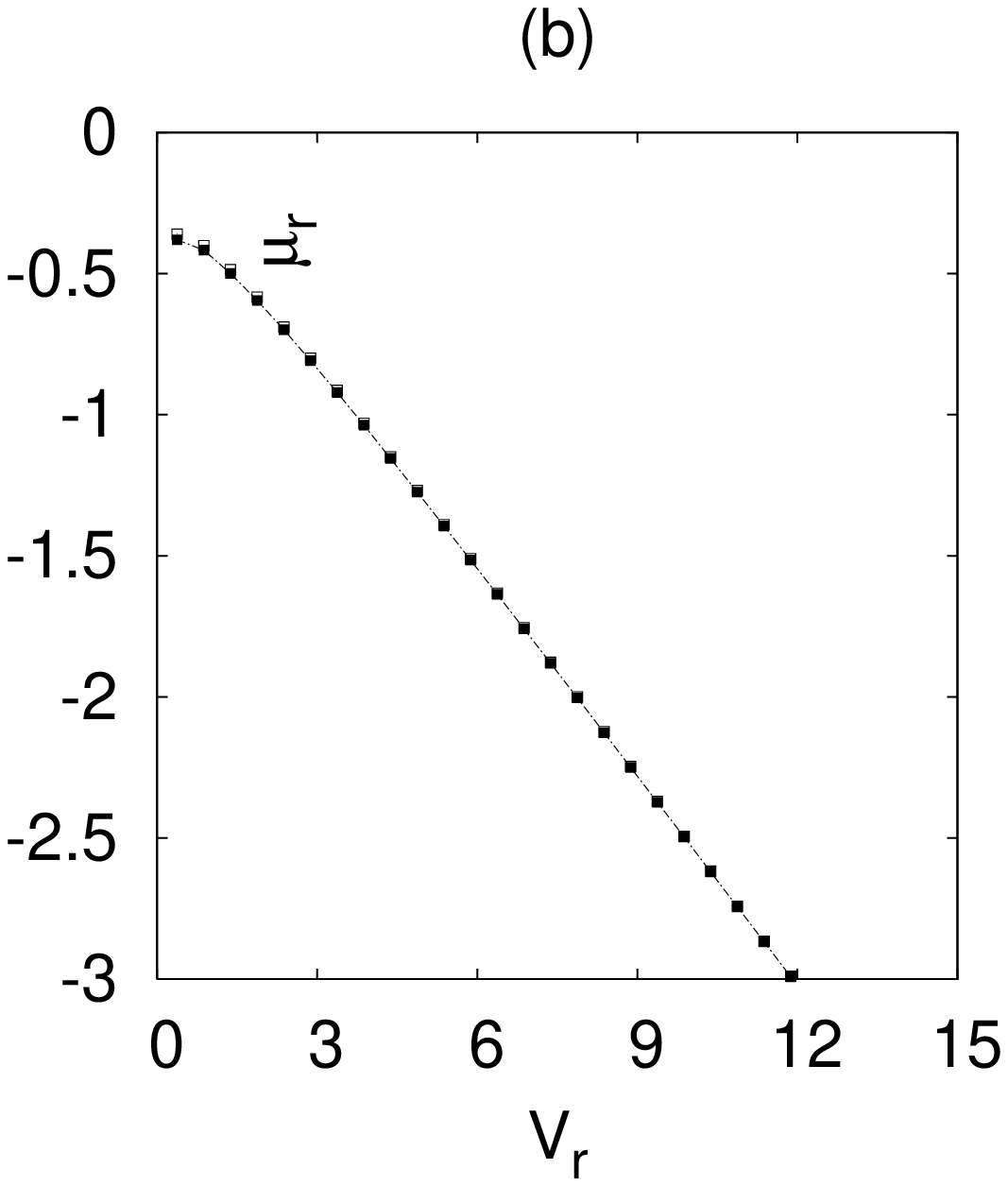}}}
\caption{\label{fig:crmu} Plots of reduced chemical potential $\mu_r = \tilde{\mu}/E_0$,  
versus reduced interaction strength $V_r = V_{0,x}/E_0$ and $V_r = V_{0,s}/E_0$ for 
a) SHS $p$-wave, and 
b) $s$-wave superfluids, 
respectively at $T = 0$ and $N_c = 0.25$.
Plots for $\alpha = 0$ and $0.1$ are shown with hollow and solid squares, respectively.
}
\end{figure}

Plots of the reduced chemical potential $\mu_r = \tilde{\mu}/E_0$ 
at low temperatures ($T \approx 0 $) and quarter filling factor ($N_c = 0.25$)
are shown in Fig.~\ref{fig:crmu} as a function of
reduced interaction $V_r = V_{0,x}/E_0$ and $V_r = V_{0,s}/E_0$ for SHS $p$-wave and $s$-wave superfluids, respectively.
The tetragonal case (I) with $\alpha = 0$ and orthorhombic case (II) with $\alpha = 0.1$ 
are shown with hollow and solid squares, respectively.
In the case of SHS $p$-wave superfluids, $\mu_r$ is a continuous function of $V_r$, but
its slope with respect to $V_r$ has a cusp
when $\tilde{\mu}$ crosses the bottom of the energy band; $\mu_r = -1$ ($\alpha = 0$: hollow squares) 
and $\mu_r = -1.1$ ($\alpha = 0.1$ :solid squares).
However in the $s$-wave case, both $\mu_r$ and the slope of $\mu_r$ with respect to $V_r$
are continuous functions of $V_r$.

\begin{figure} [htb]
\centerline{\scalebox{0.37}{\includegraphics{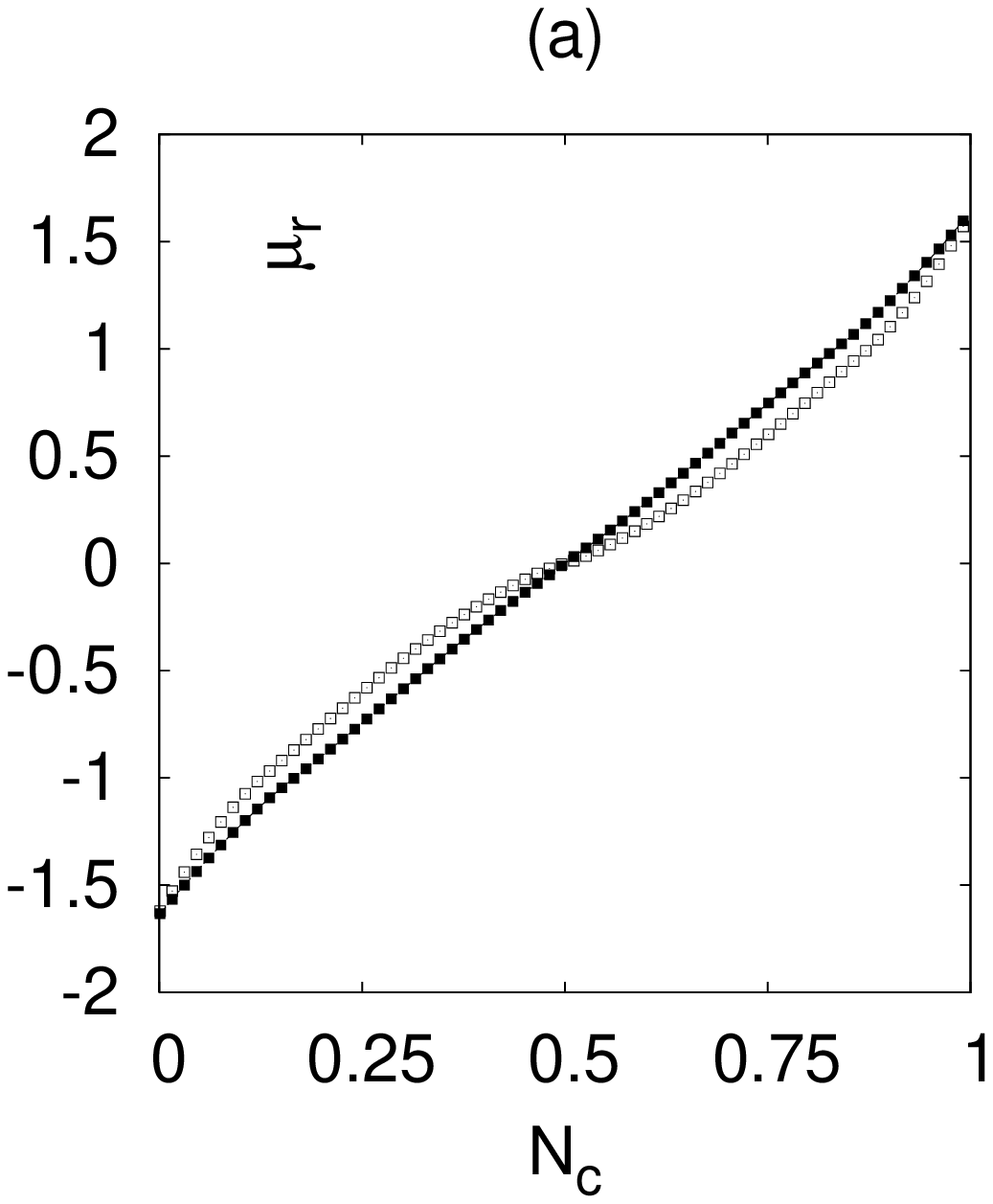} \includegraphics{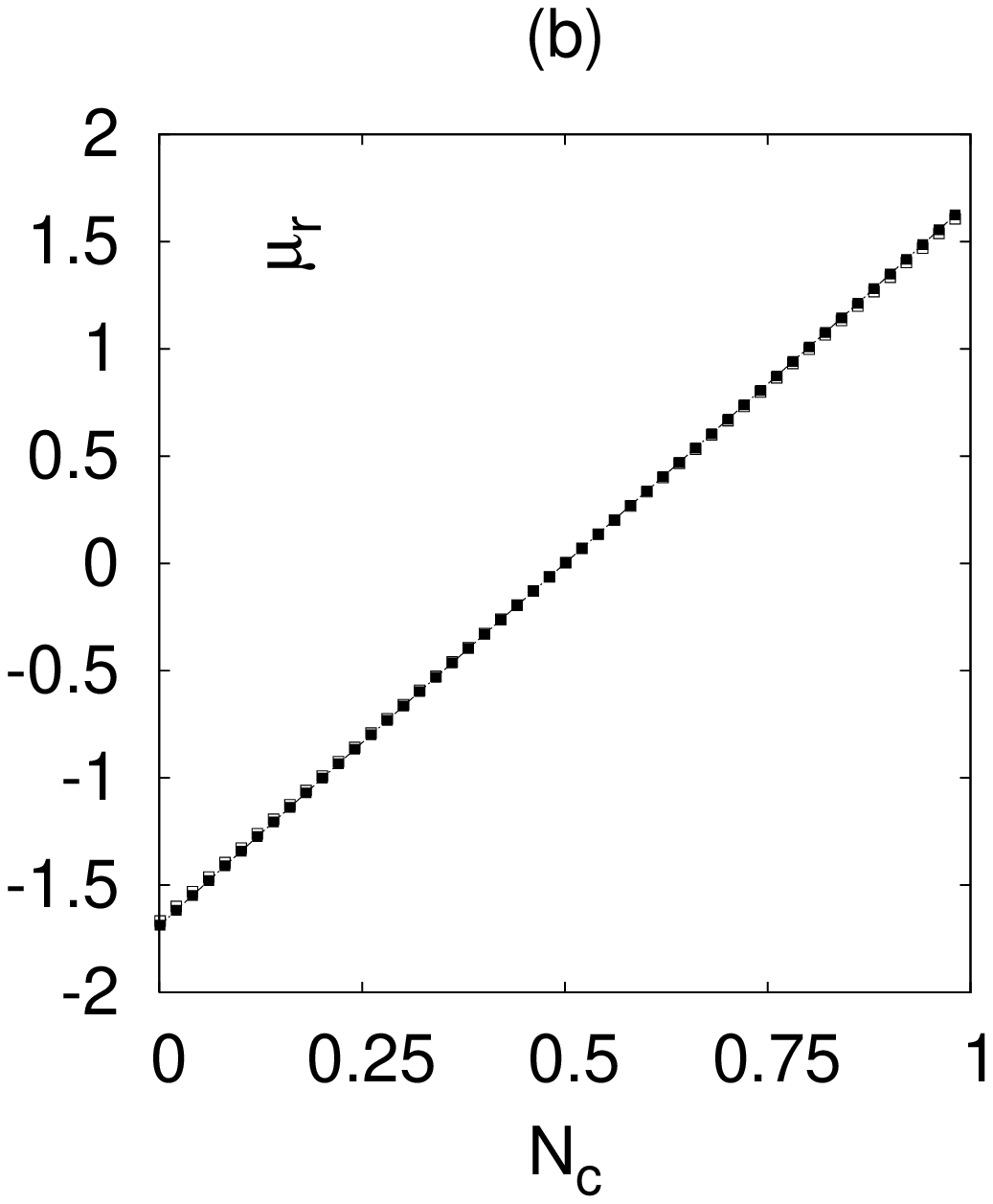}}}
\caption{\label{fig:crmu.V.const} Plots of reduced chemical potential $\mu_r = \tilde{\mu}/E_0$,  
versus filling factor $N_c$ for 
a) SHS $p$-wave, and 
b) $s$-wave superfluids
at $T = 0$ and $V_r = 6$.
Plots for $\alpha = 0$ and $0.1$ are shown with hollow and solid squares, respectively.
}
\end{figure}

Plots of the reduced chemical potential $\mu_r = \tilde{\mu}/E_0$ 
at low temperatures ($T \approx 0 $) and constant interaction strength ($V_r = 6$)
are shown in Fig.~\ref{fig:crmu.V.const} as a function of
filling factor $N_c$ for both SHS $p$-wave and $s$-wave superfluids.
Here again the tetragonal case (I) with $\alpha = 0$ and orthorhombic case (II) with $\alpha = 0.1$
are shown with hollow and solid squares, respectively.
In the case of SHS $p$-wave superfluids, $\mu_r$ is a continous function of $N_c$, but
its slope with respect to $N_c$ has a cusp
when $\tilde{\mu}$ crosses the bottom of the energy band; $\mu_r = -1$ ($\alpha = 0$: hollow squares) 
and $\mu_r = -1.1$ ($\alpha = 0.1$: solid squares). In addition, $\mu_r$ changes curvature
at $\mu_r = 0$ ($N_c = 0.5$) at the same place, where the topology of the Fermi 
surface locus $\xi ({\mathbf k}) = 0$ changes from particle-like to hole-like.
However, in the $s$-wave case, both $\mu_r$ and the slope of $\mu_r$ with respect to $N_c$
are continuous functions of $N_c$, and there is no change of slope when the topology of Fermi surface 
locus $\xi(\mathbf{k}) = 0$ changes. Thus, the slope change of $\mu_r$ for the SHS $p$-wave case
is directly related to a higher angular momentum pairing channel $(\ell \ne 0)$.

This difference between SHS $p$-wave and $s$-wave superfluids have great importance in describing
the evolution from BCS to BEC regime. As discussed above, 
the major qualitative differences between the chemical potentials of 
SHS $p$-wave and $s$-wave cases are more clearly seen in the derivative
$\partial \tilde{\mu} / \partial N_c$, which is directly related to
the atomic compressibility to be discussed next.

\subsection{Atomic Compressibility}
\label{sec2:isothermal-atomic-compressibility}

\begin{figure} [htb]
\centerline{\scalebox{0.37}{\includegraphics{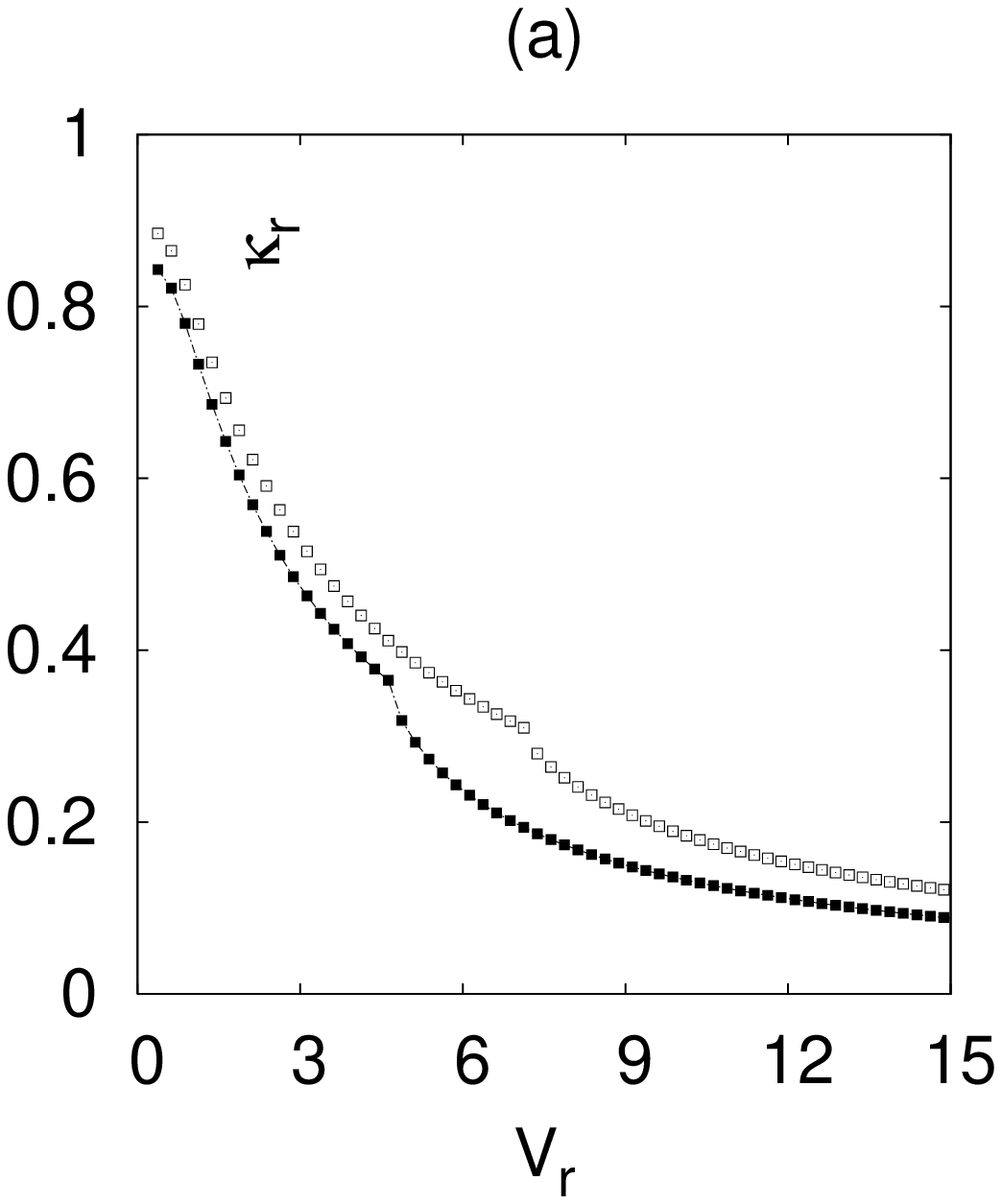} \includegraphics{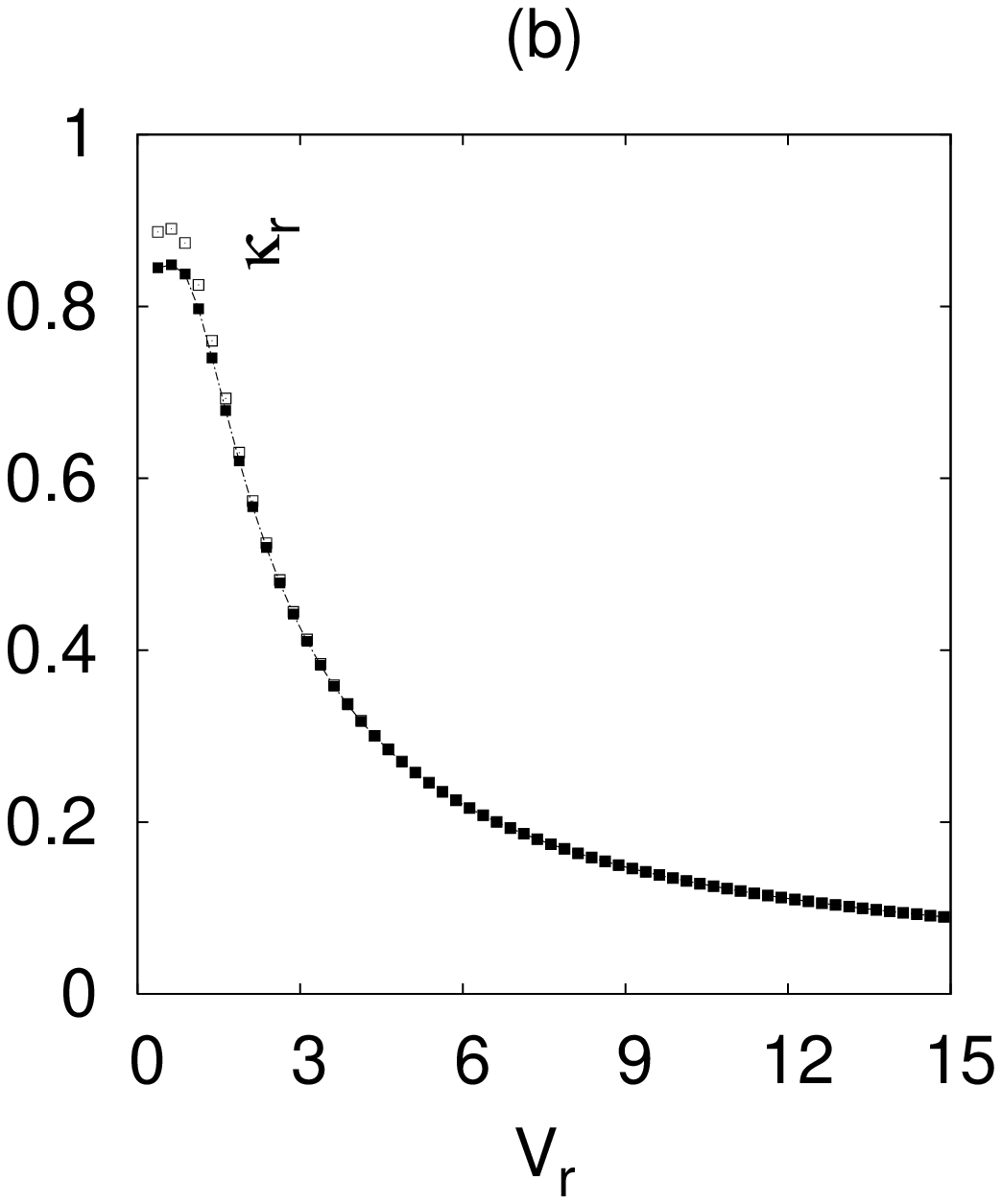}}}
\caption{\label{fig:crac} Plots of reduced
isothermal atomic compressibility $\kappa_r = \kappa_T/\kappa_0$
versus reduced interaction strength $V_r = V_{0,x}/E_0$ and $V_r = V_{0,s}/E_0$ for
a) SHS $p$-wave, and 
b) $s$-wave superfluids,
respectively, at $T = 0$ and $N_c = 0.25$.
Plots for $\alpha = 0$ and $0.1$ are shown with hollow and solid squares, respectively.
}
\end{figure}

Plots of the reduced isothermal atomic compressibility $\kappa_r = \kappa_T(T)/\kappa_0$ 
at low temperatures ($T \approx 0$) (see Eq.~\ref{eqn:zerokappa}) and quarter filling factor ($N_c = 0.25$) are shown
in Fig.~\ref{fig:crac} as a function of reduced interaction strength $V_r = V_{0,x}/E_0$ and $V_r = V_{0,s}/E_0$
for SHS $p$-wave and $s$-wave superfluids, respectively.
The tetragonal case (I) with $\alpha = 0$, and the orthorhombic case (II) with $\alpha = 0.1$
are shown with hollow and solid squares, respectively.

In the case of SHS $p$-wave superfluids, the evolution of $\kappa_r$
from BCS to BEC regime is not smooth, and a cusp occurs at critical interaction 
strengths corresponding to $\mu_r = -1$ in case (I) (hollow squares) and $\mu_r = -1.1$ 
in case (II) (solid squares). These cusps are associated with the appearance of 
a full gap in the quasi-particle excitation spectrum as the evolution
from BCS to BEC takes place.
In contrast, the evolution of $\kappa_r$ is smooth for $s$-wave superfluids, 
since the quasi-particle spectrum is always gapped during the BCS to BEC 
evolution.

\begin{figure} [htb]
\centerline{\scalebox{0.37}{\includegraphics{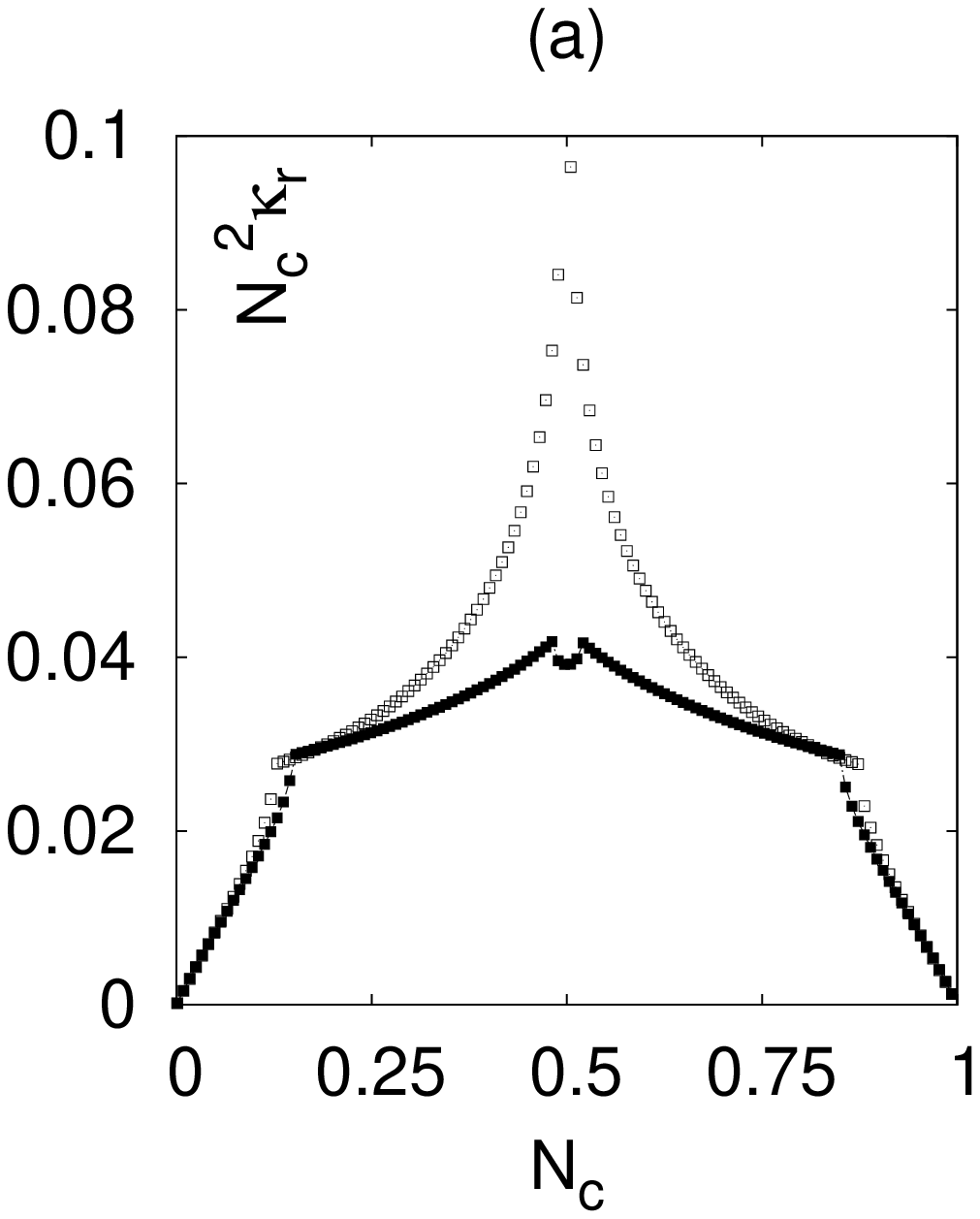} \includegraphics{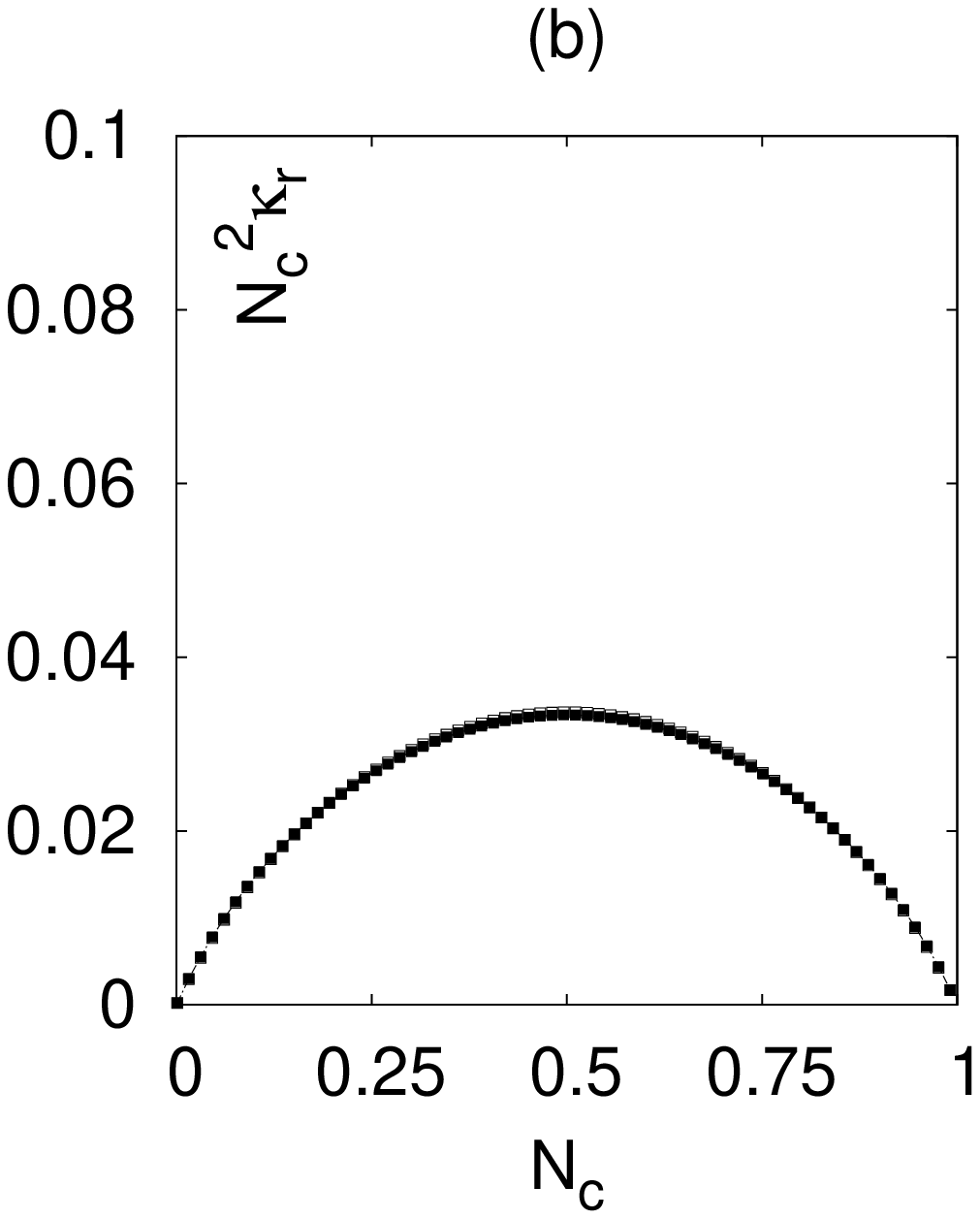}}}
\caption{\label{fig:crac.V.const} Plots of 
$N_c^2\kappa_r = dN_c/d\mu_r$ versus filling factor $N_c$ for 
a) SHS $p$-wave, and 
b) $s$-wave superfluids
at $T = 0$ and $V_r = 6$.
Plots for $\alpha = 0$ and $0.1$ are shown with hollow and solid squares, respectively.
}
\end{figure}

Plots of $N_c^2 \kappa_r = \partial N_c / \partial \mu_r$ 
at low temperatures ($T \approx 0$) and at fixed interaction strength ($V_r = 6$) are shown
in Fig.~\ref{fig:crac.V.const} as a function of filling factor $N_c$ 
for both SHS $p$-wave and $s$-wave superfluids.
The tetragonal case (I) with $\alpha = 0$ and orthorhombic case (II) with $\alpha = 0.1$
are shown with hollow and solid squares, respectively.

In the case of SHS $p$-wave superfluids, the evolution of 
$N_c^2 \kappa_r = \partial N_c / \partial \mu_r$ 
from the BCS to the BEC regime is not smooth. 
In both cases (I) (hollow squares) and (II) (solid squares), cusps occur for corresponding filling factors 
when $\mu_r = \pm 1$ and $\mu_r = \pm 1.1$, respectively.
These cusps occur again as a result of the gapless to gapped transition in the
quasi-particle excitation spectrum discussed above. Notice that the cusps appearing at
lower filling factors are associated with {\it particle} BCS-BEC transitions, 
while the cusps appearing at higher filling factors are associated with {\it hole} BCS-BEC transitions.
Furthermore, $N_c^2 \kappa_r$ has a sharp peak
at half-filling in case (I), and two small peaks in case (II) which are symmetric around half-filling.
These peaks occur inside the BCS region of the phase diagram shown in Fig.~\ref{fig:phase},
and they arise due to the same reasons discussed in Sec.~\ref{sec:atomic-compressibility},
where the BCS regime is extensively analysed.
In contrast, the evolution of $N_c^2 \kappa_r$ as a function of $N_c$ 
is again smooth for $s$-wave superfluids during the BCS to BEC evolution.

In cold Fermi gases the measurement of the isothermal compressibility $\kappa_T(T)$ 
at low temperatures is very hard, because most measurements are performed when traps are turned
off and the gas expands. However, the gas expansion process is probably close to being isentropic
and $\kappa_S$ is most likely measurable. As discussed in Sec.~\ref{sec:atomic-compressibility},
for $T \ll T_c$, the isentropic compressibility and the isothermal compressibility are essentially
proportional, and their measurements can serve as an indicator of the quantum phase transition
discussed in this section. Another important quantity that reflects such a transition is the
Cooper pair size to be discussed.

\subsection{Average Cooper Pair Size}
\label{sec2:cooper-pairsize}

The average Cooper pair size in the saddle point approximation is given by 
\begin{eqnarray}
\xi_{\rm{pair}}^2 = \langle \varphi(\mathbf{k})| (-\nabla^2_{\mathbf{k}}) |\varphi(\mathbf{k}) \rangle / 
\langle \varphi(\mathbf{k})| \varphi(\mathbf{k}) \rangle
\label{eqn:pairsize}
\end{eqnarray}
where $\varphi(\mathbf{k}) = \Delta(\mathbf{k})/2E(\mathbf{k})$ plays the role of the Cooper pair wave function. 
This quantity reflects the average size of a Cooper pair and not the coherence length $\xi_{\rm coh}$ of
the superfluid. Although these quantities are directly related in the BCS regime, they are very different
in the BEC regime~\cite{jan}. Here, however, we concentrate only on the analysis of $\xi_{\rm pair}$.

\begin{figure} [htb]
\centerline{\scalebox{0.37}{\includegraphics{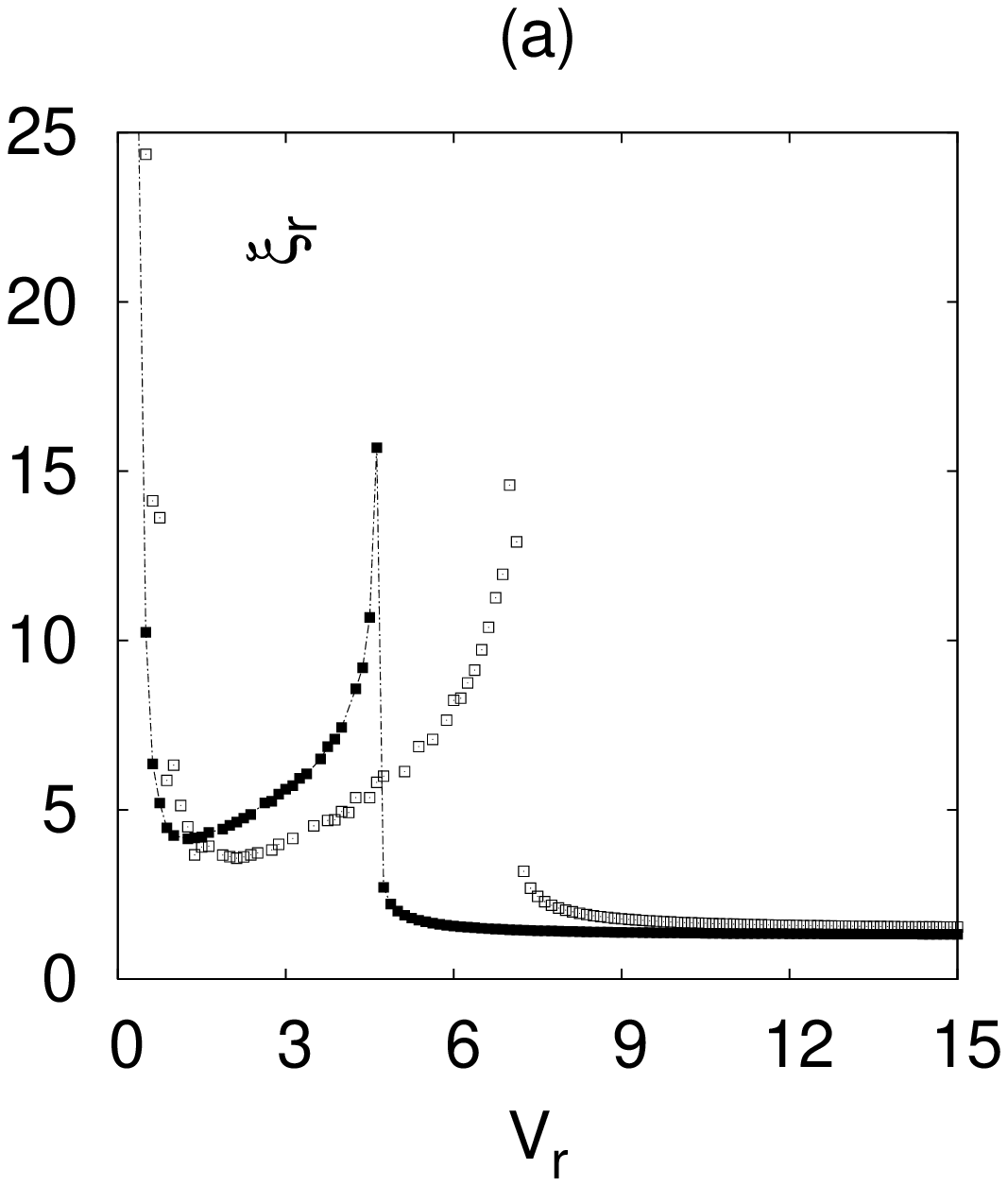} \includegraphics{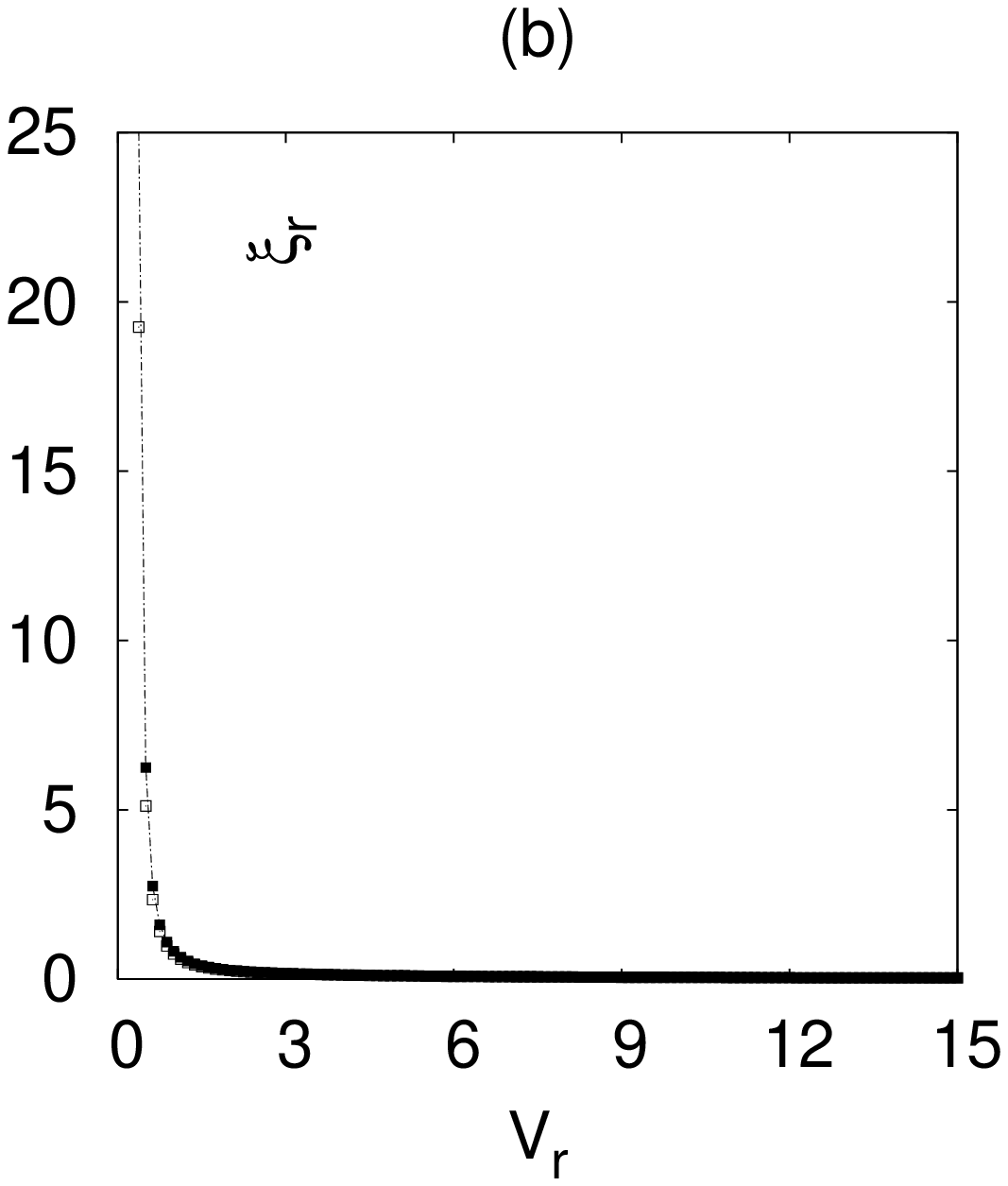}}}
\caption{\label{fig:pairsize} Plots of reduced average
Cooper pair size $\xi_r = \xi_{\rm{pair}}/a$
versus reduced interaction strength $V_r = V_{0,x}/E_0$ and $V_r = V_{0,s}/E_0$ for
a) SHS $p$-wave, and 
b) $s$-wave superfluids,
respectively, at $T = 0$ and $N_c = 0.25$.
Plots for $\alpha = 0$ and $0.1$ are shown with hollow and solid squares, respectively.
}
\end{figure}

Plots of the reduced average Cooper pair size $\xi_r = \xi_{\rm{pair}}/a$ 
at low temperatures ($T \approx 0$) and at quarter filling ($N_c = 0.25$) are shown
in Fig.~\ref{fig:pairsize}. The plots are shown as a function of reduced interaction 
strength $V_r = V_{0,x}/E_0$ and $V_r = V_{0,s}/E_0$
for SHS $p$-wave and $s$-wave superfluids, respectively.
The tetragonal case (I) with $\alpha = 0$, and the orthorhombic case (II) with $\alpha = 0.1$
are shown with hollow and solid squares, respectively.

In the case of SHS $p$-wave superfluids, the evolution of $\xi_r$
from BCS to BEC regime is not smooth, and singular behaviors occur at critical interaction 
strengths corresponding to $\mu_r = -1$ in case (I) (hollow squares) and $\mu_r = -1.1$ 
in case (II) (solid squares). These singular behaviors are associated with the appearance of 
a full gap in the quasi-particle excitation spectrum and with pairing in a
non-zero angular momentum as the evolution from BCS to BEC takes place. 

This singular behavior can be understood in terms of the vanishing of the momentum distribution
when $\tilde{\mu}$ smaller (larger) than the bottom (top) of the band for fixed filling factor $N_c$. 
For instance, in the case of $N_c = 0.25$ shown in Fig.~\ref{fig:pairsize} , when $\tilde{\mu}$ falls slightly below the bottom of
the band, suddenly $n(\mathbf{k})$ vanishes in the neighborhood of ${\mathbf k} = {\mathbf 0}$ 
which leads to unbound pairs of atoms with $({\mathbf k_1} = {\mathbf k} = {\mathbf 0 }; \uparrow)$ and 
$({\mathbf k_2} = -{\mathbf k}  = {\mathbf 0 }; \uparrow)$, and thus a rapid increase in the average pair size
$\xi_{\rm pair}$ for interaction strengths below or above the critical values. 
Beyond the critical interaction strength, $\xi_{\rm{pair}}$ decreases monotonically
for increasing $V_r$ and converges asymptotically 
(when $V_r \to \infty$ and $N_c < 0.5$) to a finite
value which is larger but of the order of the lattice spacing $a$. 
This is a manisfestation of higher angular momentum pairing and of the Pauli exclusion principle.

In contrast, the evolution of $\xi_r$ is smooth for $s$-wave superfluids,
since the momentum distribution $n(\mathbf{k})$ never vanishes at low ${\mathbf k}$, and
thus the distribution of pair sizes is well behaved. This monotonic decrease of
$\xi_{\rm pair}$ is also a reflection of a quasi-particle spectrum that is always gapped during the BCS to BEC 
evolution and of an order parameter for zero angular momentum pairing. 
Notice that, $\xi_r$ is decreasing monotonically as a function of interaction $V_r$
for $N_c = 0.25$. The limiting pair size for large $V_r$ and fixed $N_c$ is small in comparison
to the lattice spacing $a$, since the Pauli principle does not forbid atoms of opposite 
pseudo-spins to be on the same lattice site.

\begin{figure} [htb]
\centerline{\scalebox{0.37}{\includegraphics{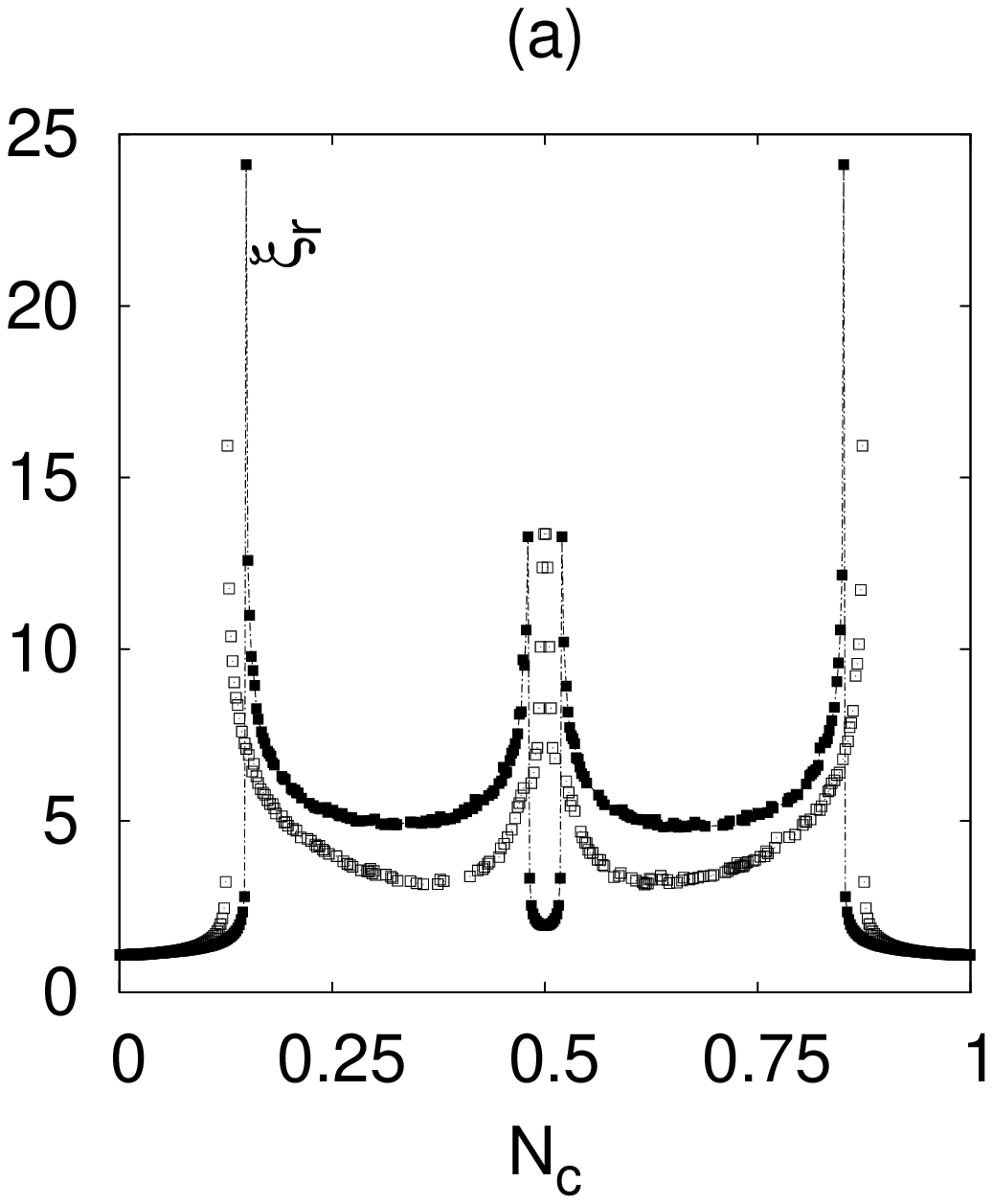} \includegraphics{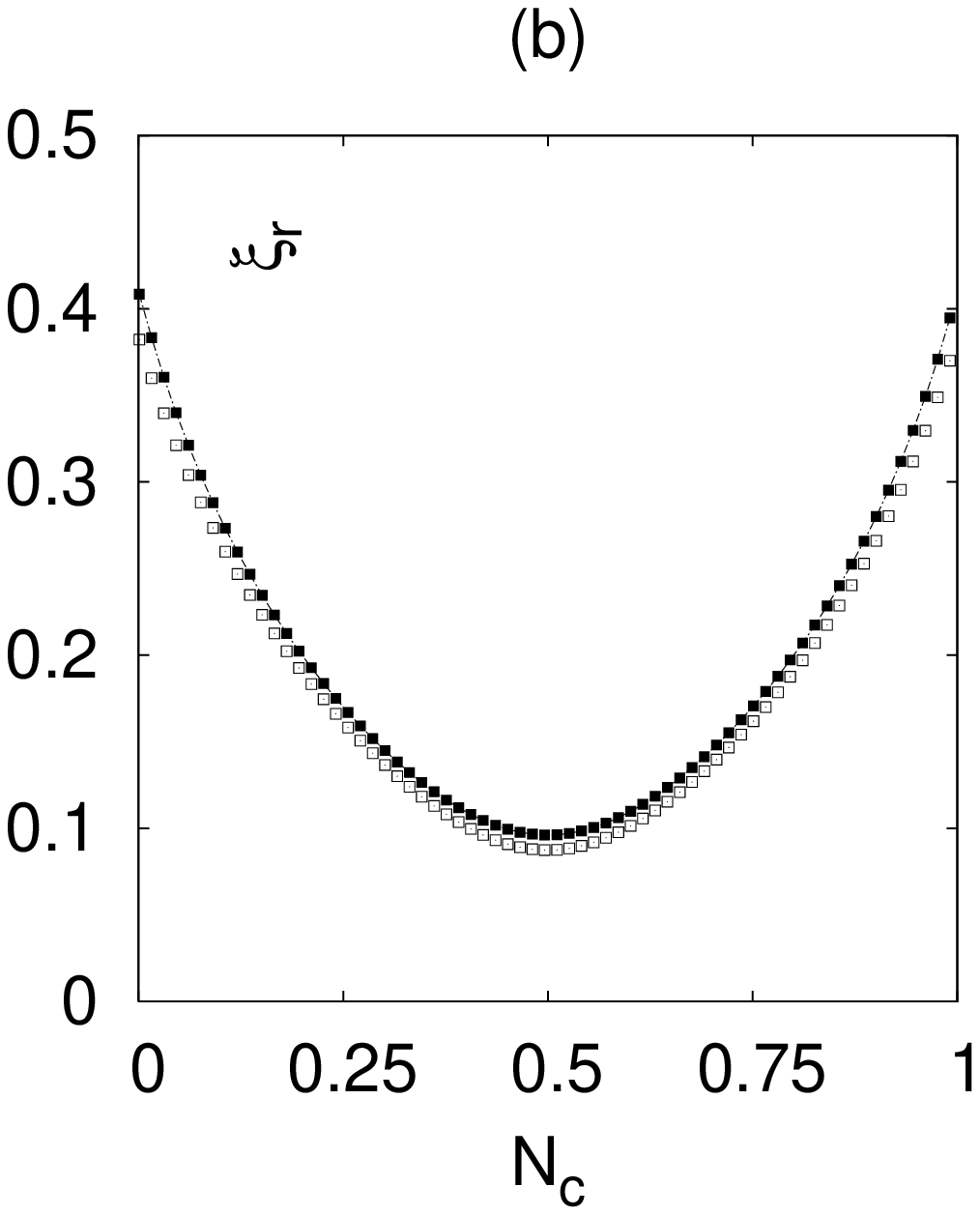}}}
\caption{\label{fig:pairsize.V.const} Plots of 
reduced average Cooper pair size $\xi_r = \xi_{\rm{pair}}/a$ versus filling factor $N_c$ for 
a) SHS $p$-wave, and 
b) $s$-wave superfluids
at $T = 0$ and $V_r = 6$.
Plots for $\alpha = 0$ and $0.1$ are shown with hollow and solid squares, respectively.
}
\end{figure}

Next, we discuss the behavior of $\xi_r$ as a function of filling factor $N_c$.
Plots of $\xi_r = \xi_{\rm{pair}}/a$ at low temperatures ($T \approx 0$) 
and at fixed interaction strength ($V_{0,x} = 6E_0$) are shown
in Fig.~\ref{fig:pairsize.V.const} as a function of $N_c$ 
for both SHS $p$-wave and $s$-wave superfluids.
The tetragonal case (I) with $\alpha = 0$ and orthorhombic case (II) with $\alpha = 0.1$
are again shown with hollow and solid squares, respectively.

In the case of SHS $p$-wave superfluids, the evolution of 
$\xi_r$ from BCS to BEC regime is not smooth. 
In both cases (I) (hollow squares) and (II) (solid squares), singularities occur for corresponding filling factors 
when $\mu_r = \pm 1$ and $\mu_r = \pm 1.1$, respectively.
These singularities occur again as a result of the gapless to gapped transition in the
quasi-particle excitation spectrum discussed above, as well as the symmetry of the order parameter,
which strongly modifies the nature of the pair wave function 
$\varphi(\mathbf{k}) = \Delta(\mathbf{k})/2E(\mathbf{k})$.
Notice that $\xi_r$ has a singularity at half-filling (BCS region) in case (I), 
and two other singularities symmetric around half-filling (BCS region) in case (II).
As the interaction is further increased, the low and high filling singularities (BCS-BEC boundaries) shift 
towards half-filling and merge with the half-filling singularity in case (I) (see also phase diagram in
Fig.~\ref{fig:phase}). This situation is analogous to the $d$-wave lattice case~\cite{hertog,andrenacci}. 
As the interaction is further increased, similar behavior occurs in case (II).
Thus, at large interaction strengths the BCS (BEC) region shrinks (expands), and the characteristic
value of $\xi_{\rm{pair}}$ is of the order of the lattice spacing $a$ except when $N_c$ is very close 
to half-filling, where the BCS-BEC phase boundary is located.

In contrast, the evolution of $\xi_r$ as a function of $N_c$ 
is again smooth for $s$-wave superfluids during the BCS to BEC evolution (see Fig.~\ref{fig:pairsize.V.const}b).
For instance, below half-filling $\xi_r$ decreases as a function of 
increasing $N_c$ (increasing DOS), while the pair wave function 
$\varphi(\mathbf{k})$ continues having the same qualitative behavior at low momenta.
Above half-filling, analogous discussions hold for holes.
Furthermore, $\xi_{\rm{pair}}$ decreases monotonically with increasing interaction for any filling
factor $N_c$, indicating that for very large interaction strengths $\xi_{\rm pair} \to 0$ for any $N_c$.

Thus far, we have investigated only momentum averaged quantities, but next we
discuss the momentum distribution, which could be measured in cold Fermi gases.

\subsection{Momentum Distribution}
\label{sec2:momentum-distribution}

The zero temperature momentum distribution 
\begin{equation}
n(\mathbf{k}) = \frac{1}{2} \left[ 1 - \frac {\xi(\mathbf{k})} {E(\mathbf{k})} \right]
\end{equation}  
is discussed in this section in the weak coupling  
BCS and strong coupling BEC regimes for both SHS $p$-wave and $s$-wave superfluids in a lattice.
We discuss here only the case of particle superfluidity with $N_c = 0.25$.
In the BCS regime, we choose $V_r$ corresponding to $\mu_r = -0.5$ for case (I) ($\alpha = 0$) 
and $\mu_r = -0.55$ for case (II) ($\alpha = 0.1$). 
Similarly in the BEC regime, we choose $V_r$ corresponding to $\mu_r = -2$ for case (I) and 
$\mu_r = -2.2$ for case (II).

\begin{figure} [htb]
\centerline{\scalebox{0.45}{\includegraphics{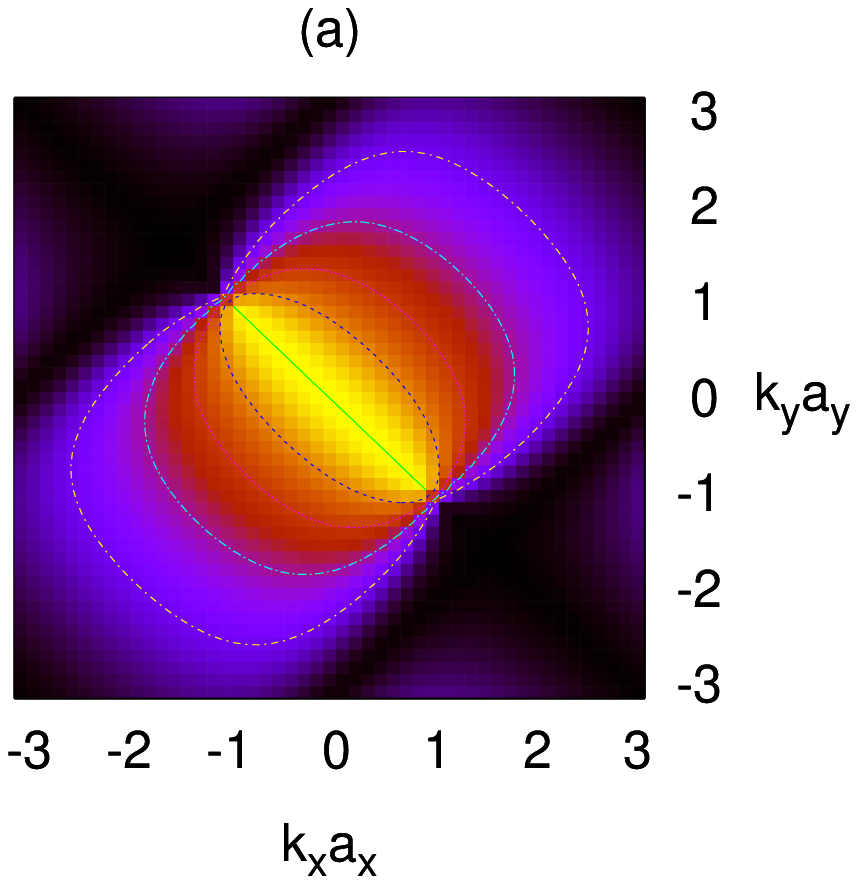} \includegraphics{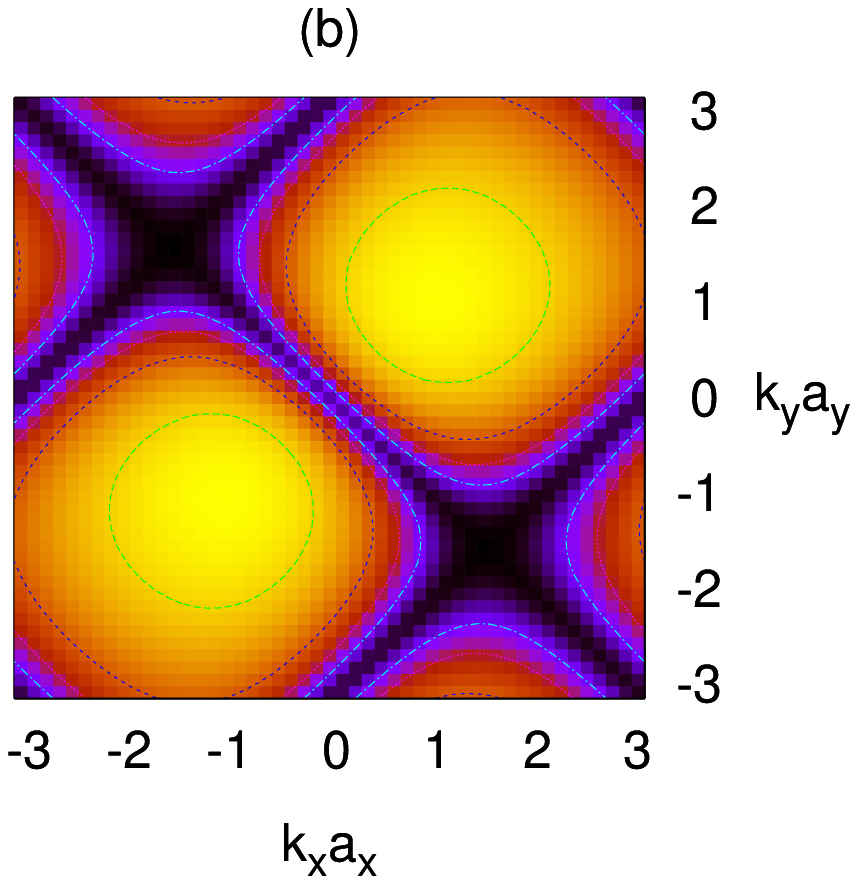}}}
\centerline{\scalebox{0.45}{\includegraphics{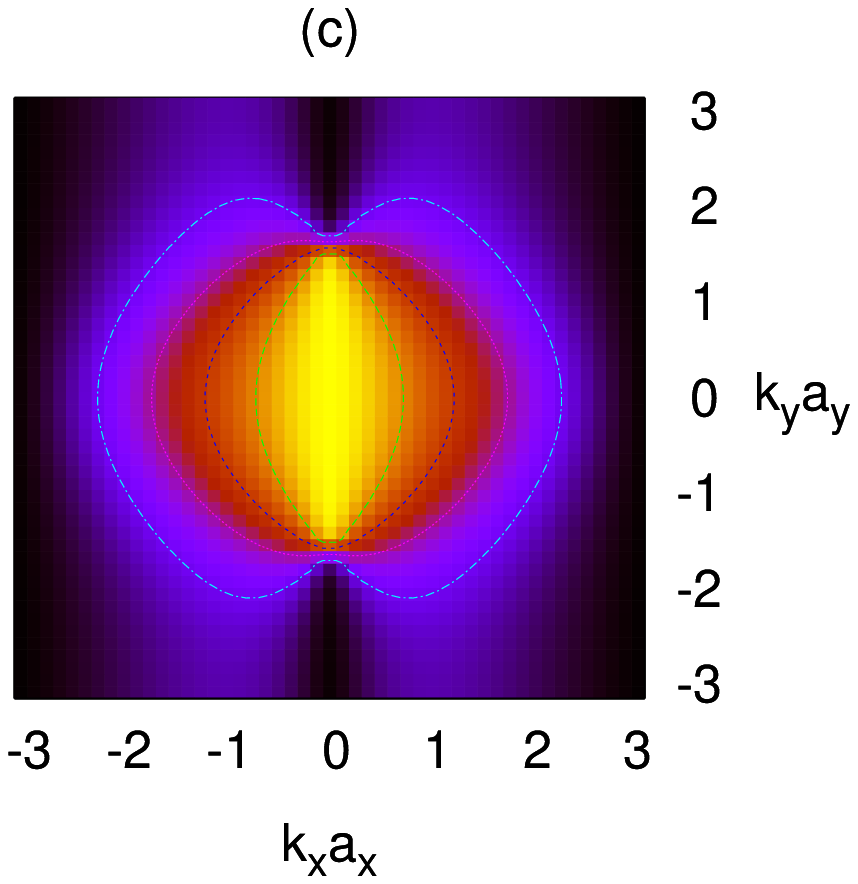} \includegraphics{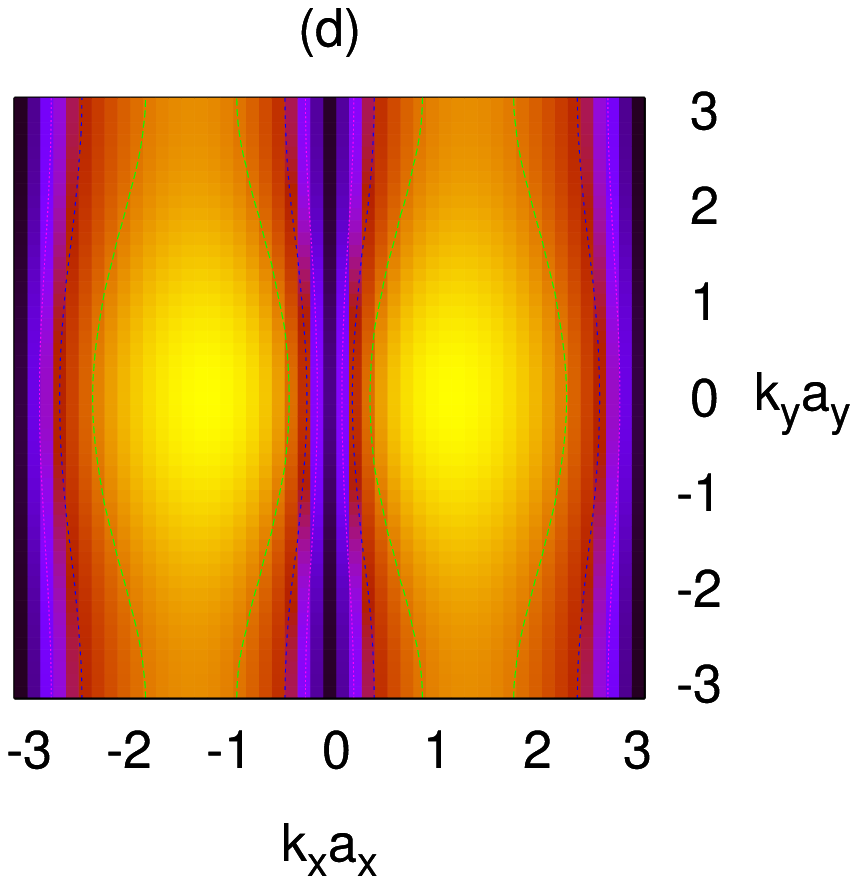}}}
\caption{\label{fig:md} (Color online) Plots of momentum distribution $n(\mathbf{k})$ 
versus $k_xa_x$ and $k_ya_y$ for an SHS $p$-wave superfluid in case (I) in
a) BCS ($\mu_r = -0.5$) and
b) BEC ($\mu_r = -2$) regimes. 
In addition plots for case (II) in
c) BCS ($\mu_r = -0.55$) and
d) BEC ($\mu_r = -2.2$)
regimes at $T = 0$.
Here the brighter the region, the higher the value of momentum distribution.
}
\end{figure}

In the case of SHS $p$-wave superfluids, the momentum distribution has a major rearrangement 
in $\mathbf{k}$-space with increasing interaction strength 
as can be seen in Fig.~\ref{fig:md}.
This rearrangement is very dramatic when $\tilde{\mu}$ crosses the bottom of the energy band 
$\tilde{\mu} = \tilde{\mu}_c$. This is a direct consequence of the change of the quasi-particle
excitation spectrum from gapless in the BCS regime ($|\tilde{\mu}| < (2 + \alpha)t$)
to fully gapped in the BEC regime ($|\tilde{\mu}| > (2 + \alpha)t$). 

For case (I), plots of $n({\mathbf k})$ (in the first Brillouin zone) for BCS 
and BEC regimes are shown in Fig.~\ref{fig:md}a and Fig.~\ref{fig:md}b, 
respectively.
Here the interaction strengths correspond to $\mu_r = -0.5$ for the BCS regime, and $\mu_r = -2$ 
for the BEC regime. The boundary line between the two regimes occurs when $\mu_r = -1$.
Notice that $n(\mathbf{k})$ is symmetric around the lines $k_ya_y = k_xa_x$ and 
$k_ya_y = -k_xa_x$ 
in both regimes as a reflection of the symmetry properties of $\xi ({\mathbf k})$ and 
$|\Delta ({\mathbf k})|$.

While the line $k_ya_y = -k_xa_x$ has the highest momentum distribution in the BCS regime, 
$n(\mathbf{k})$ vanishes along $k_ya_y = -k_xa_x$ in BEC regime. 
As the interaction strength increases, two-particle states with opposite momenta 
are taken out of the two-particle continuum into two-particle bound states with zero center
of mass momentum. As more of these tightly bound states are formed the large 
momentum distribution
in the vicinity of ${\mathbf k} = \mathbf{0}$ splits into two peaks around finite momentum 
values reflecting the $\ell = 1$ value of the angular momentum 
associated with these $p$-wave tightly bound states in the BEC regime.
For instance, in the case of $d$-wave ($\ell = 2$) superfluids the momentum distribution
in the BCS regime is centered around ${\mathbf k} = \mathbf{0}$ and splits into four peaks around finite 
momentum values reflecting the $\ell = 2$ value of the angular momentum associated 
with these tightly bound states in the BEC regime.~\cite{duncan}

For case (II), plots of $n({\mathbf k})$ for BCS 
and BEC regimes are shown in Fig.~\ref{fig:md}c and Fig.~\ref{fig:md}d, 
respectively. The figures shown correspond to interaction strengths associated 
to $\mu_r = -0.55$, and $\mu_r = -2.2$ in the BCS and BEC regimes, respectively.
The boundary line between the two regimes occurs when $\mu_r = -1.1$.
Notice that, $n(\mathbf{k})$ is symmetric around $k_xa_x = 0$ and $k_ya_y = 0$ in both regimes.
While the highest values of $n({\mathbf k})$ occur near ${\mathbf k} = {\mathbf 0}$ 
in the BCS regime, $n(\mathbf{k})$ vanishes near ${\mathbf k} = {\mathbf 0}$ 
and develop two maxima for finite value of $k_x$ in the BEC regime.

\begin{figure} [htb]
\centerline{\scalebox{0.45}{\includegraphics{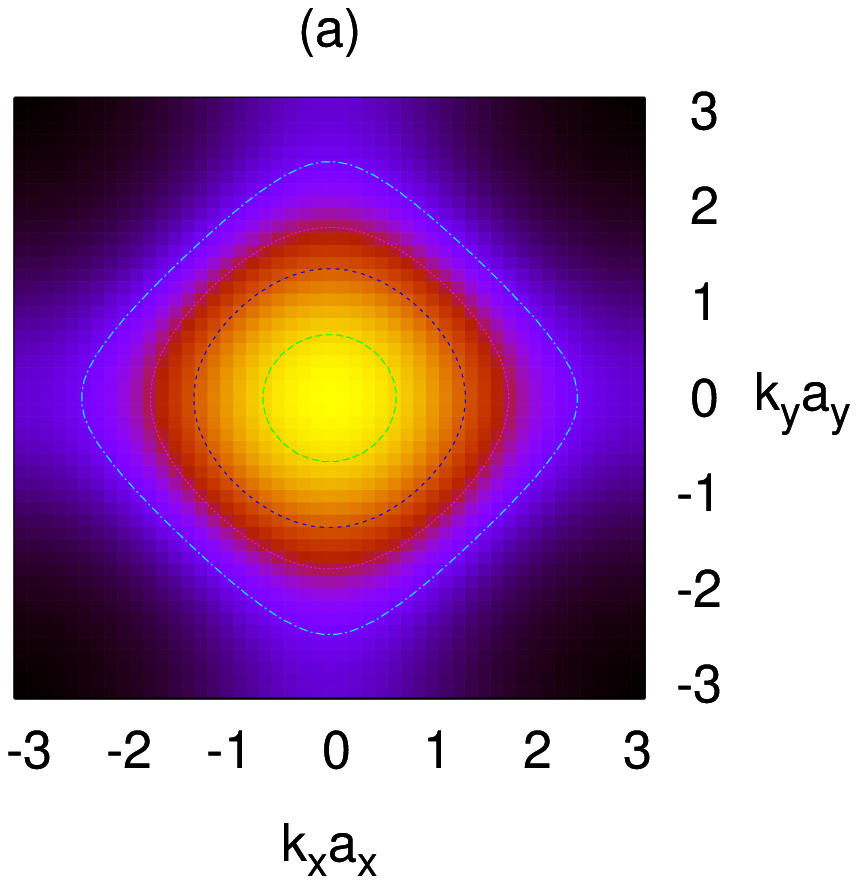} \includegraphics{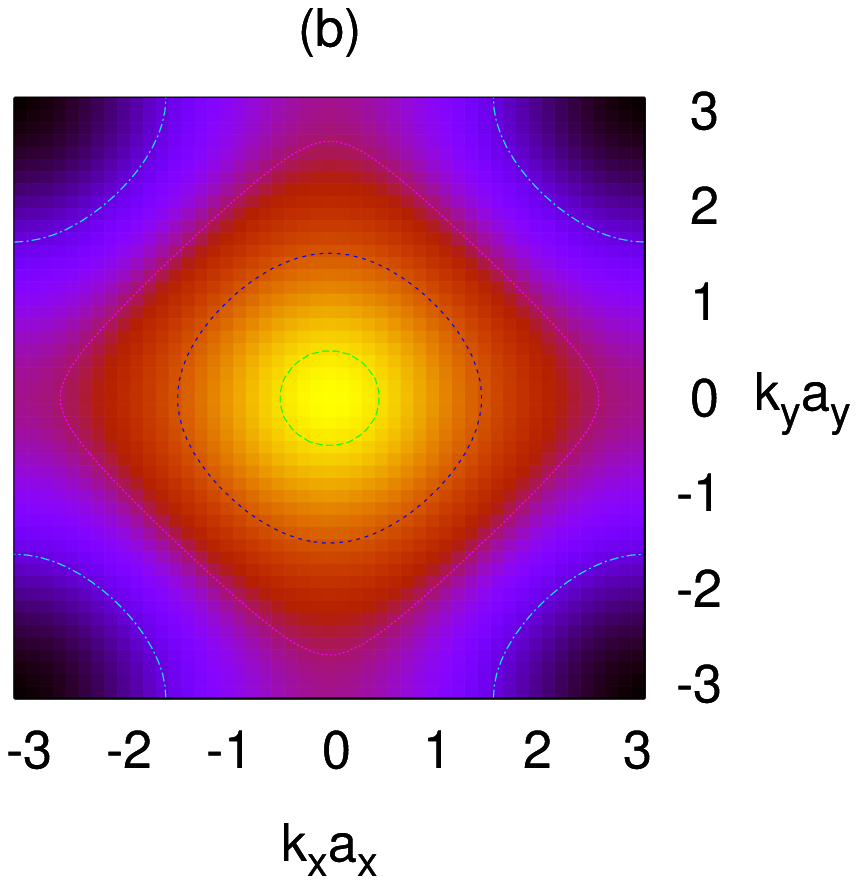}}}
\caption{\label{fig:md.swave} (Color online) 
Plots of momentum distribution $n(\mathbf{k})$ versus $k_xa_x$ and $k_ya_y$ for an $s$-wave superfluid in
case (I) in
a) BCS ($\mu_r = -0.5$) and 
b) BEC ($\mu_r = -2$)
regimes at $T = 0$.
Here the brighter the region, the higher the value of meomentum distribution.
}
\end{figure}

In contrast, the evolution of the momentum distribution from BCS to BEC regime 
is smooth for $s$-wave superfluids as shown in Fig.~\ref{fig:md.swave} for 
case (I). In case (II) the results are very similar to case (I), except for the expected 
anisotropy and we do not present them here.
Notice that for $s$-wave ($\ell = 0$) superfluids, the momentum distribution
in the BCS regime is centered around ${\mathbf k} = \mathbf{0}$,
and remains centered around $\mathbf{k} = \mathbf{0}$ even in the BEC regime.
This reflects the $\ell = 0$ value of the angular momentum associated 
with $s$-wave tightly bound states in the BEC regime.
Here, the essential qualitative difference in the BCS to BEC regimes
is that as the interaction gets stronger $n(\mathbf{k})$ broadens due to the 
formation of tightly bound states.

\section{Conclusions}
\label{sec:conclusions}

In summary, we considered SHS $p$-wave pairing of single hyperfine state 
and THS $s$-wave pairing of two hyperfine state Fermi gases in 
quasi-two-dimensional optical lattices. 
We focused mainly on superfluid SHS $p$-wave (triplet) states that break time-reversal, 
spin and orbital symmetries, but preserve total spin-orbit symmetry.

Since the paper was divided into two parts, we present our conclusions for each of these parts
separately. In the first part, 
we analysed superfluid properties of SHS $p$-wave and THS $s$-wave symmetries 
in the strictly weak coupling BCS regime. There, we calculate the order parameter, 
chemical potential, critical temperature, 
atomic compressibility, 
and superfluid density as a function of filling factor for tetragonal
and orthorhombic optical lattices.

We found that, for SHS $p$-wave superfluids, the critical temperatures 
in tetragonal and orthorhombic optical lattices
are considerably higher than in continuum model predictions, and therefore, experimentally 
attainable. In particular, we showed that a small anisotropy in the transfer energies increases the 
effective density of fermionic states of an SHS $p$-wave superfluid considerably around half-filling
which leads to higher critical temperatures.
In contrast, a small anisotropy decreases the 
density of states around half-filling, and therefore, the critical temperatures for 
an THS $s$-wave superfluid.
We also noticed that further anisotropy leads to a larger decrease in critical temperatures
for THS $s$-wave than for SHS $p$-wave superfluids.
Therefore, we concluded that a small anisotropy favours SHS $p$-wave pairing near half-filling, and 
that the critical temperatures for SHS $p$-wave superfluids are comparable and even higher than 
THS $s$-wave case for similar interaction parameters.

For SHS $p$-wave superfluids, we found a peak in the isothermal atomic compressibility at low temperatures,
exactly at half-filling of tetragonal lattices. This peak splits into two smaller peaks in 
the orthorhombic case. These peaks reflect the SHS $p$-wave structure of the order parameter 
at low temperatures and they are related to the nodes (zeros) of the quasi-particle energy spectrum. 
We also showed that the atomic compressibility peaks decrease in size as the critical 
temperature is approached from below. The peaks turn into humps at $T = T_c$. 
In contrast, for THS $s$-wave superfluids, we showed that the atomic compressibility does not show a peak 
structure at low temperatures since the quasi-particle energy spectrum is always gapped,
and that the compressibility is largely temperature independent.

We also discussed the superfluid density tensor for 
SHS $p$-wave and THS $s$-wave systems.
In the SHS $p$-wave case, we concluded that in orthorhombic lattices, 
the off-diagonal component $\rho_{xy}$ 
of the superfluid density tensor vanishes identically, while
the diagonal components $\rho_{xx}$ and $\rho_{yy}$ are different.
However, in tetragonal lattices, we showed that $\rho_{xy} \ne 0$, while $\rho_{xx} = \rho_{yy}$.
In contrast, for THS $s$-wave superfluids, 
$\rho_{xy} = 0$ and $\rho_{xx} \ne \rho_{yy}$ in the orthorhombic, while   
$\rho_{xy} = 0$ and $\rho_{xx} = \rho_{yy}$ in the tetragonal lattices.
Therefore, the presence of non-zero $\rho_{xy}$ in the square lattices is a key signature 
of our exotic SHS $p$-wave triplet state. 

In the second part of the manuscript,
we analysed superfluid properties of SHS $p$-wave and THS $s$-wave
superfluids in the evolution from BCS to BEC regime at low temperatures ($T \approx 0$). 
We discussed the changes in the quasi-particle excitation spectrum, chemical potential, 
atomic compressibility, momentum distribution and Cooper pair size
as a function of filling factor and interaction strength for tetragonal
and orthorhombic optical lattices.
We found major differences between SHS $p$-wave and THS $s$-wave superfluids 
in the evolution from BCS to BEC regime. 

In the case of SHS $p$-wave superfluids, the quasi-particle excitation spectrum,
chemical potential, atomic compressibility, Cooper pair size
and momentum distribution are not smooth functions
of filling factor or interaction strength. In particular, the singular behavior of the
atomic compressibility suggests the existence of a quantum phase transition when the chemical potential
crosses the bottom or the top of the energy band. This transition is associated with
the change in quasi-particle excitation spectrum and the higher angular momentum nature $(\ell = 1)$
of SHS $p$-wave superfluids, which are gapless in the
BCS regime, but fully gapped in the BEC regime.

In contrast, for THS $s$-wave superfluids, 
the quasi-particle excitation spectrum,
chemical potential, atomic compressibility, Cooper pair size and 
momentum distribution are smooth functions
of filling factor or interaction strength. In particular, the smooth behavior of the
atomic compressibility suggests only a crossover when the chemical potential
crosses the bottom or top of the energy band. This crossover is associated with
a fully gapped quasi-particle excitation spectrum and pairing
at the zero angular momentum channel in both BCS and BEC regimes.

These major differences between THS $s$-wave and SHS $p$-wave superfluids in an optical lattice 
suggest that SHS $p$-wave cold atoms are much richer than THS $s$-wave cold atoms,
and thus these systems may provide a new experimental direction in this field.

\section{Acknowledgement}
\label{sec:acknowledgement}

We would like to thank NSF (DMR-0304380) for financial support.

\end{document}